\documentclass[fleqn,usenatbib]{mnras}

\usepackage{times}

\usepackage{graphicx}	
\usepackage{amsmath}	
\usepackage{amssymb}	
\usepackage{adjustbox}
\usepackage{color}
\usepackage{verbatim}   
\usepackage{gensymb}
\usepackage{soul}

\usepackage{subfig}
\usepackage{cleveref}
\crefname{figure}{Fig.}{Figs.}
\crefname{table}{Table}{Tables}

\newcommand{\cmmnt}			   [1]{}

\newcommand{\ky}               [1]{#1}
\newcommand{\kyb}              [1]{#1}

\newcommand{\RTC}   {R_{\rm 200c}}
\newcommand{\MTC}   {M_{\rm 200c}}
\newcommand{\MSTARS}{M_{\rm stars}}
\newcommand{\MGAS}  {M_{\rm gas}}
\newcommand{\MHOST} {M^{\rm host}_{\rm 200c}}

\newcommand{\RGAL}  {r_{\rm 1/2}}

\newcommand{\mach}{\mathcal{M}}
\newcommand{\units}[1]{\, \mathrm{#1}}
\newcommand{\pderiv}[2]{\frac{\mathrm{\partial} #1}{\mathrm{\partial} {#2}}}
\newcommand{\rfsec}[1]{\mbox{\S\ref{sec:#1}}}
\newcommand{\kboltz}{k_{\mathrm{B}}}
\newcommand{\mode}[1]{\mathrm{Mode}\left(#1\right)}

\newcommand{\zeq}[1]{\mbox{$z=#1$}}

\newcommand{\MSUN}{\units{M_{\odot}}}

\DeclareGraphicsExtensions{.jpg,.png,.pdf}

\title[Jellyfish galaxies in TNG]
{Jellyfish galaxies with the IllustrisTNG simulations: \\ I. Gas-stripping phenomena in the full cosmological context}
\author[Yun et al.]{Kiyun Yun,$^{1}$\thanks{E-mail: yun@mpia-hd.mpg.de}
Annalisa Pillepich,$^{1}$
Elad Zinger,$^{1}$
Dylan Nelson,$^{2}$ Martina Donnari,$^{1}$ \newauthor 
Gandhali Joshi,$^{1}$ 
Vicente Rodriguez-Gomez,$^{3}$ Shy Genel,$^{4,5}$ Rainer Weinberger,$^{6}$ \newauthor 
Mark Vogelsberger$^{7}$ and Lars Hernquist$^{6}$
\\\\ 
$^{1}$Max-Planck-Institut f\"ur Astronomie, K\"onigstuhl 17, D-69117 Heidelberg, Germany\\
$^{2}$Max-Planck-Institut f\"ur Astrophysik, Karl-Schwarzschild-Str. 1, D-85741 Garching, Germany\\
$^{3}$Department of Physics \& Astronomy, Johns Hopkins University, 3400 N. Charles Street, Baltimore, MD 21218, USA\\
$^{4}$Center for Computational Astrophysics, Flatiron Institute, 162 Fifth Avenue, New York, NY 10010, USA\\
$^{5}$Columbia Astrophysics Laboratory, Columbia University, 550 West 120th Street, New York, NY 10027, USA\\
$^{6}$Harvard-Smithsonian Center for Astrophysics, 60 Garden Street, Cambridge, MA 02138, USA\\
$^{7}$Kavli Institute for Astrophysics and Space Research, Massachusetts Institute of Technology, Cambridge, MA 02139, USA\\
}

\date{Accepted XXX. Received YYY; in original form ZZZ}

\pubyear{2018}

\begin{document}
\label{firstpage}
\pagerange{\pageref{firstpage}--\pageref{lastpage}}
\maketitle

\begin{abstract}

We use the IllustrisTNG simulations to study the demographics and properties of jellyfish galaxies in the full cosmological context. By jellyfish galaxies, we mean satellites orbiting in massive groups and clusters that exhibit highly asymmetric distributions of gas and gas tails. In particular, we select TNG100 galaxies at low redshifts ($z \le 0.6$) with stellar mass exceeding $10^{9.5} \MSUN$ and with host halo masses in the range $10^{13} \leq \MTC/\MSUN \leq 10^{14.6}$. Among more than about 6 000 (2 600) galaxies with stars (and some gas), we identify 800 jellyfish galaxies by visually inspecting their gas and stellar mass maps in random projections. Namely, about 31 per cent of cluster satellites are found with signatures of ram-pressure stripping and gaseous tails stemming from their main luminous bodies. This is a lower limit: the random orientation entails a loss of about 30 per cent of galaxies that in an optimal projection would otherwise be identified as jellyfish. Furthermore, jellyfish galaxies are more frequent at intermediate and large cluster-centric distances ($ r/\RTC \gtrsim 0.25$), in more massive hosts and at smaller satellite masses, and they typically orbit supersonically. The gaseous tails usually extend in opposite directions to the galaxy trajectory, with no relation between tail orientation and position of the host’s center. Finally, jellyfish galaxies are late infallers  ($< 2.5-3$ Gyrs ago, at $z=0$) and the emergence of gaseous tails correlates well with the presence of bow shocks in the intra-cluster medium. 

\end{abstract}

\begin{keywords}
methods: numerical -- galaxies: evolution -- galaxies: clusters: general -- galaxies: groups: general -- galaxies: clusters: intracluster medium -- cosmology: theory\end{keywords}

\section{Introduction}
\label{sec1:intro}

The study of the evolutionary transformations of galaxies, particularly with large samples, has the promise to shed light on the past and fate of both the Universe as a whole, on the largest scales, as well as of our Milky Way and local neighborhood, on smaller scales. 
Despite the inherent complexity and multitude of relevant physical mechanisms, it has been recognized for several decades that the processes that transform galaxies broadly fall into two categories: internal (also known as secular) and external (or environmental).
Internal processes are mostly related to those interactions among a galaxy's constituents that happen inside a galaxy itself, e.g. feedback from the central supermassive black holes or active galactic nuclei (AGN), explosions of supernovae (SNe), star formation and stellar winds, stellar radial migration, etc. On the other hand, external processes occur because galaxies do not generally form and evolve in isolation, but can interact with other galaxies and with the cosmic dark matter and gas that permeates their surroundings. Examples of environmental processes are galaxy-galaxy mergers \citep{Toomre1972GalacticTails}, ram-pressure and tidal stripping \citep{Gunn1972OnEvolution,Dressler1983SpectroscopyCluster}, harassment \citep{Moore1995GalaxyGalaxies}, and starvation or strangulation \citep{Larson1980}.

Many environmental processes are typically in action in the high-density environments that can be found within the gravitational potential wells of massive groups and clusters of galaxies, by which we generally mean gravitationally-collapsed structures with total mass exceeding about $10^{13}\MSUN$. The pervasive gas of the intra-group and intra-cluster volumes and the proximity to other galaxies, especially the massive brightest cluster galaxies that dominate the potential, affect the lives and properties of satellite galaxies that orbit within such massive haloes. For instance, vigorous environmental interactions have been observationally suggested to result in dramatic morphological transformations, producing for example the density-morphology relation of cluster galaxies \citep{Dressler1980AGalaxies}, for which the fraction of spheroidal-like galaxies is higher towards cluster centers rather than in their outskirts. Overall, modifications to the morphology, star formation activity, color, and stellar mass and gas content of group and cluster galaxies in comparison to similar galaxies in the field has been observed both in small samples of well-resolved galaxies \citep[e.g.][]{Haynes1984TheGalaxies,Couch1987A0.31,Ellingson2001TheInfall, Boselli2006EnvironmentalClusters,Jaffe2015BUDHIESGalaxies} as well as in statistical samples of large surveys like the Sloan Digital Sky Survey \citep[SDSS, ][]{Kauffmann2004, Peng2010} and the Galaxy And Mass Assembly survey \citep[GAMA, ][]{Gama12015}.

Among observations at high spatial resolution, extreme examples of galaxies undergoing transformations because of the interaction with their environment have been identified, mostly in galaxy clusters and at low redshifts. In particular, galaxies exhibiting highly asymmetric distributions in the continuum or in the gas or clear signatures of ram-pressure stripping have been called `jellyfish' galaxies \citep[][via observations with the Hubble Space Telescope, HST, and the Multi Unit Spectroscopic Explorer, MUSE, respectively]{Ebeling2014Jellyfish:Clusters,Fumagalli2014} because of the extended gaseous `tails', `wakes' or {\it `tentacles'} stemming from their main luminous bodies. In fact, transitional galaxies in the act of being stripped of their gas had been observed across a wide range of wavelengths and gas phases: in HI \citep[e.g.][]{Vollmer1999, Kenney2004, Chung2007VirgoTails, Jaffe2015BUDHIESGalaxies}, H$\alpha$ \citep[e.g.][]{Yoshida2002, Yagi2010ACluster,Vulcani2015,Fossati2016, Sheen2017}, in cold and molecular gas \citep[e.g.][]{Vollmer1999, Jachym2014, Jachym2017}, in X-ray \citep[][]{Forman1979, Randall2008, Su2017}, and also in the optical and UV \citep[e.g.][]{ Owen2006, Cortese2007, McPartland2016Jellyfish:Clusters, Poggianti2016JELLYFISHREDSHIFT, Vulcani2018GASPFormation}. 

More recently, ambitious programs like GASP \citep[GAs Stripping Phenomena in galaxies with MUSE,][]{Poggianti2017} and VESTIGE \citep[the Virgo Environmental Survey Tracing Ionised Gas Emission, with the MegaCam at the Canada-France-Hawaii Telescope,][]{Boselli2018} are returning unprecedentedly-large samples of tens of jellyfish galaxies detected in ionized gas. Such observations are convincingly showing that ram-pressure stripping and the hot and dense intra-cluster medium (ICM) are mainly responsible for the formation of jellyfish galaxies, a hypothesis supported by theoretical studies \citep{Gullieuszik2017GASP.JO204}.  

Theoretical works based on numerical simulations \citep[e.g.][]{Roediger2007RamDisc,Bekki2009Ram-pressureEnvironments,Tonnesen2010, Smith2013TheGalaxies,Bekki2014GalacticClusters} have enabled detailed and extensive studies of the formation of jellyfish galaxies that are otherwise infeasible with observations. For instance, wind-tunnel simulations have clearly shown that a fast-moving and dense medium can strip the gas from a galaxy's body, hence producing gaseous tails similar to those seen in observations, for both disk and elliptical galaxies \citep[e.g.][]{Tonnesen2010, He2012GasStudy, Roediger2015STRIPPEDELLIPTICALS, Roediger2015STRIPPEDM89, Steinhauser2016SimulationsInteractions}. The ram pressure generated by the moving medium operates as a pressure force that causes the gas bounded by the potential well of the galaxy to move out and to be stretched asymmetrically in one direction. Simulations have shown that environmentally-driven interactions can play a significant role in deforming the distribution of gas in a galaxy, the details also possibly dependent on the magnetisation of both the medium and galaxy gas \citep{Tonnesen2014}.

Wind-tunnel tests have become progressively more sophisticated in both numerical resolution and in the physical mechanisms included within their test galaxies; furthermore, they guarantee a high-level control and fine-grained manipulation of the physical processes at play. However, they are inevitably constrained by their limited initial and boundary conditions, i.e.\@ setups that cannot practically reproduce the enormous variety of galaxy properties, ICM properties, orbital configurations, and evolving conditions of group and cluster galaxies that are observed in the Universe. To overcome these limitations, cosmological simulations that develop from the initial conditions left just after the Big Bang and Inflation and that follow the coevolution of hundreds and thousands of galaxies in different environments are of the essence. 

Modern hydrodynamical simulation campaigns of large cosmological volumes like Illustris \citep{Vogelsberger2014IntroducingUniverse,Vogelsberger2014PropertiesSimulation,Genel2014IntroducingTime}, EAGLE \citep[Evolution and Assembly of GaLaxies and their Environments, ][]{Schaye2015TheEnvironments,Crain2015TheVariations}, and earlier GIMIC \citep[Galaxies -- Intergalactic Medium Interaction Calculation,][]{Crain2009}, have been used to investigate and interpret the statistical properties of galaxies in different environments and to contrast properties of satellites vs.\@ central galaxies. However, such studies \citep[e.g.][]{Sales2015, Mistani2016, Genel2016, Bahe2012, Bahe2013, Bahe2015, Jung2018} have mostly focused on integral properties of group and cluster galaxies, without providing insight about their structural morphologies. Others, instead, have examined lower-mass halo hosts, including studies based on zoom-in simulations of Milky-Way like galaxies  \citep[e.g. with NIHAO, Auriga or APOSTLE;][]{Buck2018, Simpson2018, Fattahi2018}.
With this paper we aim to fill this gap and use the IllustrisTNG simulations \citep[Illustris-\emph{The Next Generation},][]{Pillepich2018FirstGalaxies,Naiman2018FirstEuropium, Nelson2018FirstBimodality,Marinacci2018,Springel2018FirstClustering} to provide the first comprehensive numerical study of jellyfish galaxies in groups and clusters in the full cosmological context. 

IllustrisTNG is a suite of simulations that includes gravity, magneto-hydrodynamics (MHD) and baryonic processes and that hence model the formation and evolution of galaxies within the $\Lambda$CDM paradigm. As extensively quantified in \citet{Pillepich2018FirstGalaxies}, the highest-resolution run that is currently available, TNG100, samples thousands of well-resolved galaxies orbiting in hundreds of massive group- and cluster-size haloes, hence providing a rich set of simulated galaxies -- spanning a wide range of properties and orbital configurations -- to study the effects of high-density environments and stripping phenomena. 

Among the questions we aim to address are the following. What is the frequency of satellite galaxies that exhibit highly-asymmetric gas distributions? Do jellyfish galaxies naturally emerge in the full-physics cosmological simulations and, if so, what are the physical processes responsible for their formation? What is their distribution and incidence in terms of satellite and host mass and as a function of cluster-centric distance? How does their frequency evolve with redshift? What physical conditions are needed to have such strongly ram-pressure stripped galaxies? i.e.  is the interstellar medium (ISM) and/or the circumgalactic medium (CGM) of satellite galaxies stripped? And what are the orbital conditions for satellite galaxies to exhibit gas tails at the time of observations (e.g.\@ satellite speed, infall times, and physical properties of the surrounding ICM)? 

This paper attempts to answer these questions and is structured as follows. In \rfsec{sims}, details of the cosmological simulation used in this study are provided together with the criteria adopted to select our simulated galaxy samples. We also describe the operational measurement of relevant physical quantities, like the Mach number of the satellite motions through the ICM, the ram pressure they experience, and the characterization of their gaseous asymmetries or tails. The visual identification of jellyfish galaxies in TNG100 is extensively described in \rfsec{3visual}, together with a number of galaxy stamps depicting gas density and stellar mass projections extracted from our cosmological volumes. 
In \rfsec{4. Results}, we demonstrate four main results about jellyfish galaxies: their demographics in redshift and in satellite and host mass, their distributions in the cluster phase-space and Mach number vs.\@ ram pressure diagrams, and the fundamental characteristics of their tail directions. In \rfsec{5. Discussion and Implications}, we discuss our findings and expand upon them by quantifying the limitations of random orientation for jellyfish-galaxy identification and by uncovering an observationally-viable connection between supersonic motions and bow shocks in the ICM. Finally we summarize and conclude in \rfsec{summary}.


\section{Simulations, Methods, and definitions}
\label{sec:sims}

\subsection{The IllustrisTNG simulations}
\label{sec:tng}

To search for and study jellyfish galaxies we use the outcome of the TNG100 simulation, one of the three flagship simulations of the IllustrisTNG project 
\citep[lllustris-\emph{The Next Generation},][]{ Nelson2018FirstBimodality,Marinacci2018,Springel2018FirstClustering,Pillepich2018FirstGalaxies,Naiman2018FirstEuropium}.

IllustrisTNG\footnote{www.tng-project.org} is a suite of gravity and magneto-hydrodynamical simulations of uniform large volumes, of varying size and resolution, run with the moving-mesh code {\sc arepo} \citep{Springel2010} within the framework of a $\mathrm{\Lambda CDM}$ cosmological model (with the following parameters: matter density $\Omega_{\rm m}=\Omega_{\rm dm}+\Omega_{\rm b}=0.3089$, baryonic density $\Omega_{\rm b}=0.0486$, cosmological constant $\Omega_{\rm \Lambda}=0.6911$, Hubble constant $H_0=100 h\units{km\,s^{-1}\, Mpc^{-1}}$ with $h= 0.6774$, normalization $\sigma_8 = 0.8159$ and spectral index $n_s=0.9667$). 

They follow the co-evolution of cold dark matter (DM), cosmic gas, and magnetic fields from high redshift to the current time, while including many physical processes that lead to the formation of galaxies and their co-evolution. These include gas density-threshold star formation; evolution of stellar populations represented by star particles; chemical enrichment of the interstellar medium (ISM) and the tracking of nine different chemical elements (H, He, C, N, O, Ne, Mg, Si, Fe); gas cooling and heating; feedback from supernovae in the form of galactic winds; seeding and growth of supermassive black holes and the injection of energy and momentum from them into the surrounding gas. The IllustrisTNG galaxy-formation model \citep[see][for details]{Weinberger2017, Pillepich2018SimulatingModel} is an updated version of the Illustris one \citep{Vogelsberger2013APhysics,Torrey2014AValidation}.

In these calculations, the hierarchical growth of haloes and galaxies, galaxy mergers, cosmic gas accretion into haloes, tidal and ram-pressure stripping, and dynamical friction all occur naturally, following from the solution of the equations of gravity and hydrodynamics in an expanding Universe with gravitationally-collapsing structures. The {\sc arepo} code combines both the Galilean-invariance typical of Lagrangian codes with high-order flux calculations typical of Eulerian methods: it hence leads to a more numerically-accurate treatment of the interactions between fast-moving fluids and of shocks. The improved handling of fluid instabilities in {\sc arepo}, compared to e.g.\@ classical smoothed particle hydrodynamics codes, ensures a better description of the generation of entropy due to mixing, of vorticity in curved shocks and leads to more efficient and realistic gas stripping from infalling substructures \citep{Sijacki2012, He2012GasStudy}.

The choices underlying the feedback schemes in IllustrisTNG are such that basic, fundamental, observed galaxy properties and statistics are reproduced to a reasonable degree of accuracy, e.g.\@ the galaxy stellar mass function, the total gas mass content within the virial radius of massive haloes, and the stellar mass -- size relation, all at $z=0$ \citep[see][for a full list and discussion]{Pillepich2018SimulatingModel}. In fact, the IllustrisTNG model has been shown to return reasonable, observationally consistent results beyond the ones adopted to develop the model, both from a statistical perspective as well as in the galaxy structures: e.g.\@ the emergence of a population of quenched galaxies \citep{Weinberger2018SupermassiveSimulation}, the shapes and widths of the red sequence and the blue cloud as compared to SDSS galaxies \citep{Nelson2018FirstBimodality}, the spatial clustering of both red and blue galaxies separately from tens of kpc to tens of Mpc separations \citep{Springel2018FirstClustering}, the stellar sizes of low and intermediate mass galaxies up to $z\sim2$ \citep{Genel2018ARelations}, the distribution of metals in the intra-cluster plasma \citep{Vogelsberger2018TheSimulations}, the evolution of the gas-phase metallicity vs.\@ galaxy mass relation \citep{Torrey2018SimilarRelation}, the OVI content of the circumgalactic media around galaxies \citep{Nelson2018TheIllustrisTNG}, and even the observed fraction of cool core clusters \citep{Barnes2018EnhancingConduction}.
 This makes the IllustrisTNG simulations ideal tools to further investigate specific galaxy-evolutionary scenarios, like the loss of gas in orbiting group and cluster galaxies. 

In this study, we mainly focus on the results of the TNG100 simulation, a $111\units{Mpc}$ box with a gas-cell and stellar-particle mass resolution of $1.4\times 10^6\MSUN$, \ky{kept fixed at all studied redshifts. In TNG100, the average gas cell size within the star-forming regions of galaxies is of a few hundreds of parsecs and the gravitational softening length of all resolution elements is at most 755 pc, kept constant in the $z=0 - 1$ range} \ky{\citep[see Appendix A of][for more details on the resolution characteristics of the TNG100 run]{Nelson2018TheIllustrisTNG}}. This was run with the same resolution and initial conditions (except for small changes in the cosmological parameters) as the original Illustris simulation \citep{Vogelsberger2014IntroducingUniverse,Vogelsberger2014PropertiesSimulation,Genel2014IntroducingTime} but with updated aspects of the galaxy-physics model, chiefly the modelling of magnetic fields, the inclusion of black-hole driven winds, and modifications to the galactic-scale winds driven by supernovae. At the current epoch, TNG100, contains $\sim 10$ Virgo-sized galaxy clusters ($\MTC \sim 1-4 \times 10^{14} \MSUN$) and about 2\,000 Milky Way-mass haloes as centrals, in addition to thousands of less-massive galaxies, both satellites of larger objects as well as in the field (see \citealt{Pillepich2018SimulatingModel} for an overview of the most massive objects and their stellar mass contents in TNG100).

\subsection{The galaxy sample and definitions}
\label{sec:sample}

Distinct haloes, and the subhaloes which inhabit them, are identified with the Friends-of-Friends \citep{Davis1985} and {\sc subfind} \citep{SpringelPopulating0} algorithms, respectively. 

For each FoF host halo, we define a `virial' radius, $\RTC$, as the radius within which the total mean density is equal to a factor of 200 times the critical density of the Universe, centred at the position of the most-bound resolution element within the FoF. The total mass enclosed within this radius is denoted as $\MHOST$. We call, interchangeably, groups or clusters of galaxies or massive hosts those (FoF) haloes with total mass exceeding $10^{13}\MSUN$. At $z=0$ alone, TNG100 samples 182 hosts above such total-mass limit, the most massive one reaching $3.8\times 10^{14}\MSUN$ \citep{Pillepich2018FirstGalaxies}.

Within each FoF halo, subhaloes identified by the \textsc{subfind} algorithm contain all the resolution elements (gas cells, stellar particles, DM particles, etc.\@) which are gravitationally bound to the subhalo: these elements are by extension members of the FoF host halo as well. In our framework, galaxies are subhaloes with non vanishing stellar mass: throughout this paper, a galaxy stellar mass is the sum of all stellar particles within twice the stellar half-mass radius ($\RGAL$). The position of a galaxy is the position of the most-bound resolution element in the subhalo which, often but not always, corresponds to the center of the galactic stellar distribution.  

As we are interested in examining the environmental processes acting on satellite galaxies within massive groups and clusters, we limit our study to a sample of galaxies found in host FoF haloes of mass $10^{13}\MSUN$ or greater, and at a distance of no more than one virial radius ($\RTC$) from the host center: these are called satellite galaxies and we exclude from this analysis central galaxies, namely those subhaloes whose position coincides with the FoF center. We consider only well-resolved galaxies in our sample, i.e.\@ galaxies that are resolved by at least about 3\,000 stellar particles (in addition to gas and DM elements): this corresponds to a lower limit on the stellar mass of $10^{9.5}\MSUN$.

However, to search for galaxies with signatures of gas stripping, we will restrict the sample to a sub-sample of galaxies which contain at least some gas within the galactic radius (twice the stellar half-mass radius). 

In what follows and to summarize, we define a sample of `all satellites' containing all the TNG100 satellite galaxies which obey the following criteria:

\begin{enumerate}
\item $\MHOST \ge 10^{13}\MSUN$
\item Galaxy position within host $\le \RTC$ \ky{(in 3D)}
\item $\MSTARS \ge 10^{9.5}\MSUN$.
\end{enumerate}

In addition, a sub-sample `all satellites with gas' is defined as TNG100 galaxies which satisfy the above conditions and in addition 
\begin{enumerate}
\setcounter{enumi}{3}
\item $\MGAS / \MSTARS > 0.01$,
\end{enumerate}
where $\MGAS$ is the mass of gas found within twice the stellar half mass radius of the satellite, at any given time. For the adopted stellar mass limit, this corresponds to a minimal amount of about $3 \times 10^7 \MSUN$ of  gas in a galaxy.

We focus on four snapshots of the simulation at redshifts of $z=0, 0.2, 0.4,$ and $0.6$, and enforce the above criteria on each snapshot separately. We report the sample sizes for each snapshot in \cref{tab:numbers}, top two rows, and Figure ~\ref{fig:Comparison_All_Gas_Jellyfish}. Though it stands to reason that a substantial number of the galaxies in a high-redshift snapshot are the progenitors of galaxies found at lower redshifts, we ignore any such links between the galaxies across the snapshots and treat each snapshot as an independent sample of galaxies, much as an observer would. Cumulatively, from these four snapshots the simulation provides a sample of 6\,066 galaxies, of which 2\,610 contain some gas. 

It is important to note that the two samples (all satellites vs.\@ all satellites with gas) have different distributions in stellar mass, host mass and redshifts: see \cref{fig:Comparison_All_Gas_Jellyfish}.
Over all host masses, satellites with gas are slightly less frequent (relatively to the whole population) at lower redshifts than at high redshifts, at fixed minimum galaxy stellar mass.  At all times, the number fraction of satellites with gas with respect to the whole satellite sample is a strongly decreasing function of host mass and is a non monotonic function of a galaxy's stellar mass or satellite-to-host total mass ratio.

\begin{table}
	\setlength\tabcolsep{4.5pt}
	\centering
	\begin{tabular}{lccccc} 
		\hline
		TNG100: \# of galaxies & $z=0$ &  $0.2$ & $0.4$ & $0.6$ & Total\\
		\hline
		All satellites          & 1707 & 1584 & 1507 & 1268 & 6066\\
        All satellites with gas & 511  & 641  & 746  & 712  & 2610\\
		Jellyfish galaxies      & 196  & 193  & 214  & 197  & 800\\
        \hline
        Jellyfish / all satellites & 0.11 & 0.12 & 0.14 & 0.16 & 0.13\\
        Jellyfish / all satellites with gas & 0.38 & 0.3 & 0.29 & 0.28 & 0.31\\
        \hline
	\end{tabular}
    \caption{The galaxy sample used in this study. Shown are the total number of `all satellites' which pass the sample
    criteria listed in \cref{sec:sample}, the sub-sample of satellites which contain gas within the galactic radius, and the number of confirmed
    jellyfish galaxies, for the 4 redshifts studied in this paper.
    In the bottom two rows we show the fractions of jellyfish galaxies with respect to all satellites, and to all
    satellites with gas.}
\label{tab:numbers}
\end{table}

\subsection{Property measurements}\label{sec:PropertyMeasurements}

In the following sections, we identify jellyfish galaxies from the population of satellites with gas and study their properties in the context of their environment. To this end, we calculate three physical properties that can shed light on the relationships between satellite galaxies, their gas content, and their immediate environment: the Mach number of the galaxy motion within the medium ($\mach$), the ram-pressure exerted by the medium on the gaseous component in the galaxy, and the orientation of a gaseous tail (if one is present) emanating from the galaxy. 

\subsubsection{Satellite gas, medium gas, and relative velocities}
Throughout the text, and unless otherwise noted, we define `satellite gas' as the ensemble of all gas cells that are gravitationally bound to a satellite subhalo based on the {\sc subfind} algorithm. 

We define the `medium gas' around a given satellite as all the gas cells that are part of the host halo, based on the FoF classification, found within 20 times the satellite's stellar half mass radius ($\RGAL$) and that are \emph{not} gravitationally bound to any subhalo (besides the central one) within the FoF under consideration. From the medium budget, we hence exclude both the satellite gas as well as the gas of any neighboring galaxy within the considered aperture. This medium gas represents the local ICM through which a satellite is moving: for the typical galaxy in our sample, the $20\times \RGAL$ aperture corresponds to ICM patches of about 100 kpc of diameter, within hosts that extend for many hundreds of kpc, but it can be larger for the more massive and extended galaxies.
On the other hand, the `satellite gas' includes all gas that is found within twice the galaxy's stellar half mass radius {\it and} beyond. We will comment on the impact of these particular choices in the following sections.

In general, a subhalo moves through the ICM with a certain speed. 
We define the galaxy bulk (or systemic) velocity of a satellite as the mass-weighted mean velocity of all its {\sc subfind} resolution elements: we use this in \cref{sec:phase-space,sec:tail_orientation}. Furthermore, this can be evaluated in the reference system of the whole FoF host (e.g.\@ with respect to the velocity of the central galaxy, in turn obtained as the mass-weighted velocity of its gravitationally-bound resolution elements, as in \cref{fig:phase-space diagram}) or in the reference system of the surrounding medium gas (see below). 

Importantly, for the calculations of Mach numbers and ram pressure, we use the relative velocity between the satellite gas ($\vec{v}_\mathrm{satellite ~gas}$) and the medium gas. 
We define the typical velocity of the medium gas, $\vec{V}_{\mathrm{medium}}$, by finding, for each velocity component separately, the median value of all medium gas cells around a satellite, i.e.  the FoF gas cells within 20 times the stellar half mass radius of the satellite (see above). As the gas cells are forced to have similar masses within a factor of two, this choice is quantitatively very similar to a mass-weighted mean. 

The scalar magnitude of the relative velocity of the gaseous component of a satellite with respect to its surrounding medium is measured as follows: 
\begin{equation}\label{eq:relvel}
v_{rel} = \mode{\vert \vec{v}_\mathrm{satellite ~gas}-\vec{V}_{\mathrm{medium}} \vert}.
\end{equation}
Namely, for all satellite gas cells we take the norm of their relative velocities with respect to the typical medium velocity and for the resulting distribution, we take the mode.
These choices are dictated by the fact that both medium and satellite gas are characterized by distributions of values for their velocities from which we need to distill a single, characteristic value. To do so, we elect to use the `mode' of the distribution, which is defined to be the most common value, i.e.  the peak of the histogram for the distribution. The advantage of using this measure when dealing with distributions of gas quantities, rather than the mean or median, is that it is the value that best represents the majority of the gas cells and is not sensitive to distributions which are possibly heavily skewed or multi-modal. We have verified the mode values used in this work to be stable against binning choices.

\subsubsection{Mach number}
\label{sec:machNumber}

The Mach number ($\mach$) is an important indicator of the interaction between a galaxy and the medium through which it is travelling. Galaxies moving at supersonic velocities, i.e.  $\mach>1$, will produce discontinuous features in the fluid such as shocks and contact discontinuities, whereas for subsonic motions, $\mach<1$,  the sonic waves that travel `downstream' allow for smooth changes in the properties of the medium. 
In general, the Mach number is defined as the ratio between the speed at which an object moves relative to a fluid and the sound speed of the fluid, 
\begin{equation}\label{eq:machDef}
\mathcal{M} \equiv \frac{v_{rel}}{c_{\rm s}},
\end{equation}
where the sound speed is given by
\begin{equation}
	c_{\mathrm{s}}\equiv\sqrt{\pderiv{P}{\rho}\Bigg|_S}=\sqrt{\gamma\frac{P}{\rho}}=
    \sqrt{\gamma \frac{\kboltz T}{\mu m_\mathrm{p} }},
	\label{eq:Sound Speed}
\end{equation}
where $P$, $\rho$, and $T$ are the pressure, density, and temperature respectively. The second and third equalities are valid for an ideal gas equation of state, which is used in our simulation with an adiabatic index of $\gamma = 5/3$ for a mono-atomic gas.  The factor $\kboltz$ is the Boltzmann constant and $\mu m_{\mathrm{p}}\simeq 0.59~m_{\mathrm{p}}$ is an average particle mass appropriate for the ICM ($m_{\mathrm{p}}$ being the proton mass). In these estimates of the gas Mach number, we ignore the modifications due to magnetic fields.

To calculate the Mach number of the gas in a satellite galaxy we must identify the velocity of said gas with respect to the ambient gas and determine the sound speed of the medium. We take the relative velocity as defined in \cref{eq:relvel}. In addition, we also calculate the sound speed for each gas cell in the medium \ky{by means of \cref{eq:Sound Speed}}. 
The definition of the Mach number in our study is therefore 
\begin{equation}
   \mach \equiv \frac{ v_{rel} }{\mode{c_{\mathrm{s,medium}} }},
	\label{eq:MachNumber}
\end{equation}
where we again take the `mode' of the distribution of the sound speed of the gas cells in the medium surrounding a satellite, $c_{\mathrm{s,medium}}$. We adopt the same operational definition of sound speed in \cref{eq:Sound Speed} whether the gas is star forming or not: in fact, our definition of medium gas is such that essentially no star-forming gas is included in the measurement of the sound speed of the medium.

\begin{figure*}
	\centering
	\includegraphics[width=0.95\textwidth]{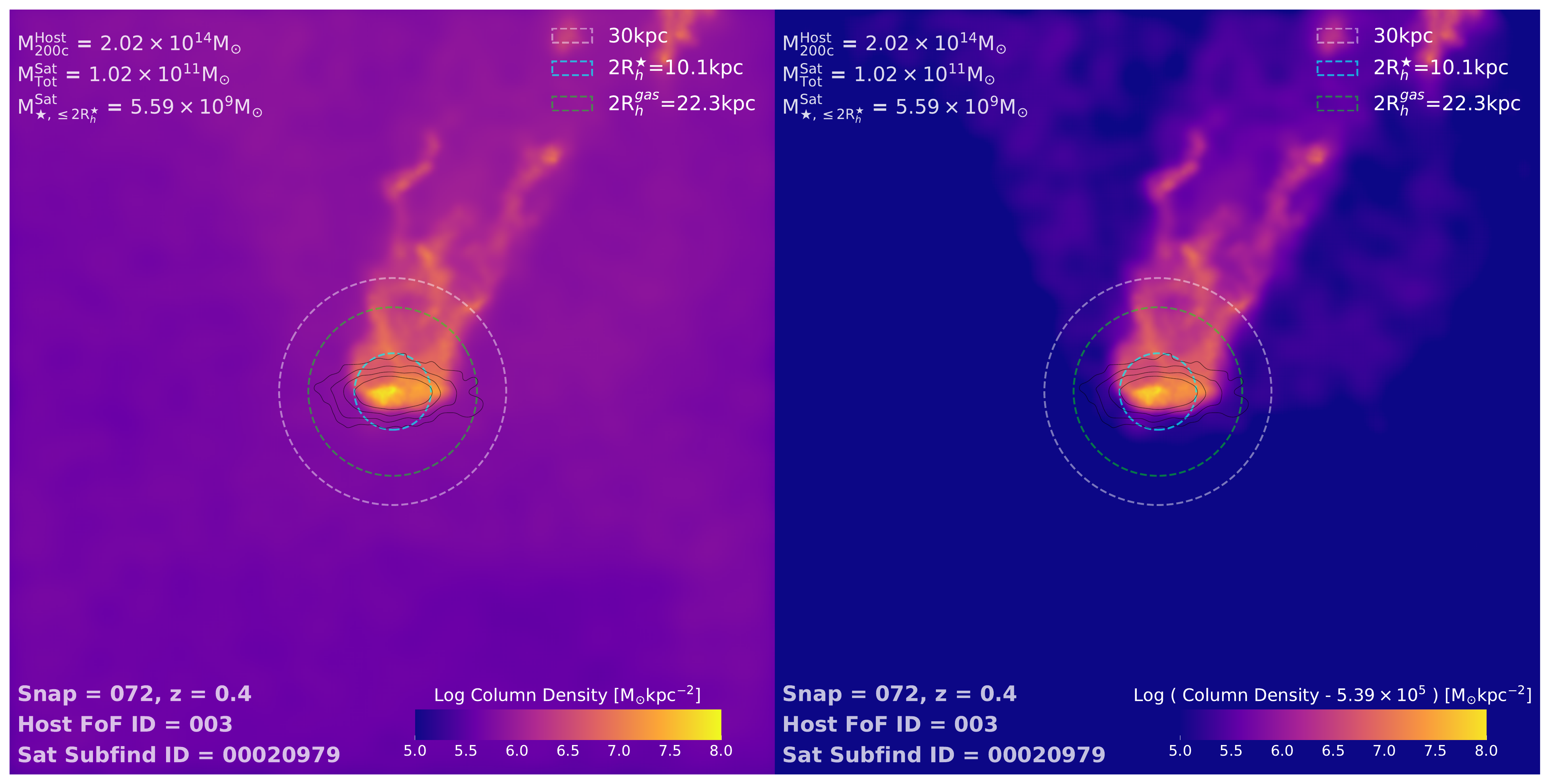}
    \caption{A sample image of one of the simulated galaxies used for the visual inspection and representing a poster-child jellyfish galaxy.
    The left-hand image shows the gaseous column density (color-map)  and stellar column density (black contours), projected along a randomly selected direction. The right-hand image shows the same galaxy with the background density subtracted. The image size is 40 times the stellar half mass radius. Black contour lines show the 85, 80, 75, and 70 per cent levels of the maximal stellar column density. The radius of the dashed cyan circles \ky{is} twice the stellar half mass radius, the dashed green circles denote the twice gas half mass radius, and the white dashed circles have a radius of $30\units{kpc}$ (comoving) from the center of the galaxy. The virial mass of the host, $\MHOST$, the stellar and gas mass of the galaxy are shown in the top-left corner while the redshift and information on the simulation snapshot and ID numbers of the host and galaxy in the simulation catalogues are shown on the bottom-left.}
        \label{fig:Image Sample}
\end{figure*}

\subsubsection{Ram pressure}\label{sec:ramPressure}

The physical process for gas removal most commonly linked to the formation of jellyfish galaxies is stripping due to the ram-pressure exerted by ICM due to the motion of the galaxy within the medium. 

We obtain a measure for the ram pressure acting on the galaxies in the simulation based on the equation \citep{Gunn1972OnEvolution}: 
\begin{equation}
     P_\mathrm{ram}= \mode{\rho_\mathrm{medium} }\times v_{rel}^{2},
	\label{eq:Ram Pressure}
\end{equation}
where $\rho_\mathrm{medium}$ is the density of the medium gas cells and the relative velocity is defined above. Once again we use the `mode' of the density and velocity distributions to find a representative value for these parameters.

\subsubsection{Orientation of the gaseous tails}
\label{sec:tailDefinition}
One of the defining characteristics of a jellyfish galaxy is an extended gaseous `tail' emanating from the galaxy. In our analysis we wish to compare the tail orientation to other `vector directions of interest', e.g.\@, the direction of the satellite motion, the direction to the host, the galaxy spin orientation etc.\@ (see \rfsec{tail_orientation}). In fact, any asymmetric gas distribution can be characterized by an orientation vector, even if a pronounced tail is not present.

We hence define the orientation of gas tails or asymmetries by considering, for any given galaxy, the \emph{satellite} gas in a shell between four and twenty times its stellar half mass radius and calculating its density-weighted mean position with respect to the galaxy center:
\begin{equation}\label{eq:tailDefinition}
\vec{x}_\mathrm{tail}\equiv {\sum_{|\vec{x}_i|\in[4,20] \RGAL } \rho_i \vec{x}_i }  \Big/ {\sum_{|\vec{x}_i|\in[4,20] \RGAL } \rho_i },
\end{equation}
where $\rho_i$ and $\vec{x}_i$ denote the density and position of individual gas cells, respectively.
Since the gas stripped from a galaxy is usually much denser than the medium, this measure provides a good estimate of the orientation of a tail. We have also considered different choices, as for example by increasing the minimum boundary for the tail from $2\times \RGAL$ to a larger multiple and by weighting the average differently: we will comment on these in the following sections. Furthermore, we have decided to consider only the gravitationally-bound gas for the tail-direction measurement. Gaseous tails may include material that is not (anymore) gravitationally-bound to the satellite they stem from; nevertheless, we find, through visual verification, that the estimates of the tail direction are more stable and meaningful if only the gravitationally-bound gas is considered in \cref{eq:tailDefinition}.
Finally, it is important to note that, even in the absence of a tail, the above measure may give a non-zero result since this can occur only for a perfectly-spherical density distribution around the satellite. For almost-symmetric gas distribution, the ``tail'' vector will essentially assume a small value with random orientation.

\section{Jellyfish visual identification}
\label{sec:3visual} 

\begin{figure*}
	\includegraphics[width=\textwidth,height=0.92\textheight, keepaspectratio=true]{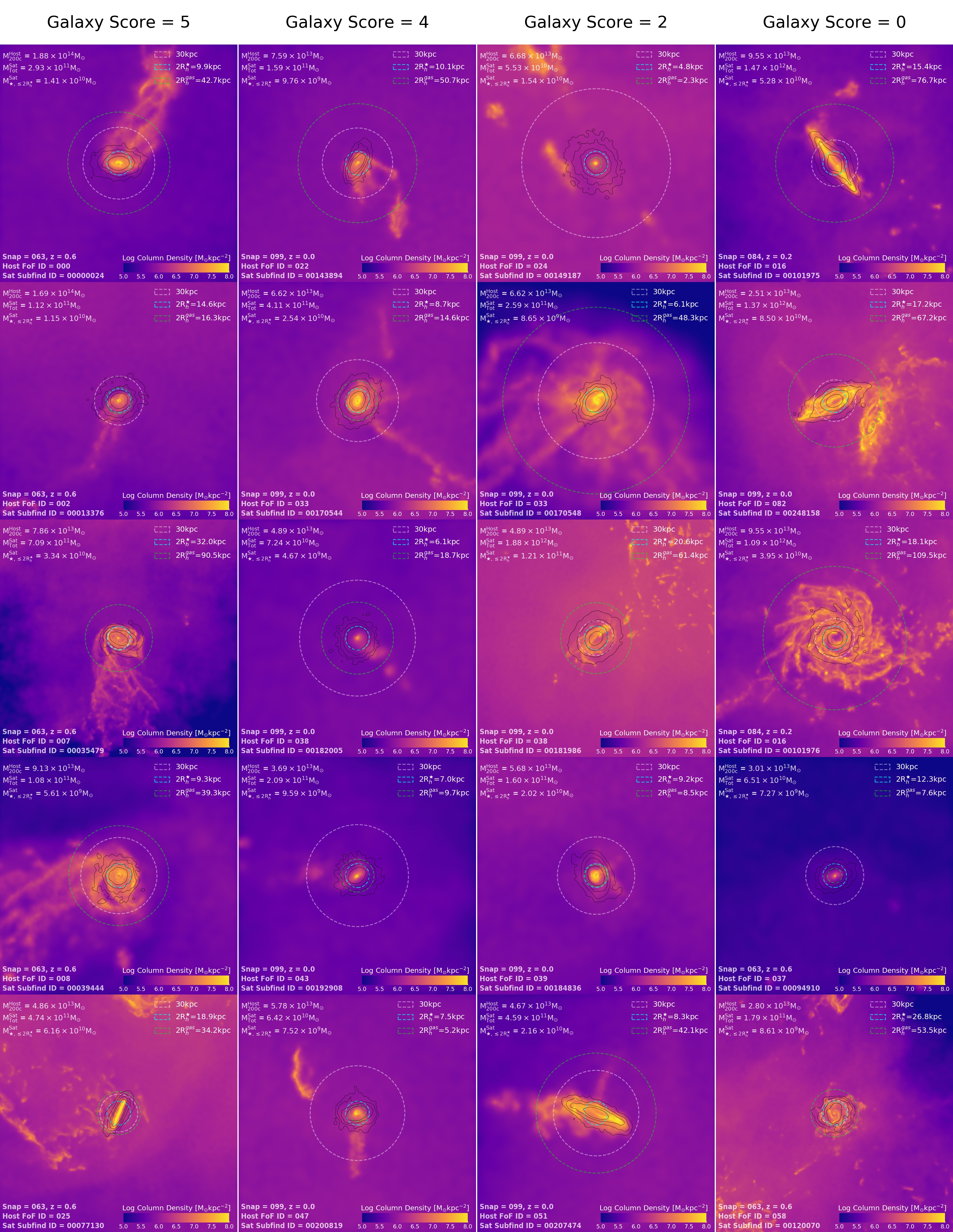}
   \caption{Gallery of some of our simulated galaxies, arranged in columns according to scores of 5, 4, 2, and 0, from left to right, according to our visual inspection. The images are all un-subtracted maps of the gas column density, overlaid with stellar density contours as shown in the left-hand side of \cref{fig:Image Sample}, with identical designation of lines and circles. The galaxy score shows how many independent Inspectors deemed the galaxy to be a jellyfish galaxy based on the visual classification criteria described in \rfsec{3visual}. 
   }
    \label{fig:Images with Scores}
\end{figure*}

We identify jellyfish galaxies via a visual inspection of galaxy images produced for each object in the `all satellites with gas' sample  (see \rfsec{sample}). 

Here the meaning of what a jellyfish galaxy is should be interpreted loosely: in practice, we are interested in galaxies with strong signatures of ram-pressure stripping i.e.\@ gas distributions that are clearly asymmetric in one direction around a galaxy. Moreover, we want to mimic what an observer would do, by inspecting the distribution of the gas in a random projection after having identified a satellite galaxy possibly via optical imaging, i.e.\@ via its stellar light. This is why our parent sample comprises satellite galaxies with a non-vanishing stellar mass. However, in what follows, we will consider all gas, without distinguishing between cold, cool, or star-forming gas, as would be the case for molecular, HI or H$\alpha$ observations, respectively.
Note that visual inspection is common practice in observational astronomy and there are no reasons not to attempt it with simulation data. Visual classifications of optical galaxy features have been successfully applied to thousand-strong galaxy samples (e.g.\@ with the Galaxy Zoo project\footnote{http://zoo1.galaxyzoo.org/Project.aspx}) and it is thus far the main identification approach for jellyfish galaxies in observational data, e.g.\@ within the GASP and VESTIGE programs (see Introduction).

For each galaxy in the sample, we extract a box of 40 times the stellar half-mass radius on a side, centered on the galaxy, and create color-maps of the gas column density, overlaid with contours representing the stellar density ($85$, $80$, $75$, and $70$ per cent of peak stellar density), by projecting these values along the z-axis of the simulation box -- hence a random projection for each galaxy. The host and galaxy mass, as well as information on the ID numbers of the host and galaxy in the simulation catalogues is also shown. Each image consists of 800 $\times$ 800 pixels and thus each pixel size for the average galaxy in our sample corresponds to about $0.25-0.35$ kpc a side \ky{(in comoving units, as throughout the paper)}. The gas density field, sampled by point-like mass elements with coordinates placed at the Voronoi mesh-generating points, is smoothed via a convolution with a simple Gaussian kernel whose size is linearly proportional to the mean radius of the gas cells over the closest 10 pixels. \ky{In practice, the spatial resolution of the galaxy images is adaptive and depends on the imaged density field and galaxy's extent. However, the choices are implemented unchanged across redshifts and the above-mentioned pixel size of about $0.25-0.35$ comoving kpc represents a fixed ``rest-frame'' spatial resolution of the images at both $z=0$ and $z=0.6$. We hence put ourselves in the best possible conditions to identify jellyfish galaxies, without mimicking any possible bias affecting observational studies across redshifts.}

\cref{fig:Image Sample} shows an example of what we consider a poster-child jellyfish galaxy in TNG100, with a gaseous disk out of which a tail longer than $50-60$ kpc is clearly manifest with an otherwise unperturbed stellar body. In \cref{fig:Images with Scores} other examples of galaxies that went through the visual classification are given. 

In addition to images in which we show all the gas in the volume (\cref{fig:Image Sample}, \emph{left}) we also produce images in which the background density is subtracted (\cref{fig:Image Sample}, \emph{right}). This is done by subtracting from the image the mean gas column density for all gas cells in the box found beyond twice the \emph{gas} half mass radius (shown as a green circle in \cref{fig:Image Sample}). 
As can be seen in the figure, the gas stripping features are much more prominent in the subtracted image, and in some cases the subtraction reveals features which would be otherwise indiscernible in the un-subtracted images. 

Perusal of the thousands of galaxies in our sample reveals a wide variety of gaseous features of different shapes and sizes. Galaxies with prominent stripping features and `tails' are easily identified as jellyfish galaxies, as can be seen in the leftmost column of \cref{fig:Images with Scores}, while galaxies which lack any such features can easily be excluded from the jellyfish sample (rightmost column of \cref{fig:Images with Scores}). However, visual inspection by its nature is subjective and depends on a given Inspector\footnotemark, and thus there are many galaxies in which the jellyfish classification is not straightforward and may be subject to Inspector bias (middle columns of \cref{fig:Images with Scores}). In fact, in the chaotic environment of groups and clusters of galaxies, complex gas distributions are very frequent and here the aim is to avoid those cases where the gas is stripped because of tides rather than the ram pressure or where the satellite gas distribution is caused by an ongoing merger or a companion.

To overcome this issue and achieve a classification not prone to bias from a single Inspector, all galaxies were inspected by 5 different Inspectors, who performed the classification independently and gave each galaxy in the sample a score of 1 (jellyfish) or 0 (non-jellyfish). Before classifying the galaxies, a set of general criteria were established for giving to a galaxy the label ``jellyfish'':
\begin{itemize}
\item An asymmetric gas distribution stretched in one preferred direction is manifest in the stamp.
\item Gas tails/wakes (i.e.\@ portions of asymmetric gas) should be ``attached'' to the main body of the galaxy.
\item No galaxy companion within the field of view is obviously interacting with the subject.
\end{itemize}
In addition, a few test cases were discussed jointly by the Inspectors to achieve a common baseline for the classification.

For each galaxy, the final score was obtained by adding up the classification score of the 5 Inspectors, resulting in a score ranging between 0 for a galaxy which none of the Inspectors judged to be a jellyfish galaxy, and 5 for a galaxy that was unanimously identified as a jellyfish galaxy. 
In all the quantitative analysis that follows, the term `jellyfish galaxy' or `jellyfish sample' refers to the 800 galaxies (out of the 2\,610 galaxies in the sample) which have a score of 5, i.e.  deemed to be a jellyfish galaxy by all Inspectors. 

\begin{figure}
   \centering
    \adjustbox{trim={0.07\width} {0.077\height} {0.03\width} {0.09\height},clip}%
	{\includegraphics[width=0.90\columnwidth]{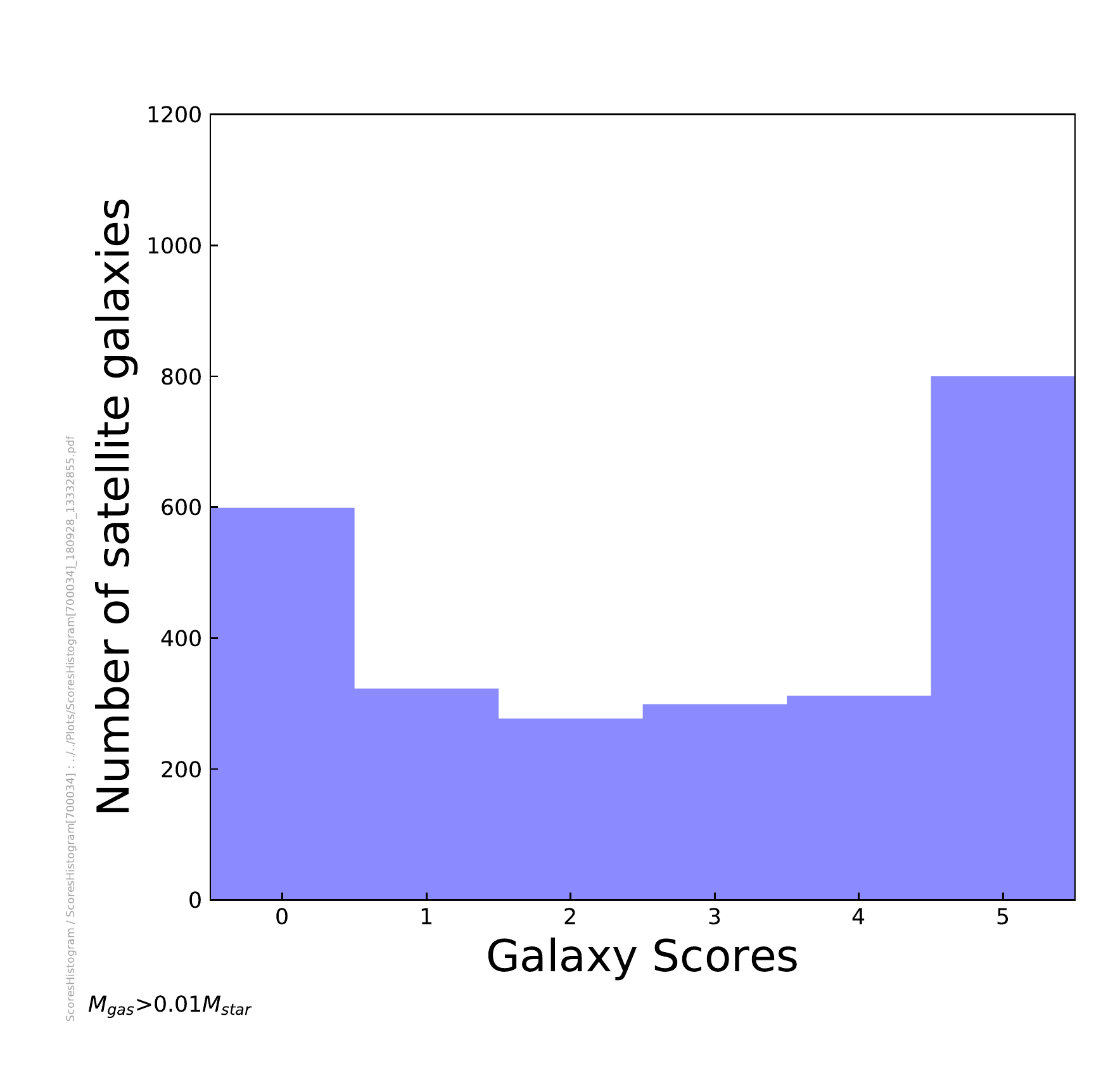}}
      \caption{Histogram of the `jellyfish' scores for the 2\,610 galaxies in the sample that we denote `All satellite with gas', i.e.\@ with at least 1 per cent gas mass fraction and spanning the four redshift snapshots analyzed in this paper.  The scores represent the number of Inspectors who visually classified a given galaxy as a jellyfish galaxy, for a total of five Inspectors. Throughout the paper, ``jellyfish'' galaxies are those with score equal to 5. The peaks at score 5 (jellyfish galaxies) and 0 (certainly not-jellyfish galaxies) show a general consensus among the Inspectors as to what does and does not constitute a jellyfish galaxy.  }
    \label{fig:Galaxy Scores}
\end{figure}

In \cref{tab:numbers} we report the number of jellyfish galaxies at each of the 4 redshifts we analyzed as well as the fraction of jellyfish galaxies compared to the `All satellites' sample and the `All Satellites with gas' sample.
In \cref{fig:Galaxy Scores}, we show the distribution of scores for the entire sample of `All satellites with gas', spanning all 4 redshifts considered here and peaking at the lowest and highest scores. The bimodal distribution shows that there is general consensus (among the Inspectors) on what constitutes a jellyfish galaxy and also what does not. Indeed, this evident bimodal distribution shows that we call jellyfish galaxies those galaxies which are unequivocally asymmetric on scales larger than the visible bulk stellar mass or gas mass distribution. Yet, much of the bimodality is the result of some of the choices we imposed a priori, e.g.\@ the exclusion of ongoing interactions and mergers and of those cases where the distribution of gas within and around a satellite was too confused or disturbed.
However, there is a considerable number of galaxies (about 30 per cent, cumulatively) of intermediate scores, which attest to the complex and subjective nature of visual classifications, when performed by humans. Interestingly and reassuringly, we have found that no single Inspector is particularly biased in finding asymmetric gaseous tails or distributions.

In \cref{fig:Images with Scores} we present 20 representative examples of galaxies of various scores to highlight the choices and challenges associated with the visual classification. The images are arranged in 4 columns of score 5, 4, 2, and 0, from left to right. All the images in the leftmost column, of score 5, show well-defined tails which extend out from the galaxy's main body. 
In the images in the second from left column, of score 4, while some of the galaxies do show tail-like features, it is hard to determine whether they are indeed tails or irregular features as in the top and bottom images. In some cases there is more than one tail-like feature (second from top) while in some, the features are small or appear to be disconnected from the main body of the galaxy (third and fourth from top). In the images in the third column from the left, of score 2, one still finds asymmetric gas features but one would be hard put to define a direction of motion based on them. The right-most column, of score 0, shows galaxies with either a largely symmetric gas distribution which attest that external forces are not disturbing the gas, or with galaxy companions that hence have been excluded by construction, or even with a very small, negligible gas content for which no spatial distribution is manifest. \ky{The galaxy on the top right panel of \cref{fig:Images with Scores} is, for example, a possibly controversial case. In our approach, it does not satisfy the second criterion, i.e. the need for the asymmetric tails to be ``attached'' to the main body of the galaxy: indeed, none of the visual Inspectors has dubbed it jellyfish. However, we note that several such examples in the literature have been commonly called jellyfish galaxies. This attests at the complexity of visual classifications and possibly suggests that the criteria adopted here are more conservative than others in the literature.} 

\footnotetext{Even for a given Inspector, the classification of a galaxy may be affected by different factors such as the time of day, number of galaxies previously inspected, classification experience, etc.} 

Additional examples of jellyfish galaxies from TNG100 (i.e.\@ score = 5 galaxies) can be found in Figs.~\ref{fig:Jellyfish Galley 1} and \ref{fig:Jellyfish Galley 2}. Note that we do not make any distinction based on a galaxy (stellar) morphology, and indeed our jellyfish sample contains both disk-like (e.g.\@ in \cref{fig:Images with Scores} and \cref{fig:Jellyfish Galley 1}, bottom row, first and third stamp from the left) and seemingly spheroidal galaxies (\cref{fig:Jellyfish Galley 1}, right column, second and third from the top). Note, moreover, that jellyfish galaxies exhibit very different gaseous morphologies: for example, in our sample we include both galaxies with thin tails that extend far beyond the main galaxy body, for many times the gaseous or stellar size of the galaxy (e.g.\@ \cref{fig:Jellyfish Galley 1}, top row, middle two galaxies); as well as disk galaxies where the gaseous wakes depart from the main galaxy body like a table cloth: e.g.\@ in \cref{fig:Jellyfish Galley 1}, top right stamp. 

\section{Results}
\label{sec:4. Results}

\subsection{TNG jellyfish galaxies: basic demographics}
\label{sec:4.1. TNG jellyfish galaxies: basic demographics}

\begin{figure*}
\adjustbox{trim={0.01\width} {0.00\height} {-0.01\width} {0.0\height},clip}%
{\includegraphics[width=0.91\columnwidth]{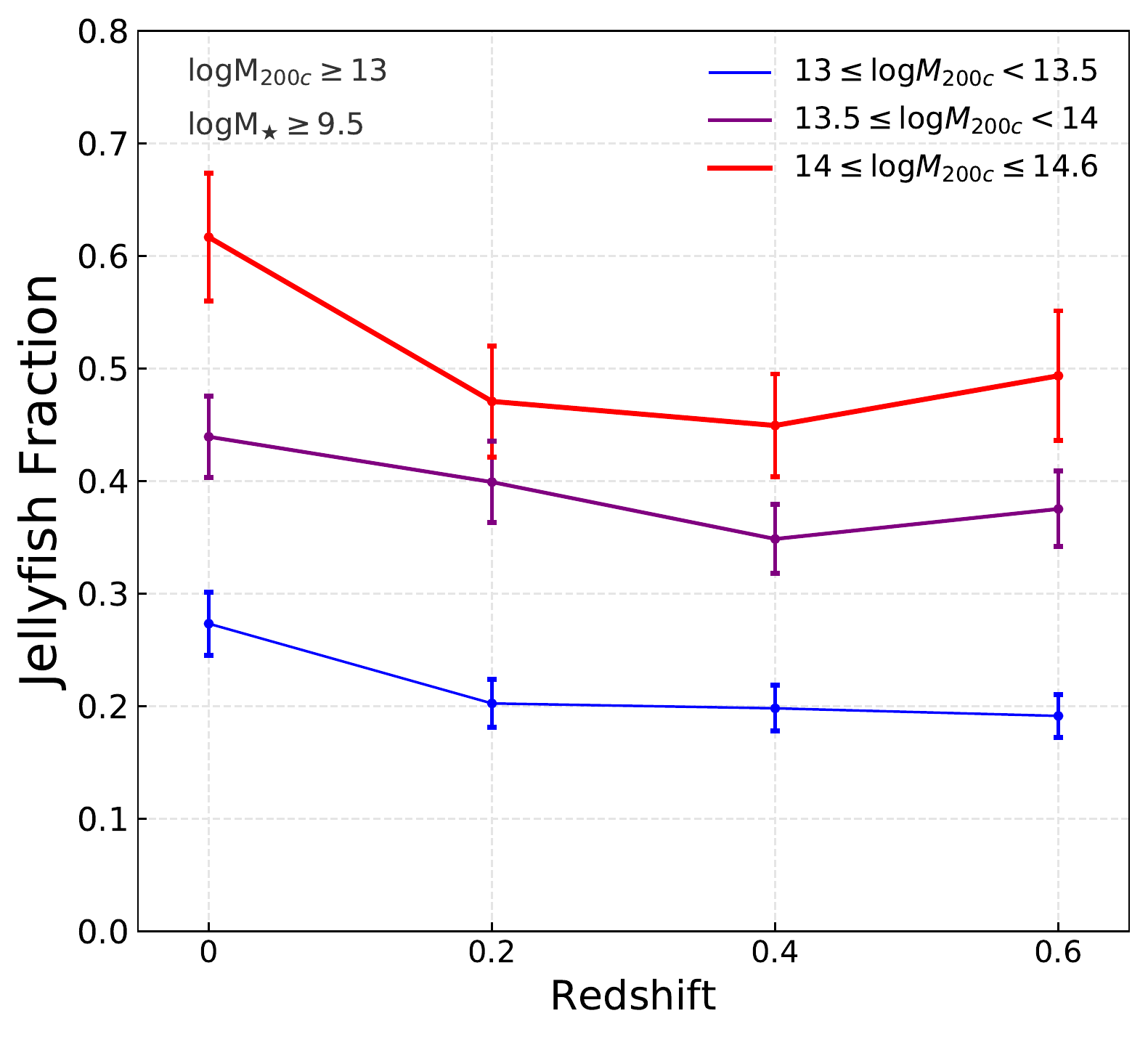}}
\adjustbox{trim={0.0\width} {0.00\height} {-0.035\width} {0.0\height},clip}%
{\includegraphics[width=0.91\columnwidth]{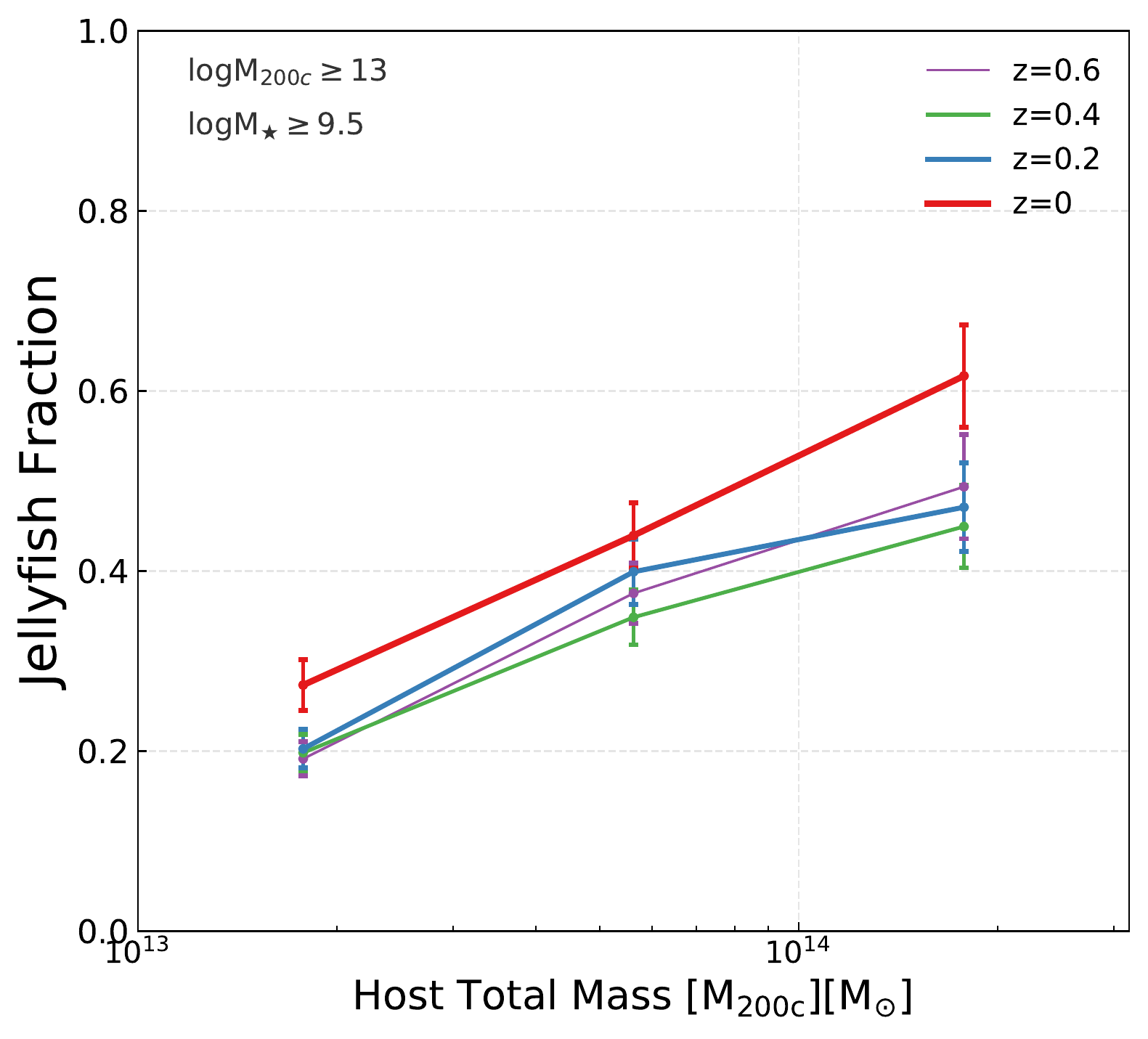}}\\
\adjustbox{trim={0.01\width} {0.00\height} {0.01\width} {0.0\height},clip}%
{\includegraphics[width=0.94\columnwidth]{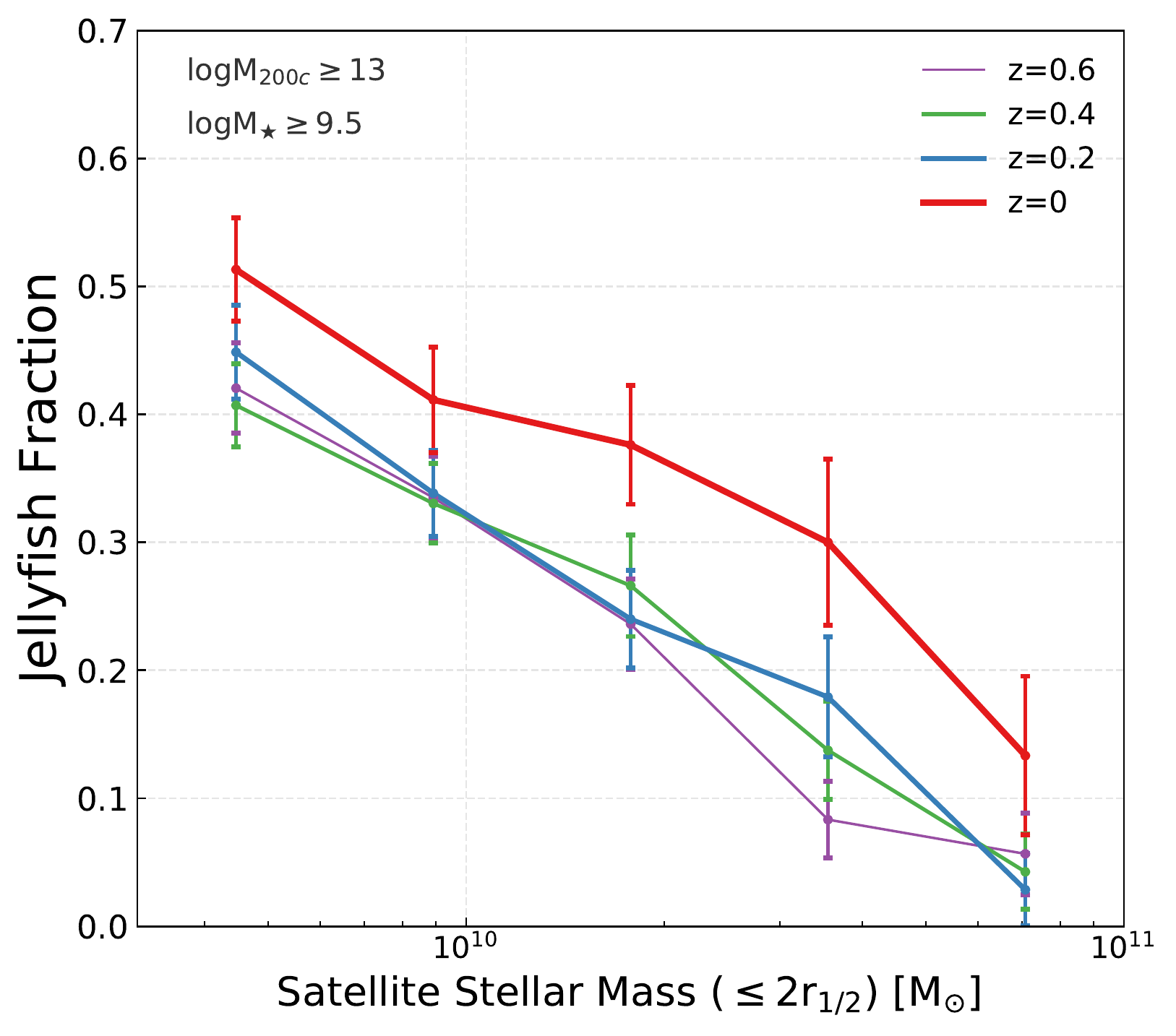}}
\adjustbox{trim={0.01\width} {0.00\height} {0.01\width} {0.0\height},clip}%
{\includegraphics[width=0.95\columnwidth]{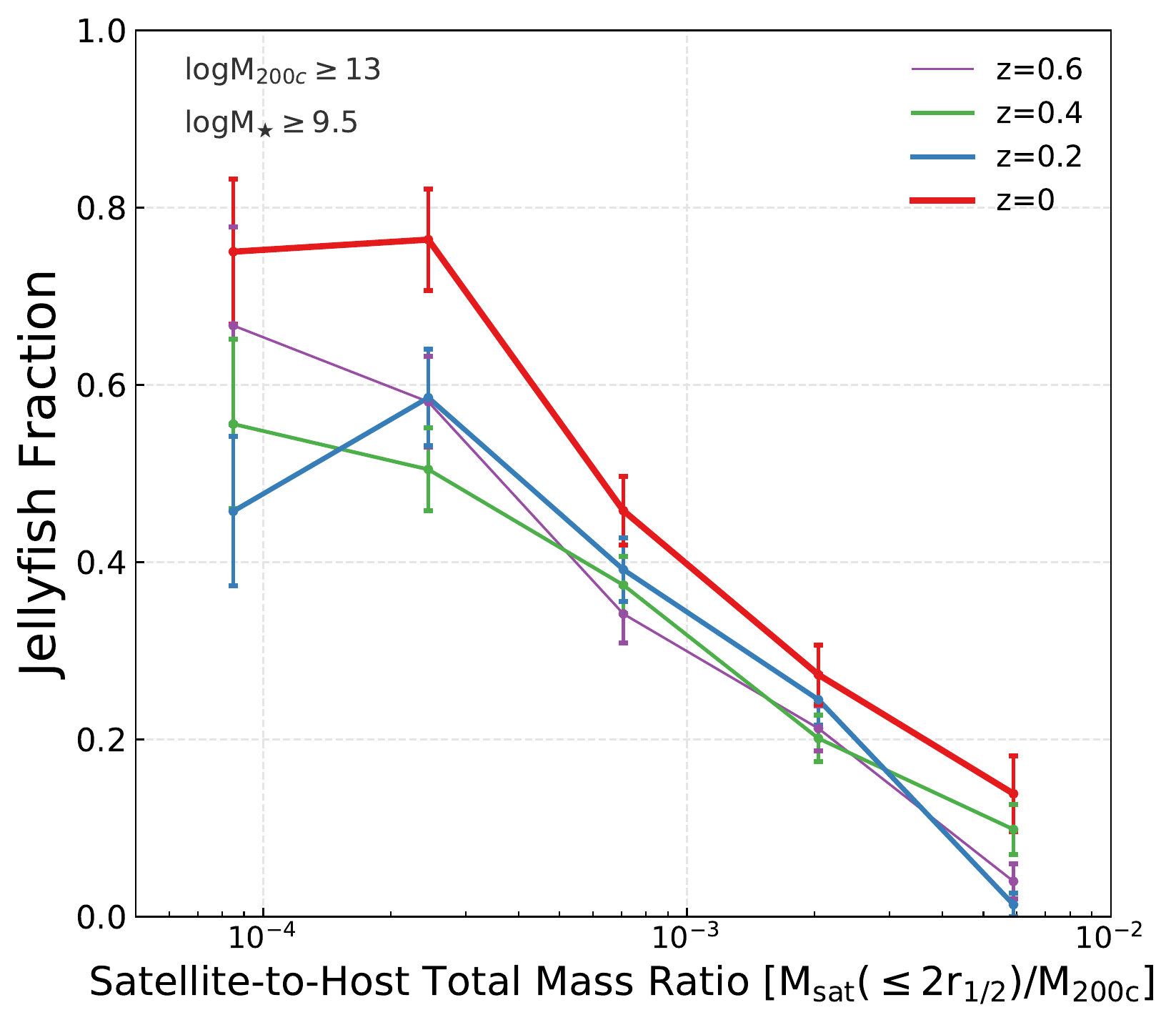}}
\caption{The frequency of jellyfish galaxies among those of the sample that we denote `All satellites with gas'.
The fractions are given as function of redshift, divided into host mass bins (top left). The frequency of jellyfish galaxies is also given as a function of host mass (top right), satellite stellar mass (bottom left), and mass ratio between the total mass enclosed within twice the stellar half mass radius and the virial mass $\MHOST$ of the host (bottom right), for four redshifts. Here we only consider satellite galaxies with stellar mass larger than $10^{9.5} \MSUN$ that are located within the virial radius of their host with total mass larger than $10^{13} \MSUN$. Error bars are estimated via binomial statistics.
   There is no strong trend in the jellyfish fraction as function of redshift in all three bins of the cluster host mass, whereas the jellyfish fraction increases with increasing host mass and decreasing satellite stellar mass and satellite-to-host total mass ratio, at all redshifts.}
   \label{fig:Jellyfish fraction}
\end{figure*}

With a catalogue of jellyfish galaxies in hand (defined by our visual classification scheme), we now study the demographics of the jellyfish galaxy population in our sample. In \cref{tab:numbers} we show the number and fraction of jellyfish galaxies in the 4 chosen redshift bins. We find that there is little evolution of the jellyfish fraction with redshift, with 38 per cent of all satellites with gas identified as jellyfish galaxies at \zeq{0}, and a $28\--30$ per cent fraction at the higher redshifts up to $z\sim 0.6$. Of the entire sample of galaxies, regardless of their gas content, the fraction remains roughly constant with time at $11\--16$ per cent. 

We extend our analysis by dissecting the jellyfish fraction into bins of redshift, host mass, stellar satellite mass, and satellite-to-host total mass ratio, and show the results in \cref{fig:Jellyfish fraction}. Here we give the fraction of jellyfish galaxies within the sample of `All satellites with gas', including error bars estimated via binomial statistics. 

We find that the consistency of the jellyfish fraction across redshift remains largely valid even when examined in different host masses, as can be seen in the top left panel of \cref{fig:Jellyfish fraction}: as a reminder, this is the case for satellites and hosts sample that are both mass limited. The exception to this finding is that, in high-mass hosts of $10^{14}\MSUN \le \MHOST \le 10^{14.6}\MSUN$, the fraction of jellyfish galaxies increases from roughly $50$ per cent in the high-redshift snapshots to $62$ percent at \zeq{0}. In fact, the median host mass within our large host-mass bins changes slightly at different redshifts \ky{(see the host halo mass functions at the four studied redshifts in \cref{fig:Host mass distribution}).} Being extracted from a fixed comoving volume and the most massive haloes assembling more recently, the median host mass becomes larger at progressively lower redshifts, even within a bin. For example, at $z=0$, the median (maximum) host mass in the highest bin is $2.1 \times 10^{14}~ (3.8 \times 10^{14}) \MSUN$, while it is $1.7 \times 10^{14} ~(2.7 \times 10^{14})\MSUN$ at $z\sim 0.2$. 
If the ram-pressure stripping effects are more pronounced in more massive hosts (see subsequent Sections), this population change across redshifts may suffice to explain the increase in jellyfish fraction at recent times and should be taken into account when interpreting the redshift trends in general. 

Conversely, the jellyfish fraction in our mass-limited satellite sample is a strong function of the host mass, at all redshifts, increasing from about 20 to 40 per cent from groups to clusters at e.g.\@ $z\sim0.6$ and up to $\sim 60$ per cent in the $10^{14.6}\MSUN$ clusters at $z=0$ (top right). This hints at the possibility that, due to the stronger environmental effects, high-mass hosts are more efficient in converting satellites into jellyfish galaxies. This indeed may come as no surprise since, due to the higher density of the medium in these systems and the higher typical velocities of satellites, which are set by the deeper potential well, the ram pressure exerted on these galaxies is stronger (\cref{eq:Ram Pressure} -- see following Sections).  The counterpart to this finding is shown in the bottom left panel of \cref{fig:Jellyfish fraction}, where we find that the jellyfish fraction increases with decreasing satellite mass -- smaller galaxies, with shallower potential wells, are more susceptible to the environmental stripping processes. 

We define a mass ratio between the total mass (dark matter, stars, gas, etc.\@) within twice the stellar half mass radius of the satellite and the host virial mass $M_\mathrm{200,c}$, and plot the dependence of the jellyfish fraction on this mass ratio in the bottom right panel of \cref{fig:Jellyfish fraction}. Similarly, we find that, overall, jellyfish fractions increase with decreasing mass ratios at all redshifts, except in the lowest bin of mass ratio of $10^{-4}$ at $z=0$ (red) and $z=0.2$ (blue). In this mass ratio bin, our sample is incomplete in low mass hosts of $\gtrsim 10^{13}\MSUN$ since we only considered galaxies with \emph{stellar} mass above $10^{9.5}\MSUN$. This shows that less massive galaxies are more likely to be jellyfish galaxies, especially in high-mass hosts, which stands to reason since their gravitational binding energy is lower, making them more susceptible to ram-pressure stripping. 

Importantly, these trends are {\it not} the same when one considers all satellite galaxies.
As shown in \cref{fig:Jellyfish fraction All}, the frequency of jellyfish galaxies within the population of all satellites is not a monotonic function of host mass nor of mass ratio. This is because the population of all satellites with some gas is already a biased population (in redshift, satellite mass, and host mass) with respect to the more numerous population of all luminous satellites, as is demonstrated explicitly in \cref{fig:Comparison_All_Gas_Jellyfish}. It is important to keep this in mind when comparing jellyfish fractions across different simulation and observation compilations.

Instead, similar qualitative trends as in \cref{fig:Jellyfish fraction} are found even when only more massive satellites are selected, e.g.\@ for $\MSTARS \ge 10^{10} \MSUN$. In this case, the frequency of jellyfish galaxies is slightly lower: for example, at $z=0$, about 30 per cent instead of the 38 per cent jellyfish galaxies with our lower fiducial minimum stellar mass.



\subsection{Environmental properties}

\subsubsection{Location in phase-space within the hosts}
\label{sec:phase-space}

\begin{figure*}
	\centering
	\adjustbox{trim={0.10\width} {0.07\height} {0.01\width} {0.005\height},clip}%
	{\includegraphics[width=\textwidth]{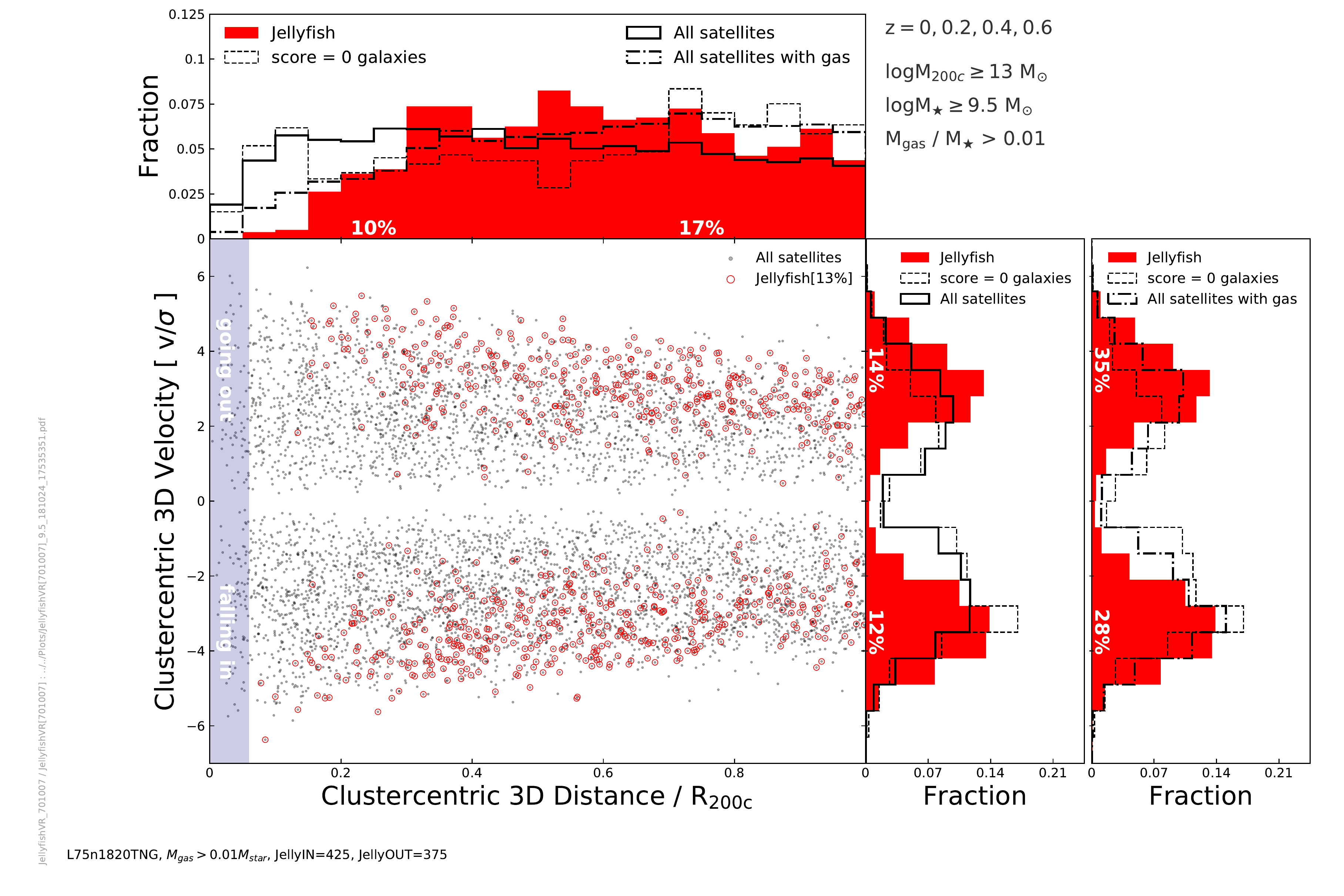}}
    \caption{The distribution of the jellyfish galaxies (red circles) and the entire satellite population (`All satellites', with or without gas: black dots) from all four redshift snapshots in the 3D bulk velocity and \ky{3D} cluster-centric distance plane. The distance is shown in units of the host virial radius $\RTC$ and the bulk velocity of each satellite is normalized by its host velocity dispersion.
A histogram of the distribution of \ky{3D} cluster-centric distance is given for a number of sub-populations in the top panel: jellyfish galaxies in red, all satellite galaxies in solid black, all satellites with gas in dash-dot black, and galaxies that are certainly not jellyfish galaxies (score = 0) in dashed black stairs histograms. All histograms are shown as a fraction of the total number of objects in each sub-sample. The percentage in white corresponds to the fraction of jellyfish galaxies of all galaxies found in the regions  $<0.5\RTC$ and $>0.5\RTC$ (left and right, respectively). To the right of the main panel, the 3D velocity distribution for the jellyfish and the entire sample is shown (mid panel) alongside the distribution for the jellyfish and the satellites with gas sample (rightmost panel). The histograms show the fraction of the population in each sub-sample. The percentage of jellyfish galaxies out of all infalling satellites (radial velocity $v_\mathrm{r}<0$) and all outgoing satellites (radial velocity $v_\mathrm{r}>0$) are shown in white for the two sample histograms. There is a relative deficiency of jellyfish galaxies in the central regions of the host $\lesssim 0.2-0.25\RTC$, and overall, jellyfish galaxies travel faster than the global population. In addition we find there is a small overabundance of jellyfish galaxies on outgoing trajectories, i.e.  positive radial velocities. 
    }
    \label{fig:phase-space diagram}
\end{figure*}

We study here the kinematic and spatial distributions of jellyfish galaxies compared to the general population of satellite galaxies. \cref{fig:phase-space diagram} shows a sub-sector of the satellite phase-space defined by the \ky{3D} cluster-centric distance and the 3D bulk velocities of the satellites in all hosts of $10^{13}\MSUN$ and above from the 4 redshift snapshots, cumulatively. Accompanying the phase-space plot are histograms detailing the separate distributions of the cluster-centric distance (top) and the 3D velocity (right). The histograms for the jellyfish galaxies (red) are compared to those of the whole sample (``All satellites'', black solid), of the sample of galaxies which contain gas (``All satellites with gas'', dash-dot black) and those galaxies that are surely not jellyfishes (``score = 0 galaxies'', black dashed histograms). 

In general, satellites are distributed evenly in cluster-centric distance within the host. Jellyfish galaxies also follow this distribution except in the central regions of the hosts, within $0.25\RTC$, where a dearth of jellyfish galaxies is apparent compared to the general population. While they account for 17 per cent of all satellites beyond half the virial radius, jellyfish galaxies only make up 10 percent of satellites found within $0.5\RTC$. In fact, the distribution of jellyfish galaxies seems to peak slightly at intermediate cluster-centric distances and at small cluster-centric distances only satellites with very high velocities can be jellyfishes.

The absence of jellyfish galaxies in the inner regions suggests that, indeed, ram-pressure stripping rather than tidal stripping is the process responsible for their formation since tidal stripping is only effective at small radii due to its strong radial dependence.
Both stripping mechanisms, in fact, are generally more effective in the inner regions: ram-pressure stripping due to the higher density of the ICM (see \cref{eq:Ram Pressure}), and tidal stripping since the tidal forces depend on distance from the potential center as $\propto r^{-3}$. Possibly, the stripping mechanisms may have already depleted the gas from the galaxies that are found at the time of observation at small cluster-centric distances. This can be seen in the top panel of \cref{fig:phase-space diagram}. While score = 0 galaxies have a more similar distribution in cluster-centric distance to the whole population of luminous galaxies, in general satellites that still contain some gas are biased towards larger distances from the host center: see black, dash-dot histogram. Namely, satellites in the central regions are found not to retain their gas as frequently as they retain their stars, and this bias may in itself explain the lack of jellyfish galaxies at small cluster-centric distances.

When examining the distribution of the velocities of satellite galaxies, we normalize the satellites's bulk velocities by the (3D) velocity dispersions of their host, calculated as the standard deviation of the bulk velocities of all subhaloes in that FoF halo (luminous and dark, with at least  32 resolution elements), in the reference system of their host.
We find that jellyfish galaxies typically have larger velocities than the general population, in qualitative agreement with observational findings based on projected velocities and distances (see \citealt{Jaffe2018GASP.Clusters}, their Fig. 7). This is true at all cluster-centric distances both for galaxies on infalling (radial velocities, $v_\mathrm{r}<0$) and outgoing ($v_\mathrm{r}>0$) trajectories.

When considering infalling and outgoing galaxies separately, we find slightly more infalling jellyfish galaxies: about 53 per cent of all the jellyfishes are on an infalling orbit. When we extend this division and compare infalling vs.\@ outgoing jellyfish to their parent samples, we find jellyfish galaxies constitute a higher fraction in the outgoing population. This is more evident when comparing to the `All satellites with gas' sample, where nearly a third of  all galaxies on outward trajectories are jellyfish galaxies compared to a fraction of less than a fourth of all infalling  galaxies. In other words, a satellite which manages to retain its gas after (at least) one pericentric passage is more likely to be a jellyfish galaxy (see \rfsec{5. Discussion and Implications}). 

We explored the phase-space distribution of satellites for the 4 redshift bins separately and found that the above trends hold at all studied times. The same is true when examining the distribution in separate host-mass bins and when restricting the stellar mass of the satellites to be greater than $10^{10}\MSUN$.

\subsubsection{Properties of the satellite gas vs. the ICM gas}
\label{sec:MaRamPressure}

\begin{figure*}
	\centering
	\adjustbox{trim={0.10\width} {0.09\height} {0.01\width} {0.03\height},clip}%
	{\includegraphics[width=\textwidth]{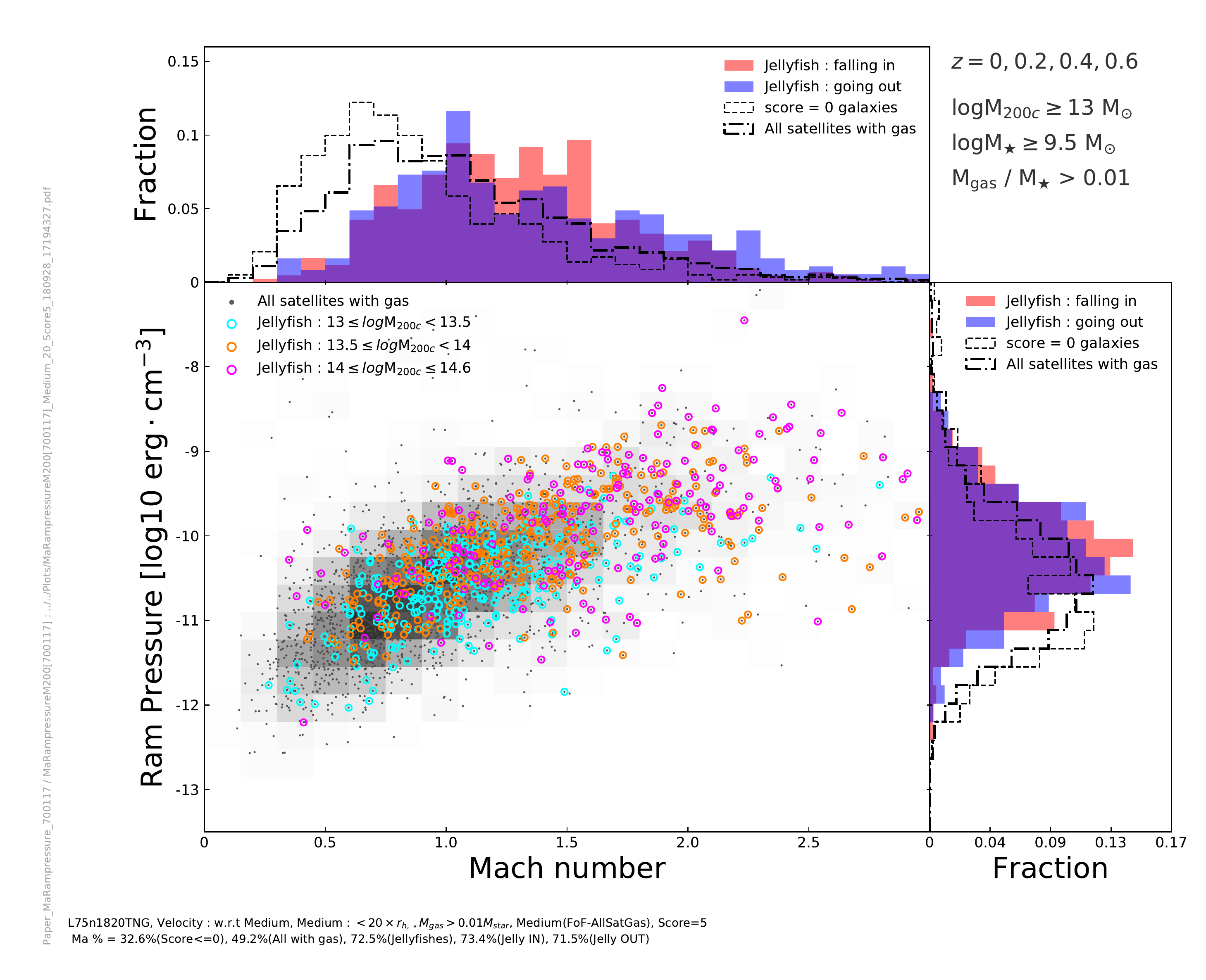}}
    \caption{The distribution of jellyfish galaxies (colored circles) and all satellite galaxies which contain some gas (black dots) in the ram pressure vs.\@ Mach number ($\mach$) plane. The jellyfish galaxies are color-coded according to three bins of host mass. The 2D histogram of the entire satellite population in the plot is shown in shaded gray, with darker gray denoting larger number densities of galaxies. On the top and to the right of the main panel we show the corresponding histograms of Mach numbers and ram pressures for the jellyfish galaxies -- blue and red for infalling and outgoing satellites, respectively, -- the population of satellites with gas (dash-dot black), and the galaxies that are not classified as jellyfishes (score = 0, dashed black). All histograms are shown as a fraction of the total number of galaxies in each sub-sample. Jellyfish galaxies are more likely to be supersonic ($\mach>1$) and also experience stronger ram pressure. Jellyfish galaxies in more massive hosts experience stronger ram pressure on average.}
    \label{fig:Mach number vs Ram pressure}
\end{figure*}

To link the jellyfish galaxies to the environmental processes which form them, we examine the Mach number of the galaxies, defined in \rfsec{machNumber}, and calculate the ram pressure exerted on them, as detailed in \rfsec{ramPressure}, and plot the results in \cref{fig:Mach number vs Ram pressure}. The full sample of galaxies which contain gas (for which the Mach number and ram pressure can be calculated) is shown as black dots and the jellyfish galaxies in this sample are further marked with colored circles, where the colors correspond to the virial mass of the host. Histograms of the distribution of the Mach number and ram pressure are shown above and to the right of the main plot, respectively. The histograms of jellyfish galaxies are separated into infalling (red) and outgoing (blue) trajectories while the distribution of the entire sample is plotted in dash-dot black and the subset of score = 0 galaxies in black dashed. 

Higher Mach numbers correspond to larger ram pressure, which one would expect since both properties depend on the velocity of the satellite (\cref{eq:machDef,eq:Ram Pressure}). When examining the jellyfish galaxies,  we find that satellites which reside in more massive hosts experience, on average, stronger ram pressure since the ICM is denser and the typical velocities are higher in these hosts. In terms of Mach numbers, we find a weak dependence of the Mach number distributions on host mass, with more massive clusters hosting satellites with Mach number distributions shifted towards larger values. 

The distribution of Mach numbers shows that jellyfish galaxies have higher Mach numbers than the general population with gas or than galaxies that are decisively not jellyfish galaxies (score = 0). For jellyfish galaxies, the Mach number distribution peaks at values of about $\mach \sim 1.2-1.4$,  while all satellites (and score=0 galaxies) peak at Mach numbers of about $\mach \sim 0.7$. A significant fraction of jellyfish galaxies (73 per cent) travel supersonically in the ICM, but not all of them and there are also satellites with supersonic motion that have not been classified as jellyfish galaxies (33 per cent). Interestingly, the Mach distributions of the infalling and \ky{outgoing} satellites are quite different, with larger values for the bulk of the galaxies falling in, but a more pronounced tail for satellites on outbound orbits at the high end of the Mach number range. We speculate that such high Mach numbers are related to the fact that galaxies in outgoing orbits are effectively traveling against the gas flow that may arise because of the ongoing gravitational collapse of their massive hosts or because of gas inflows from the large scale structure. As we will expand upon later, it turns out that satellite galaxies in outbound orbits and with high Mach numbers are in fact \ky{so-called} fly-bys\ky{, i.e. galaxies that are not directly falling towards the host center or may not be gravitationally-bound to the host but have} very large impact parameters and/or travel on very circular orbits in the cluster outskirts, further supporting the above interpretation. 

Jellyfish galaxies experience higher ram pressure than non-jellyfish galaxies, as one would expect, with infalling and outgoing jellyfish galaxies having indistinguishable distributions in ram pressure values. 

An interesting finding of this plot is that while the upper-right regions of the plot are dominated by jellyfish galaxies, one can still find satellites with high Mach numbers and that experience strong ram pressure, but are not classified as jellyfish galaxies. This could be due to the challenges of the visual classification of the satellites. A misclassification by even one of the Inspectors can result in tagging satellite as non-jellyfish since we based our identification on an unanimous vote from all visual Inspectors (\rfsec{3visual}). In addition, since we based our examination on images created in a random viewing angle, misclassification can occur when a jellyfish satellite is viewed from a direction that obscures its characteristic features (see discussion of this in \rfsec{optimal}). 
The complementary question -- why are there jellyfish galaxies with low values of ram pressure -- is also an interesting one, albeit one which is easier to answer. One must bear in mind that the extent of gas stripping in a satellite is determined by the balance between the ram pressure and the gravitational binding force. It is therefore quite possible for a satellite with a shallow potential well to experience extended stripping due to weak ram pressure, a line of analysis that will be pinned down in future work.
Finally, we also find a number of galaxies with low Mach numbers but large values of ram pressure that are not classified as jellyfish (top-left corner in the main plot of \cref{fig:Mach number vs Ram pressure}). 

We have checked all these cases and we find that, as mentioned above, galaxies with high ram pressure and high Mach number that are not jellyfish galaxies are often either fly-bys or on circular orbits at large cluster-centric distances: half of these (half a dozen in total) have actually a non-zero score, i.e.\@ a score between 2 and 4, and therefore could possibly be dubbed as misclassifications. On the other hand, the majority of those galaxies with high ram pressure but low Mach number and that are not jellyfish galaxies has indeed a score of 0; i.e.\@ all Inspectors think they are not jellyfish galaxies. Upon individual inspection, we find them close to the cluster centers or to other more massive galaxies (hence the high ram pressure values)  but with complicated and very disturbed gas distributions in their surroundings that have led them not to be classified as jellyfishes.

Now, in the main panel of \cref{fig:Mach number vs Ram pressure}, we find that for more massive hosts, jellyfish galaxies progressively occupy regions at higher values of ram pressure and Mach numbers: see trends in \cref{fig:Mach number vs Ram pressure}, with colors shifting towards the top-right corner of the panel for more massive hosts. Are such trends distinctive features of jellyfish galaxies compared to all satellites with gas? 
To confirm this, we compare (albeit do not show) the histograms of the Mach number and the ram pressure of both jellyfish galaxies and all satellites with gas in distinct bins of host mass. For both subpopulations of satellites, the range of ram pressure and Mach number values increases with increasing host mass. Importantly, however, such an enhancement is more pronounced for jellyfish galaxies: the deviations between the medians of the distribution of the two subpopulations increase with increasing host mass. This implies that as the mass of host increases, the jellyfish galaxies have a tendency to possess higher Mach numbers and experience larger ram pressures in comparison to the general population of satellites, hence making them effectively more prone to be ram-pressure stripped. This also confirms that the physical conditions in more massive hosts are responsible for the higher fractions of jellyfish galaxies in more massive hosts that we see in top right panel of \cref{fig:Jellyfish fraction}.

Finally, the results shown thus far may depend somewhat on the choices we make when determining the Mach number and ram pressure (sections \rfsec{machNumber} and \rfsec{ramPressure}). To explore this issue, we vary these choices and repeat our analysis in a number of ways, as we describe below. 
\begin{enumerate}
\item Definition of the medium: We change the definition of the medium so that medium properties are defined by the gas that is not gravitationally bound to the satellite subhalo and is found within 10 times the stellar half mass radius rather than 20 times in the fiducial analysis. This affects the typical velocity of the medium \mbox{$\vec{V}_\mathrm{medium}$}, the typical sound speed $c_\mathrm{s,medium}$ as well as the typical density of the medium $\rho_\mathrm{medium}$. 
Employing such a reduced physical aperture to determine the medium properties results in a smaller velocity difference between satellite and medium which in turn leads to a shift towards lower values of Mach number and ram pressure (\cref{eq:MachNumber,eq:Ram Pressure} ), but does not affect the qualitative conclusions reported above.

\item Definition of the medium: In the fiducial analysis, the medium gas is cleaned of all gas elements that are gravitationally bound to {\it any} subhalo within a given aperture, either the satellite itself under analysis or any neighboring one. Alternatively, we could have allowed the medium gas to include the gas of other galaxies in the same FoF host that might have occasionally be placed within the medium aperture: such galaxy gas would necessarily exhibit very different properties from those of the surrounding ICM. Through a one-to-one comparison of the Mach numbers and ram pressures of galaxies with such different operational definitions of medium gas, we find negligible differences: in fact, only a handful of satellites show deviations of up to 25 per cent while the majority of galaxies are labeled with Mach number and ram pressure values that are consistent to sub-percent accuracy. This is thanks to the fact that, even if large-density gas cells are included erroneously in the medium, the value of the `mode' for the medium properties (typical velocity, sound speed, and density) remains the same.

\item Definition of the satellite velocity: In the fiducial choice, the satellite gas velocity is determined with respect to the typical velocity of the surrounding medium gas. We have checked a plausible alternative: i.e.\@ we assume the typical velocity of the medium to be zero and simply use the satellite velocity in the frame of reference of the central galaxy in the host. The choice of the central galaxy frame of reference, rather than the host medium, is made in an attempt to mimic what observations can achieve and what is usually done in phase-space diagrams as the one in \cref{fig:phase-space diagram} \footnotemark. Under this assumption, the satellite velocities are found to be greater than in the fiducial choice, leading to higher Mach numbers and ram pressure values. For example, the fraction of jellyfish galaxies that move in supersonic motion changes from about 70 per cent in the fiducial case to about 90 per cent. We hence note that observer-friendly estimates are destined to overestimate the Mach numbers of orbiting satellites. However, also in this case, the distributions of jellyfish galaxies in comparison to all satellites with gas or score=0 galaxies are clearly different. 
\end{enumerate}
\footnotetext{Note that, exploring the relation between the center-of-mass velocity of central galaxies and their host halo, we find that while most central galaxies indeed move with their host haloes, a significant fraction ($\sim 13$ per cent at \zeq{0}) have a velocity difference with their host halo greater than 10 per cent. This is the case for which the most massive galaxy of a FoF is not dubbed the central, as it is not located at the absolute minimum of the potential.}

In summary, the trends we uncover in \cref{fig:Mach number vs Ram pressure} are preserved under different operational definitions for the measurements of the physical conditions: jellyfish galaxies exhibit higher Mach numbers and experience higher ram pressure than the rest of the satellite population. However, statements concerning what fractions of galaxies are sub- or super-sonic can change considerably according to the frame of reference and the way the medium is defined.

\subsection{Orientation of the tails}
\label{sec:tail_orientation}

\begin{figure*}
	\centering
    {\label{fig:tailVel}
    \adjustbox{trim={0.10\width} {0.09\height} {0.01\width} {0.05\height},clip}%
	{\includegraphics[width=1.12\columnwidth]{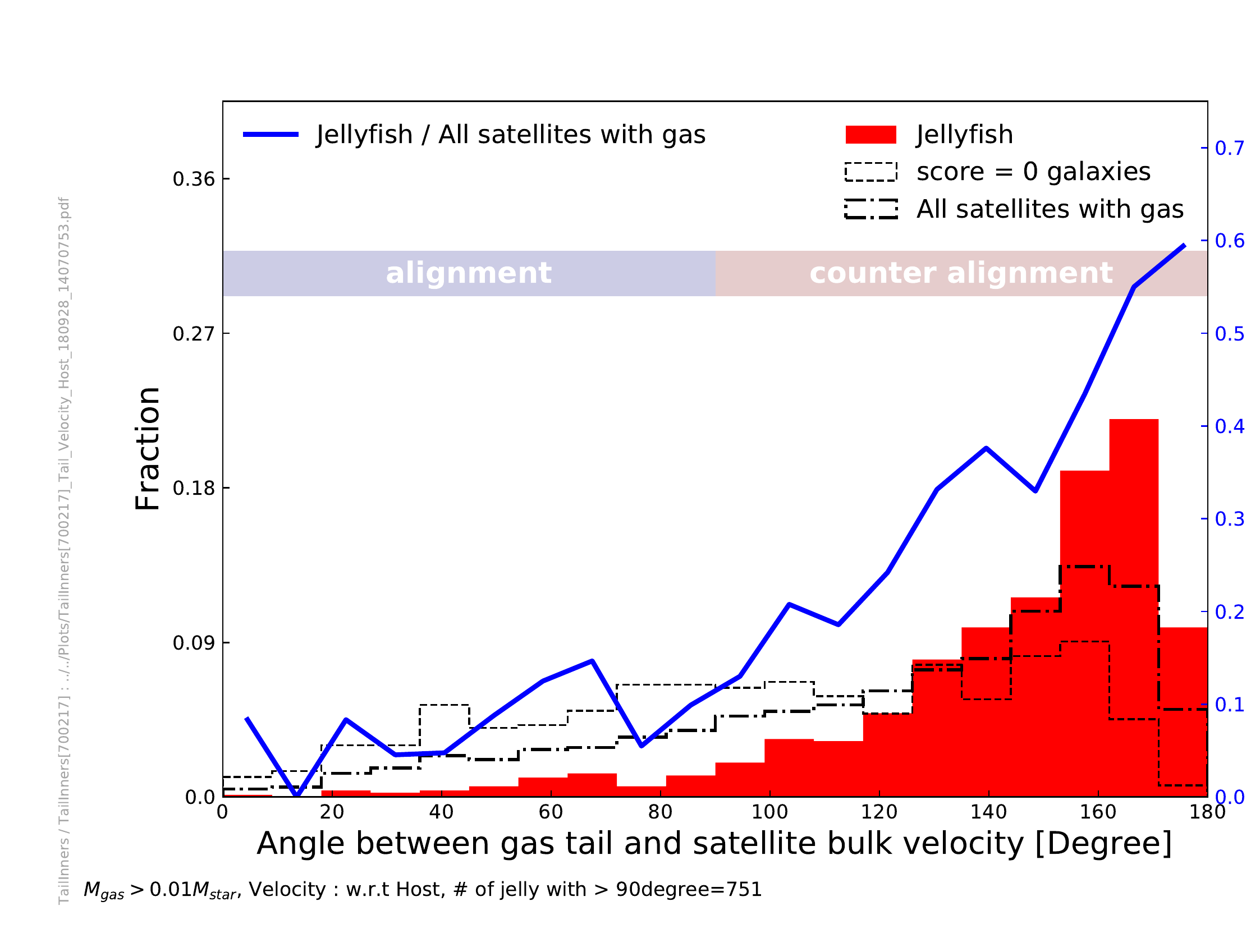}}}
    {\label{fig:tailHost}
    \adjustbox{trim={0.07\width} {0.09\height} {0.01\width} {0.05\height},clip}%
    {\includegraphics[width=1.12\columnwidth]{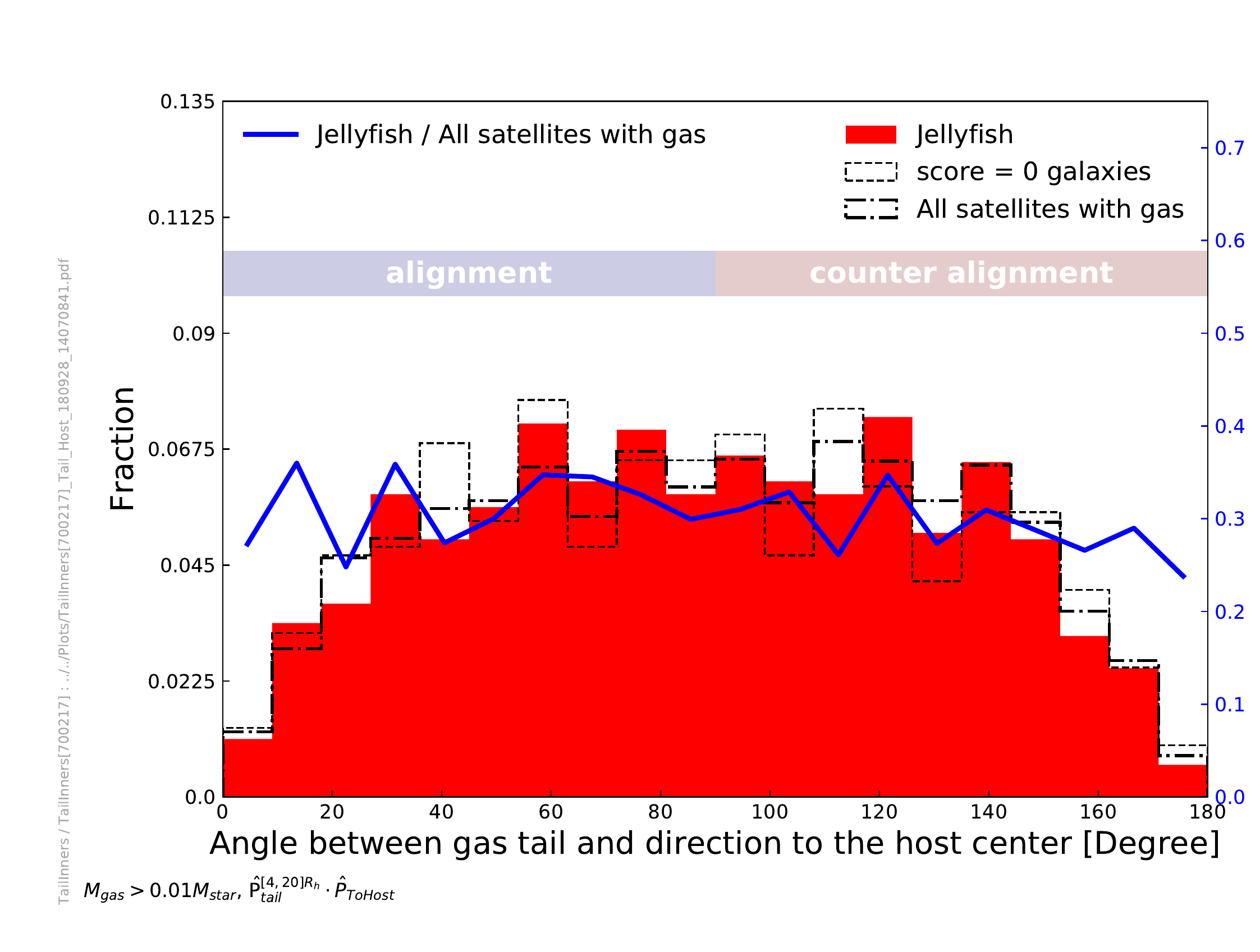}}}
    \caption{The distribution of angles between the vector of the gas tail of a galaxy (or more generally of the asymmetric gaseous distribution of a galaxy) and the vector of its bulk velocity (left) and the vector pointing towards the host center (right). The three histograms show the distribution of jellyfish galaxies (red), all satellites with gas (dash-dot black step histogram), and score = 0 galaxies (dashed back) as a function of the angles. The values below $90\degree$ signify alignment between the two vectors and values above $90\degree$ show anti-alignment. The fraction of jellyfish galaxies to all satellites with gas for each bin is shown as a blue line with values indicated by the right-hand y-axis. As always, in each histogram the normalization is made by accounting for all galaxies in the given sub-sample. \ky{These are ``true'' angle distributions, i.e. projection independent.}
The skewness of the jellyfish distribution to higher angles in the left panel, as compared to the general population, shows that the gaseous asymmetries of most jellyfish galaxies are anti-aligned with the velocity vector of the satellite. Conversely, there seems to be no connection between tail and host center in the jellyfish galaxies when compared to the general population of satellites.}    
    \label{fig:Tail}
\end{figure*}

In this section, we study the orientation of the gaseous tails (or asymmetric gas portions) that emanate from a galaxy with respect to both the direction of the satellite motion and the direction to the center of the host halo\ky{, all in the 3D space}. To do so we use our definition of the tail vector as defined in \rfsec{tailDefinition}. As detailed in \cref{sec:PropertyMeasurements}, the bulk velocity vector of a satellite is defined to be the center-of-mass velocity of all matter in the subhalo, in the frame of reference of the central galaxy in the host. The vector pointing from the satellite to the host center is defined as the direction to the host center. 

We calculate the angles subtended by the tail vector and the velocity vector for all satellites, and show the distribution in \cref{fig:Tail} (left), separated into the jellyfish galaxies (red), the entire satellite population with gas (dash-dot black), and non-jellyfish galaxies (score = 0, dashed black). \ky{These are true angle distributions, i.e. projection independent.} The ratio between the first two classes is also plotted (in blue, corresponding to the right-hand y-axis). Values of \mbox{$<90\degree$} imply an alignment between the two vectors and values of \mbox{$>90\degree$} signify anti-alignment. 

The distribution of the jellyfish galaxies is markedly skewed to larger angles, with about 94 per cent of jellyfish galaxies possessing angles greater than $90\degree$, clearly favoring a gas tail which is anti-aligned with the satellite's velocity vector. This result is consistent with ram-pressure stripping being the physical mechanism that produces galaxies with asymmetric gas distributions, i.e.\@ jellyfish galaxies. In the general population of satellites with gas, a similar trend is also present but in a much weaker fashion, furthering the finding that this is a strong feature of the jellyfish galaxy population alone. Indeed, for score=0 galaxies, the distribution is flat, consistent with a random orientation of the gas (almost symmetric, with gas orientation vector with very small module) with respect to the direction of the motion.

Though very few in number, we do find some jellyfish galaxies with gaseous tails that are somewhat aligned with the velocity vector. Upon closer study, we find that some of them show very low velocity even at large cluster-centric distance, implying that they are at the apocenter from their host. Upon further study of their trajectories, we find that the rapid change of the velocity direction at the apocenter is the cause for which the angle between the velocity vector and the previously-formed tails are small. 

An alignment of the tail in the direction of the host centre would point to the influence of the gravitational field of the host on the tail direction. In \cref{fig:Tail} (right), we find that values of the orientation angles with respect to the cluster centers are distributed fairly evenly and similarly so for all the various sub populations. This is evident by the nearly flat blue line which denotes the ratio between jellyfish and all satellites with gas.

\section{Discussion and Implications}
\label{sec:5. Discussion and Implications}

\subsection{On the physical conditions that produce jellyfish galaxies}

In the previous Sections, we have shown that a non-negligible fraction of simulated satellite galaxies that orbit in massive galaxy groups and clusters exhibit markedly asymmetric distributions of gas, or tails. We have dubbed them ``jellyfish'', following the tradition in the observational and theoretical literature. We associate this phenomenon with ram-pressure stripping, a working hypothesis that has been confirmed by a series of findings, chiefly: 1. jellyfish galaxies are more frequent at intermediate and large cluster-centric \ky{3D distances (in comparison to a stellar mass- i.e. magnitude-limited sample of galaxies)}; 2. on average they move at higher speeds and with larger Mach numbers in comparison to the general satellite population with the same minimum stellar mass; 3. they experience on average slightly larger ram pressures; and 4. their tails are generally oriented opposite to the direction of motion (i.e.\@ at about 180$^\circ$ angles), with no connection between the orientation of the tails and the position of the host centers. 

From visual inspection, we tentatively state that it is the interstellar medium that is stripped into the highly asymmetric gas distributions or tails of the jellyfish galaxies \citep[consistent with the findings of][]{Zinger2018QuenchingClusters, Bahe2016TheHoles}. The gaseous tails exhibit surface densities larger than $10^{6.5-7} ~{\rm kpc}^{-2}\MSUN$ (see images), decreasing without obvious discontinuities from surface densities in excess of $10^{7.5-8} ~{\rm kpc}^{-2}\MSUN$ within the innermost regions of the depicted galaxies, i.e.\@ in their interstellar medium. In fact, in this paper we consider only satellites found within the virial boundary of group and cluster hosts and \citealt{Zinger2018QuenchingClusters} has shown that the galaxy gas is stripped from satellites predominantly located inside the virial radius. Future studies will explore stripping in galaxies also at larger cluster-centric distances.
Furthermore, we find jellyfish galaxies with a broad range of stellar morphologies, both well-defined disks as well as spheroidal and rounder stellar-mass distributions.

\ky{For observational programs, we hence suggest that, for a parent sample of satellite galaxies selected at fixed magnitude limit in stellar light, jellyfish galaxies should be searched at intermediate cluster-centric distances and in more massive hosts. We expand on these ideas in what follows.}

\ky{In the previous Sections, we have uncovered a relative deficit} of jellyfish galaxies in the innermost regions ($\lesssim 0.25 \RTC$ in 3D clustercentric distances) of clusters: \ky{such deficit is to be intended {\it in comparison to a stellar mass- i.e. magnitude-limited control sample of satellites} and we think we understand it.} In general, satellites that retain their gas (and hence can still appear as jellyfish) are preferentially found at larger distances in comparison to the entire population of luminous satellites, suggesting that by then the gas has been removed efficiently. In fact, regions at small cluster-centric distances and small velocities are preferentially occupied by satellites that fell into the current host a long time ago, possibly many Gyrs before the time of observation (see \citealt{Rhee2017Phase-spaceLoss}, their Fig. 2). On the other hand, in more massive hosts, satellites experience larger ram pressures because the ambient gas (the host ICM) is, on average, denser than in low-mass hosts and because galaxies move overall faster. 

\ky{Importantly, we find that our} result for higher probabilities of having jellyfish galaxies at intermediate and large cluster-centric distances also holds when the phase-space diagram of the satellite population (\cref{fig:phase-space diagram}) is studied in 2D projected cluster-centric distance. \ky{The fractions of jellyfish galaxies at small 2D projected distance are slightly higher (\kyb{7.4 per cent}) than those in 3D cluster-centric distance (\kyb{4.1 per cent}), in both cases for regions within $0.25 \RTC$ and in comparison to all luminous satellites in our parent sample. These fractions correspond to samples of jellyfish galaxies of the order of more than 100 jellyfish examples within $\leq 0.25 \RTC$: namely, although there is a relative deficit, we do still find jellyfish galaxies in the cluster cores.} 
\ky{On the other hand, \citealt{Jaffe2018GASP.Clusters} find the jellyfish galaxies with the longest gas tails to reside very near the cluster centers, namely within $0.5 \RTC$ in projection. Unfortunately, a comparison of the frequency of jellyfish examples with that work is not straightforward, as the parent sample of galaxies studied within the GASP program is different from ours, with e.g. different stellar-mass cuts. However, we have checked whether also in our simulations we find the} most striking stripped tails to reside at small projected cluster-centric distances. \ky{Based on a visual inspection of jellyfish galaxies within and outside $0.5\RTC$, we find exceptionally long and clear ram-pressure stripped tails both near ($\leq 0.5\RTC$) as well as far ($\geq 0.5\RTC$) from the host centers, in similar fractions. Understanding such discrepancy requires, in our view, an automated classification of jellyfish galaxies, identically applied to both observed and simulated galaxies.}

Interestingly, we find that satellites in outgoing trajectories present slightly larger fractions of jellyfishes than those in infalling orbits. Clearly, the former must have passed at least once through pericenter. Furthermore, jellyfish galaxies in outgoing orbits can reach larger Mach number values than their infalling counterparts (see \cref{fig:Mach number vs Ram pressure}, top panel, blue vs.\@ red histograms): this difference however does not hold in the bulk of the two populations -- the median Mach numbers of the red and blue distributions are essentially indistinguishable -- but only in the high-Mach number tails. A similar description applies to the Mach-number distributions of satellites with gas that are falling in and going out, not just jellyfish galaxies (not shown). All these considerations make an argument based on the different velocities for falling-in vs. outgoing-in jellyfish galaxies a not viable culprit. Instead, we find that timing arguments are the interpretation key.

\begin{figure*}
   \centering 
    \adjustbox{trim={0.08\width} {0.08\height} {0.02\width} {0.01\height},clip}%
	{\includegraphics[width=0.5\textwidth]{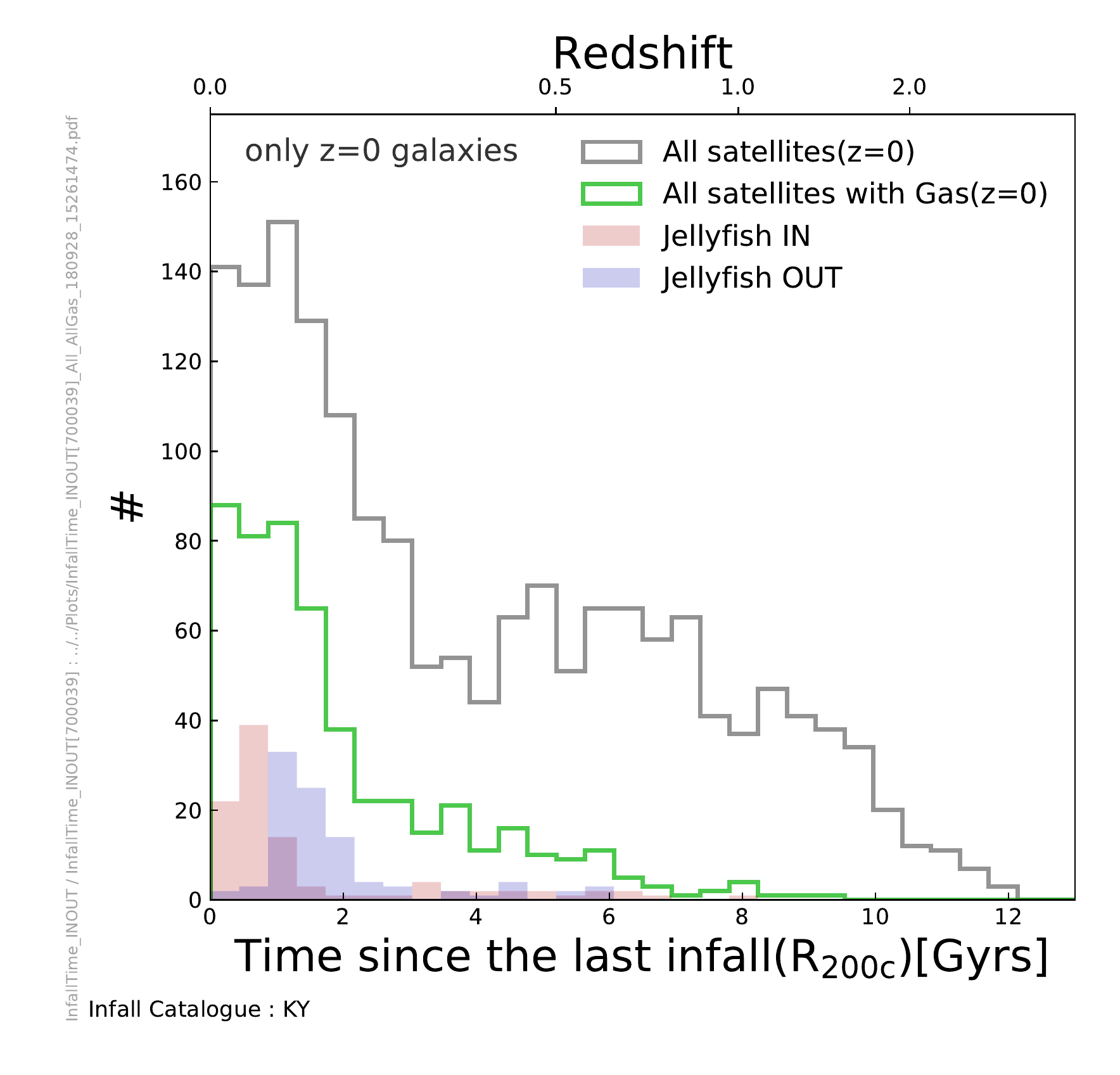}}
    \adjustbox{trim={0.08\width} {0.08\height} {0.02\width} {0.01\height},clip}%
    {\includegraphics[width=0.5\textwidth]{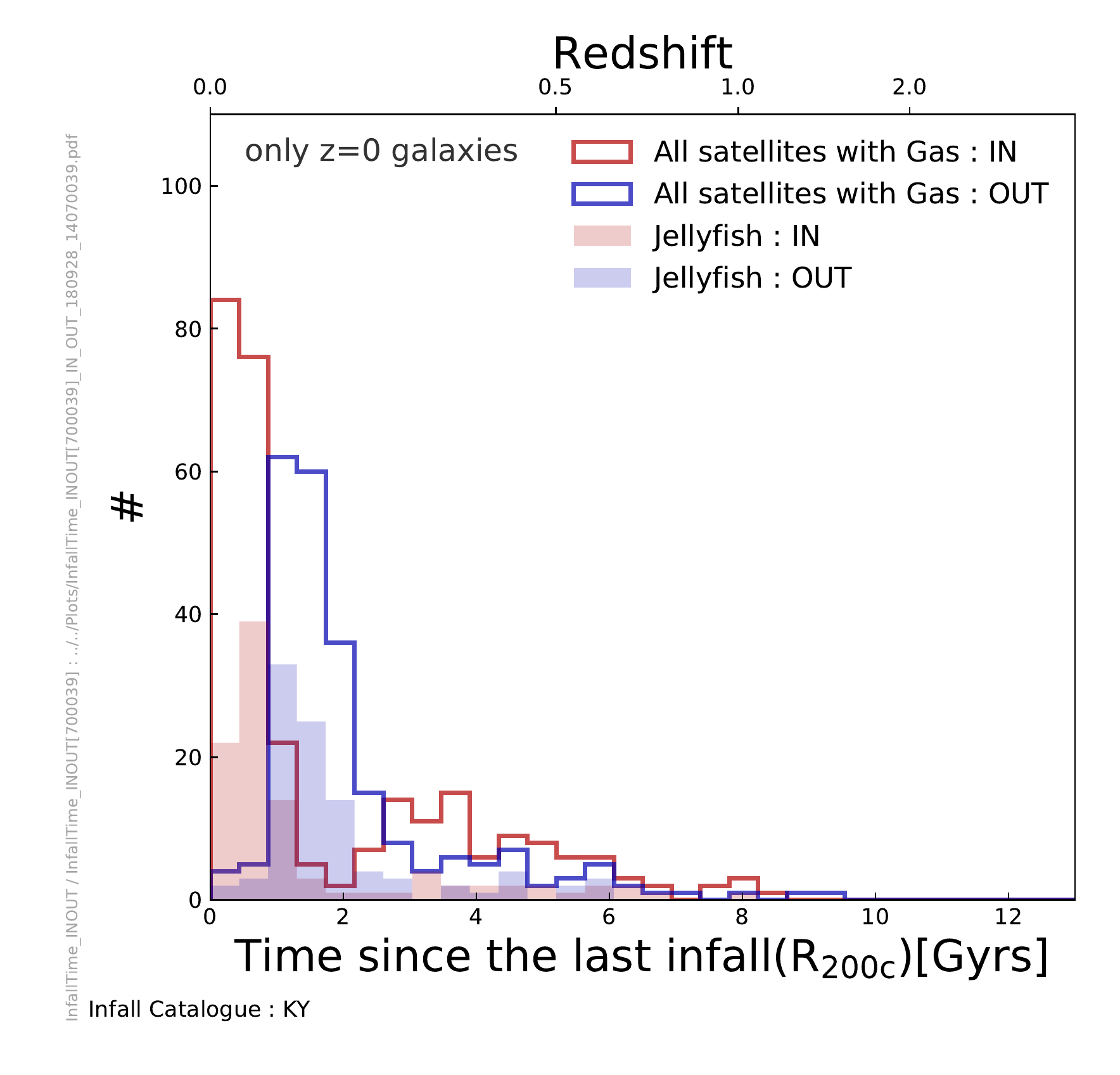}}
    \caption{Distributions of the infall times of various subsets of satellite galaxies selected at $z=0$. Here by infall we mean the moment when a satellite crosses the virial radius ($\RTC$) of its current host for the last time. In both panels, jellyfish galaxies `IN' (red filled histograms) and `OUT' (blue filled histograms) denote score=5 galaxies that are infalling or going away from their host center based on the sign of their radial bulk velocity. We note that the great majority of jellyfish galaxies are recent infallers ($\leq 2.5$ Gyrs ago); instead  95 per cent of satellites with gas have fallen into their last host up to 6 Gyrs ago.}
    
    \label{fig:infallTimes}
\end{figure*}

In \cref{fig:infallTimes}, we show the distributions of the infall times of all the $z=0$ satellite galaxies. Here infall is defined as the moment when a satellite crosses the virial radius of its current host for the last time \citep{Rodriguez-Gomez2015TheModels,Chua2017SubhaloVariation}. Firstly, all satellites in our parent samples (both black and green curves, left panel) have somewhat bimodal distributions of infall times, separated at about the value of 3 Gyrs ago. This bimodality is highly reduced if we were to include also \ky{back-splash} satellites, i.e.\@ satellites that have already fallen into the $z=0$ hosts in the past, but are at the moment beyond their virial radius and hence are not part of our samples. They would fill up the distribution of ancient and intermediate ($> 3$ Gyrs ago) infallers. Secondly, the satellites with gas are biased compared to all satellites in the sample: 95 per cent of satellites identified at $z=0$ to have some gas have fallen into their current clusters more recently than 6 Gyrs ago and overall their infall-time distribution is skewed towards more recent infall than the general population. This means that gas is efficiently stripped within the first 3-4 Gyrs of life in the dense environments, on average, and at least for our mass-limited satellite sample (see also Donnari et al.\@ in prep). Thirdly, and importantly, jellyfish galaxies are markedly {\it recent infallers}, and more so than any of the parent samples: there are very few jellyfish galaxies that have started to orbit in their current host more than 2.5 Gyrs ago. Very similar results have also been inferred from observations (\citealt{Jaffe2018GASP.Clusters}). In contrast, we find that the distribution of infall times for score=0 galaxies is indistinguishable from that of all satellites. 

Finally, the bulk of the jellyfish galaxies on outgoing trajectories are on average orbiting in massive hosts for about 1 Gyr longer than their in-falling peers (red and blue shaded histograms in \cref{fig:infallTimes}). The longer (but not too long) permanence in massive hosts is not however the reason for which outgoing satellites have higher jellyfish fractions. In fact, in the right panel of \cref{fig:infallTimes}, we compare the infall-time distributions of all galaxies with gas (not just jellyfish ones) by distinguishing between in-going and out-going satellites: thick red vs.\@ thick blue empty histograms. The formers (red empty histogram) exhibits a more populous tail at ancient infall times (longer than 3 Gyrs ago)
\footnote{This population discrepancy between in-going and out-going satellites can be largely attributed to \ky{back-splash} galaxies that preferentially replenish the number of in-going satellites. The majority of \ky{back-splash} galaxies that go back into the host cluster are more likely to be classified as $z=0$ in-going satellites because of their infalling trajectories.}.
These correspond to galaxies that, selected at $z=0$, have by now gone through multiple pericenter passages. Of those, only a negligible fraction are jellyfish, as it can be seen by comparing the red empty vs.\@ the red shaded histograms: this makes the overall fraction of jellyfish galaxies in incoming trajectories smaller than for satellites on outgoing orbits.

\subsection{On the effects of numerical resolution}
So far, we have studied satellite galaxies in only one IllustrisTNG box realization: the $\sim100$-Mpc on-a-side volume at its best resolution \ky{and we have} prevented dramatic resolution issues by selecting, a priori, only satellite galaxies that are sufficiently massive to host more than e.g.\@ $10^{9.5}\MSUN$ in stars, i.e.\@ with more than about three thousand stellar particles in total per galaxy. How are our results affected by numerical resolution?

We have repeated the visual inspection of $z=0$ satellites with gas (same selections as in the fiducial case, only one Inspector) for the lower resolution versions of the flagship simulation, TNG100-2 and TNG100-3, at respectively 8 and 64 times worse particle mass resolution than TNG100. In TNG100 and TNG100-2 the trends of the fractions of jellyfish galaxies with host mass, stellar mass, and satellite-to-host mass ratio are remarkably similar to one another, with on average a fraction of $1.3-1.5$ more jellyfish galaxies (here dubbed with score = 1, given the single inspector) in the lower resolution version. This is reasonable as lower resolution implies, among other effects, shallower mass distributions, that are hence more prone to stripping than at better resolution. This is also consistent with the fact that at lower resolution the fractions of quenched satellites is larger than at higher resolutions (see Donnari et al.\@ in prep.\@). \kyb{However we note that the fractions of jellyfish galaxies are not always higher at lower resolution. If we compare the jellyfish fraction with a more conservative mass cut (i.e. $\MSTARS \ge 10^{10.5}\MSUN$) so that also for TNG100-2 galaxies are resolved with at least a thousand stellar particles, then the fractions from both TNG100 and TNG100-2 are well within the 1-sigma statistical error bars of \cref{fig:Jellyfish fraction}.} 

On the other hand, the findings for TNG100-3 are very difficult to interpret, as the gas distributions are sampled with just too few cells for the gaseous projections and the visual inspection to be reliable. We believe that the TNG100 vs.\@ TNG100-2 comparison reasonably suggests that the qualitative (and quantitative) findings of this paper are not (dramatically) affected by numerical resolution. \kyb{An ultimate resolution test of the findings of this paper will be achieved with the upcoming TNG50 volume, at about 15 (2.5) times better mass (spatial) resolution than TNG100.}

\subsection{On the effects of magnetic fields and galaxy-physics model}
The TNG100 box adopted throughout this analysis is a magneto-hydrodynamical simulation, for which ideal magneto-hydrodynamics allows us to follow the coevolution of gas and magnetic fields starting from a small magnetic field seed in the initial conditions (see \citealt{Pillepich2018FirstGalaxies} for details). Suggestions have been made that the survival of ram-pressure-stripped tails may be different if magnetic fields are included or not. 

We anticipate that MHD is not a pre-requisite for the presence of jellyfish galaxies, as in fact many galaxies with extended gaseous tails can be found also in the original Illustris simulation\footnote{E.g.\@ by simply inspecting the Explorer at www.illustris-project.org.}. Illustris is the realization of the same patch of synthetic universe as TNG100 but with a different underlying galaxy-physics model and in particular, among other differences, without MHD and with a different implementation of the BH feedback at low-accretion rates.

We have analyzed the $z=0$ galaxies in Illustris with the same procedure and selection as done for TNG, but with only one visual Inspector. It turns out that the fraction of galaxies being ram-pressure stripped at $z=0$ is lower in Illustris compared to TNG100 (24 per cent vs.\@ 38 per cent at $z=0$) but the trends with the host-to-satellite mass ratio and the virial mass are similar in both simulations. In Illustris, quenching mechanisms are less efficient for both centrals as well as satellites in groups and clusters (Donnari et al.\@ in prep). \ky{The} original Illustris simulation is affected by a lack of gas within the virial radii of massive groups and clusters \citep[][their Fig. 8]{Genel2014IntroducingTime, Pillepich2018SimulatingModel} because of the effects of the violent AGN feedback. These findings are all consistent with the idea that the intra-group and intra-cluster media in Illustris and TNG are different, namely, the gas content of the ICM is lower in Illustris. The deficit of the ICM in turn influences the effectiveness of ram-pressure stripping and overall leads to a lower number of satellites that exhibit tails of stripped gas, i.e.\@ jellyfish morphologies. 

To isolate the possible effects of MHD alone, we have compared the outcome at $z=0$ of two smaller simulations of the same cosmological volume (25 $h^{-1}$ Mpc on a side) at the same TNG100 resolution, with and without magnetic fields. The statistics is necessarily very limited, just about 30 satellite galaxies with gas in a handful of group-like hosts, admitting as low as one gas cell per galaxy. We find that the fraction of jellyfish galaxies (upon visual classification of two Inspectors) is about $25-30$ per cent in both cases, similar to the fiducial results if the gas condition was modified to $\MGAS > 0$. We also find that the overall trends of increasing (decreasing) jellyfish frequency with host mass (satellite-to-host mass ratio) are reproduced in both cases. However, a more detailed comparison of analog galaxies across the two simulations and of their gaseous morphologies needs to be postponed to a targeted study as it is beyond the scope of this paper. 

\subsection{On the effects of orientation for jellyfish identification}
\label{sec:optimal}

\begin{figure}
\hbox{{
\adjustbox{trim={0.0\width} {0.006\height} {0.0\width} {0.0\height},clip}%
{\includegraphics[width=0.99\columnwidth]{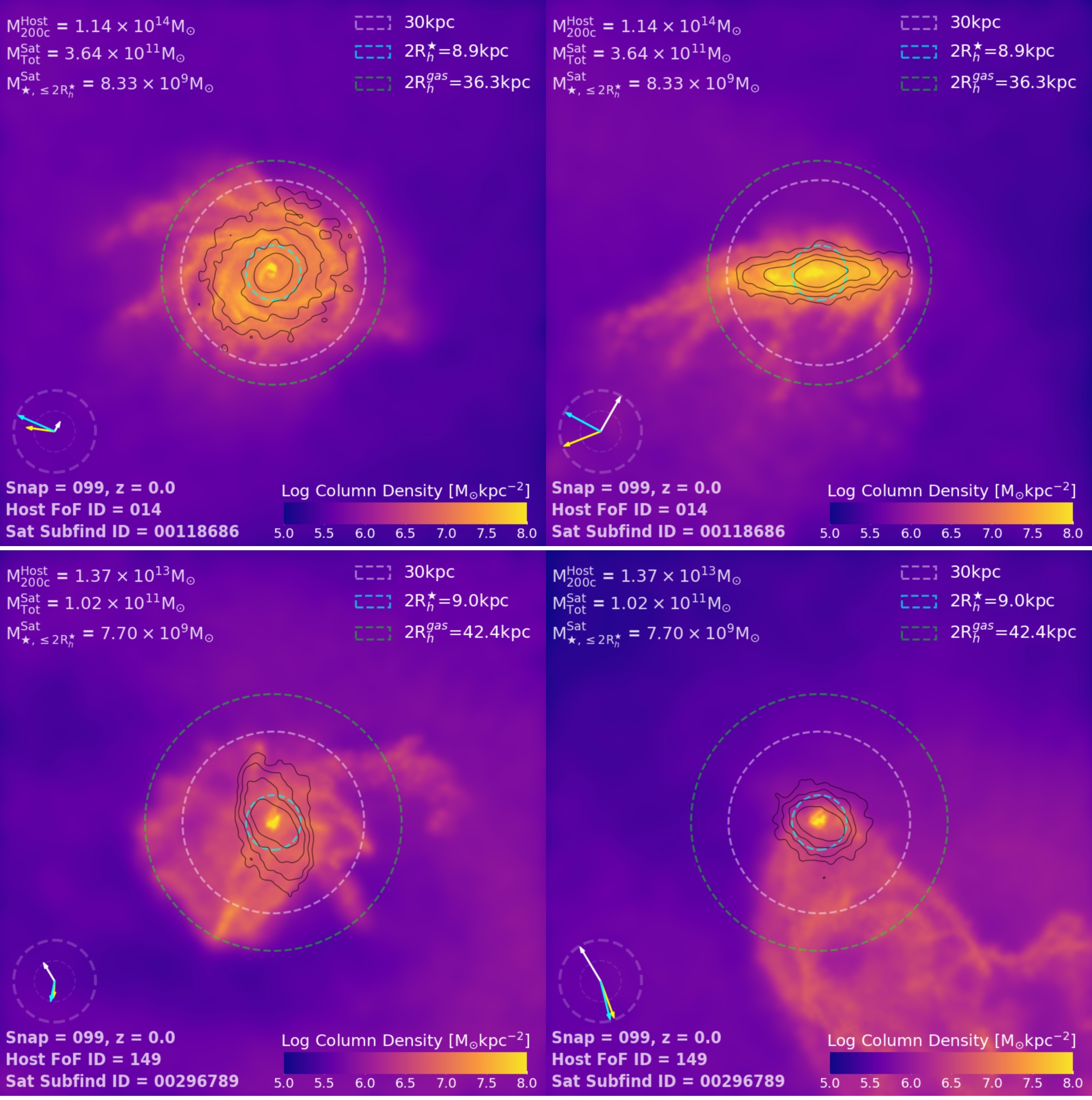}}}}
\hbox{{
\adjustbox{trim={0.0\width} {0.005\height} {0.0\width} {0.0\height},clip}%
{\includegraphics[width=0.99\columnwidth]{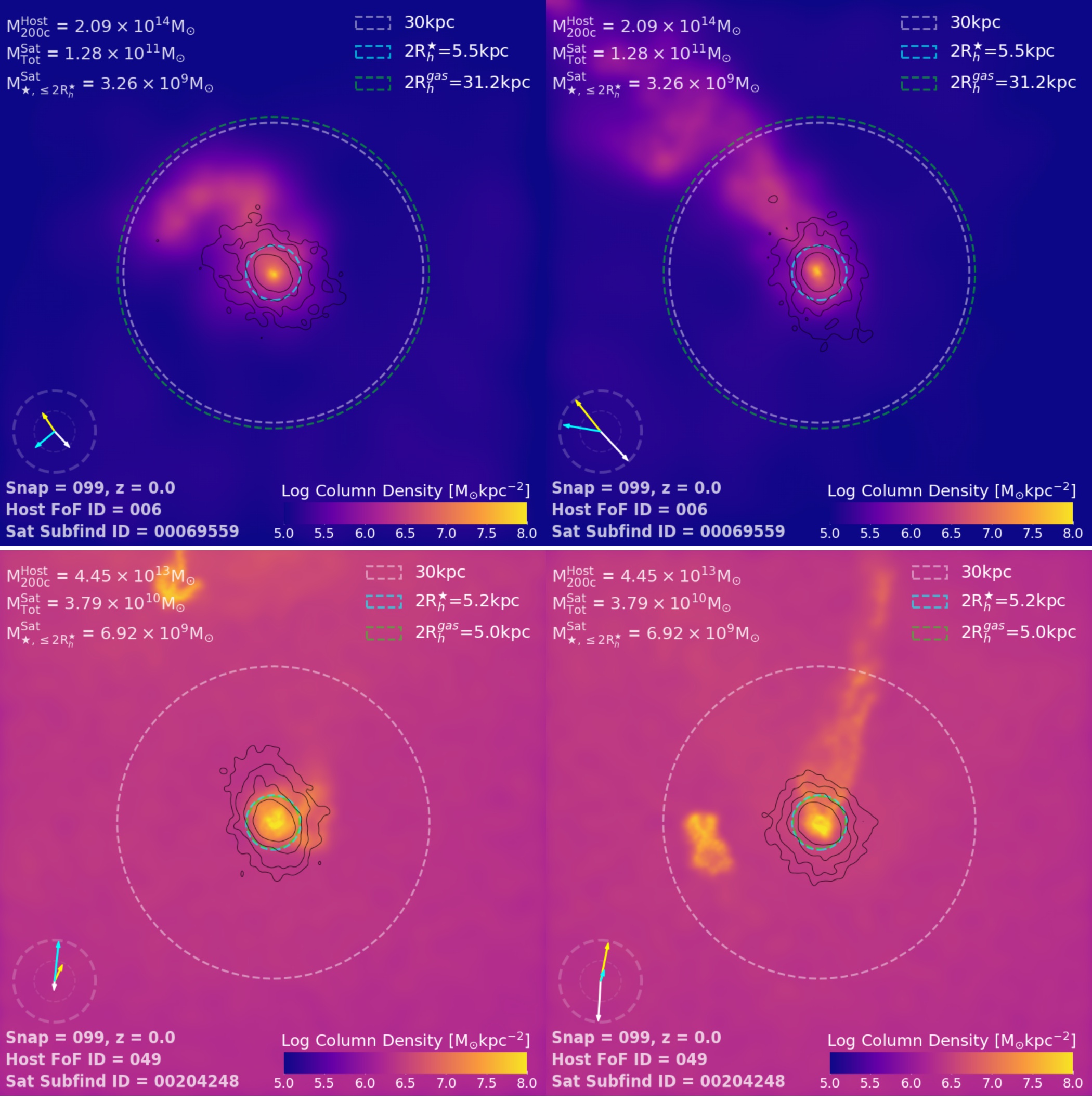}}}}
\caption{Four examples of TNG100 satellite galaxies (from top to bottom) depicted in random (left panels) and optimal (right panels) projections. As in \cref{fig:Images with Scores}, we show both the gas column densities along the line of sight as a color map and the stellar mass density contours in black solid curves. The optimal projection is defined as the one where the bulk satellite velocity and the gas tail vectors both lie in the plane perpendicular to the line of sight. In each panel, in the left-bottom corner, the projected vectors of the satellite velocity (white), of the gaseous tail (yellow), and of the direction to the host center (cyan) are shown in relation to one another. The thick-dashed and thin-dashed white circles provide a measure of the 100 and 50 per cent, respectively, of the 3D norm of each vector.}
   \label{fig:optimalprojection}
\end{figure}

So far we have analyzed the demographics and properties of jellyfish galaxies that have been identified in random projections in the simulation box, i.e.\@ by replicating the situation an observer would be facing, with galaxies and their tails in random projections on the sky. However, in \cref{fig:Tail}, we have demonstrated that ram-pressure stripped gas asymmetries and tails are preferentially oriented in opposite direction with respect to the galaxy's bulk velocity. This suggests that favorable configurations exist for gaseous asymmetries and tails to be more easily identified. Here we therefore study the impact of the projection effects on our statistical results by studying TNG galaxies in optimal projections.

In practice, we now project the gas column densities of our simulated galaxies on a plane that enhances the appearance of possible gaseous tails. Namely, we rotate the view direction in order to make the galaxy's bulk velocity vector lie on the image plane. 
Examples of the comparison between random and optimal projections are given for four galaxies in \cref{fig:optimalprojection}. Each pair of stamps represents one TNG satellite, in random (left) and optimal (right) projection, i.e.  in the latter case, with the velocity vector perpendicular to the line-of-sight.
In each panel, on the lower left corner, three vectors are represented in their respective projections: the satellite's bulk velocity (white), the satellite tail direction (yellow), and the pointer to the host center (cyan). The thick and thin dashed white circles represent the 100 and 50 per cent of the 3D norm of each vectors. As one can notice, in the randomly-projected images (left column), the vectors are smaller than their 3D 100 per cent norm would indicate, whereas in the optimal projections the velocity vector (white arrow) always extends in the plane of the image for 100 per cent of its 3D norm, as per construction.

The re-orientation of the galaxies clearly uncovers gaseous asymmetries and tails that might have gone unnoticed in the random projections. In particular, the top two galaxies of \cref{fig:optimalprojection} are actually examples of satellites that none or only one Inspector would have labeled as jellyfish in the random projections (scores = 0 or 1). The bottom two are cases of scores between 2 and 4 in the random visual inspection. In all these examples, the optimal orientation would have obviously brought the observed galaxies to be considered highly asymmetric, i.e.\@ jellyfish. 

By repeating the visual analysis (one Inspector) in the optimal orientation, we find that the fraction of jellyfish galaxies increases from 38 to 55 per cent between the random- and optimal-projection approach. In other words, 30 per cent of actual jellyfish galaxies are missed or mis-classified because of the limitations of random orientations. Hence, all the numbers quoted thus far in the paper with the random orientations shall be considered lower limits, that can be corrected by a factor of about 1.4 to get true values. This applies in similar terms to any straightforward demographic statistic about asymmetric gas distributions that could be produced observationally. 
Importantly, we find that the vast majority of galaxies that are jellyfish in the optimal orientation but are not in the random one are those with high random-projection scores. In fact, only a handful of galaxies that are originally labeled as score = 0 galaxies are in fact jellyfish galaxies. 

Interestingly, the trends in jellyfish fractions as function of host total mass, satellite-to-host total mass ratio, and satellite stellar mass are all consistent with those shown in \rfsec{4.1. TNG jellyfish galaxies: basic demographics}. Projection effects do not impact in any biased way our ability to determine general demographics and environmental dependencies. In other words, the mechanisms that drive the formation of gaseous tails depend sufficiently strongly on the environmental interactions and galaxy properties that the effects of random orientations can not wash them out, at least on a statistical sense and given a large-enough sample of galaxies. Yet, small effects can be noticed in the Mach number and ram pressure distributions: the jellyfish galaxies that are newly discovered in the optimal projections occupy regions of the parameter space that extend to lower values of Mach numbers and ram pressures. The visual classification that uses random projections is destined to miss mild or weak gaseous asymmetries as those that could be caused by mild or weak environmental configurations. \ky{On the other hand, among those galaxies with gas and with supersonic motions at $z=0$ that have not been tagged as jellyfish in the random projection, only 49 have been recovered as jellyfish galaxies in the optimal projection. Namely, our recovery fraction for galaxies with Mach number larger than 1 is about 37 per cent and many satellites with supersonic speeds are not-jellyfish galaxies, regardless of orientation and probably because they are fly-bys (see \cref{sec:ramPressure}). In fact, among the satellites that move supersonically, the fraction of jellyfish galaxies goes from about 67 per cent (see \cref{sec:MaRamPressure}) in random orientations to about 80 per cent in the optimal configuration. }

\subsection{On the role of supersonic motion: bow shocks}
\begin{figure*}
	\includegraphics[width=0.99\textwidth,height=0.9\textheight, keepaspectratio=true]{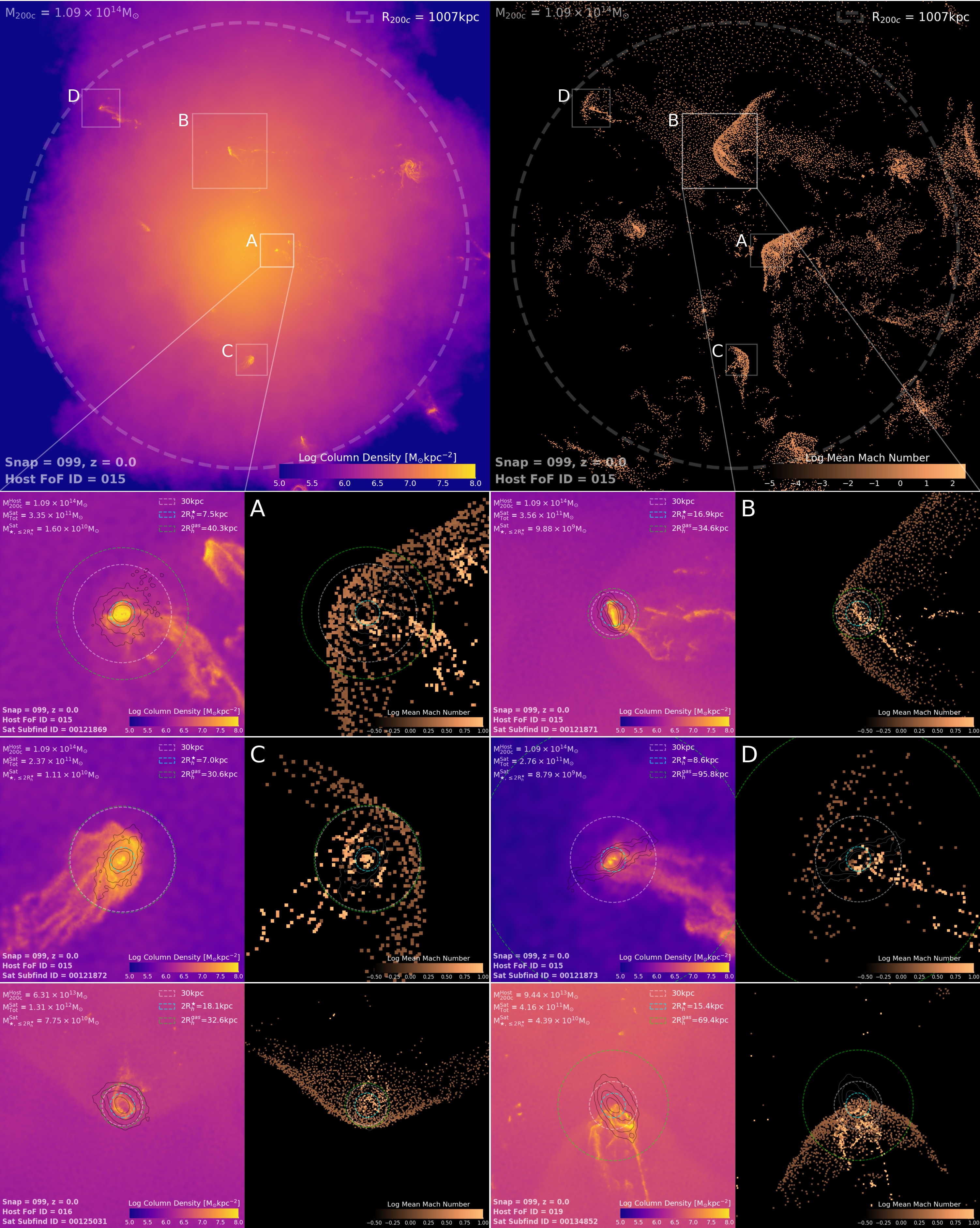}
   \caption{Gas mass density projections (left panels) and mass-weighted mean Mach numbers in the gas cells along a random line of sight (right panels) in a group-mass halo (top) and a selection of its and others satellite galaxies. The Mach number projections reveal well-defined bow shocks in front of galaxies with clear evidences of ram-pressure stripping. The spatial scales of the bow shocks are remarkably larger than those of any other galaxy feature: they cover vast areas in the projected cluster volumes (top-right panel), highlighting features otherwise not as vivid in the gas mass density maps (top-left panel).}
    \label{fig:BowShocks}
\end{figure*}

In the previous Sections, we have demonstrated that jellyfish galaxies exhibit larger 3D velocities and hence higher Mach numbers than the general population of satellites. More than 70 per cent of jellyfish galaxies move at supersonic motions (Mach numbers $> 1$) with respect to their surrounding intra-cluster medium. This is not a necessary condition. However, supersonic motions are usually associated with the formation of bow shocks in front of the moving body. How are jellyfish galaxies connected with the presence of bow shocks?

To answer this question, we have constructed images of the spatial distribution of the Mach numbers of individual gas cells in and around the satellite galaxies in our sample. In particular, we have adopted the same field of view and choices of the gas-density projections of \cref{fig:Image Sample,fig:Images with Scores,fig:Jellyfish Galley 1,fig:Jellyfish Galley 2} (random projection) and we have projected the line-of-sight mass-weighted mean of the Mach numbers of the gas cells, with a pixel size fixed to 1 comoving kpc per side and without any smoothing kernel. To do so we use the output of the shock finder algorithm implemented in {\sc arepo} in \citealt{Schaal2015} and applied in post processing to e.g.\@ the Illustris simulation in \citealt{Schaal2016ShockUniverse}. After finding hydrodynamical shocks in the cosmic gas, the shock finder (applied now on-the-fly on the TNG data) labels each gas cell with a Mach number value, with respect to surrounding gas. 

Beautiful examples of these projections can be found in \cref{fig:BowShocks}, for both the gas in a $10^{14}\MSUN$ cluster as well as in the zoomed-in stamps of some of its satellite galaxies and others. Strong gas discontinuities can be identified in the spatial projections of the gas-cell Mach numbers in and around jellyfish galaxies: these are effectively bow shocks, sometimes visible also in the gas mass density projections. The spatial extent of such bow shocks is remarkable, as they often appear larger than any other satellite feature, as can be appreciated in the zoomed-out cluster images (top panels of \cref{fig:BowShocks}). Note that the Mach numbers in the bow shocks are never, on average, supersonic. This is due to the fact that the speed of the gas within the bow shocks is significantly lower than the sound speed due to its high temperature: bow shocks are stationary-wave fronts where the transferred hydrodynamical quantities are stacked, so that the speed decreases whereas the pressure and temperature increase accordingly. However, interestingly, important discontinuities (higher Mach numbers of the gas) can be found also within the gaseous tails themselves, where the Mach numbers are higher than in the bow shocks: something to be further investigated in future works.

\begin{figure}
	\centering
    \adjustbox{trim={0.09\width} {0.08\height} {0.01\width} {0.10\height},clip}%
    {\includegraphics[width=1.00\columnwidth]{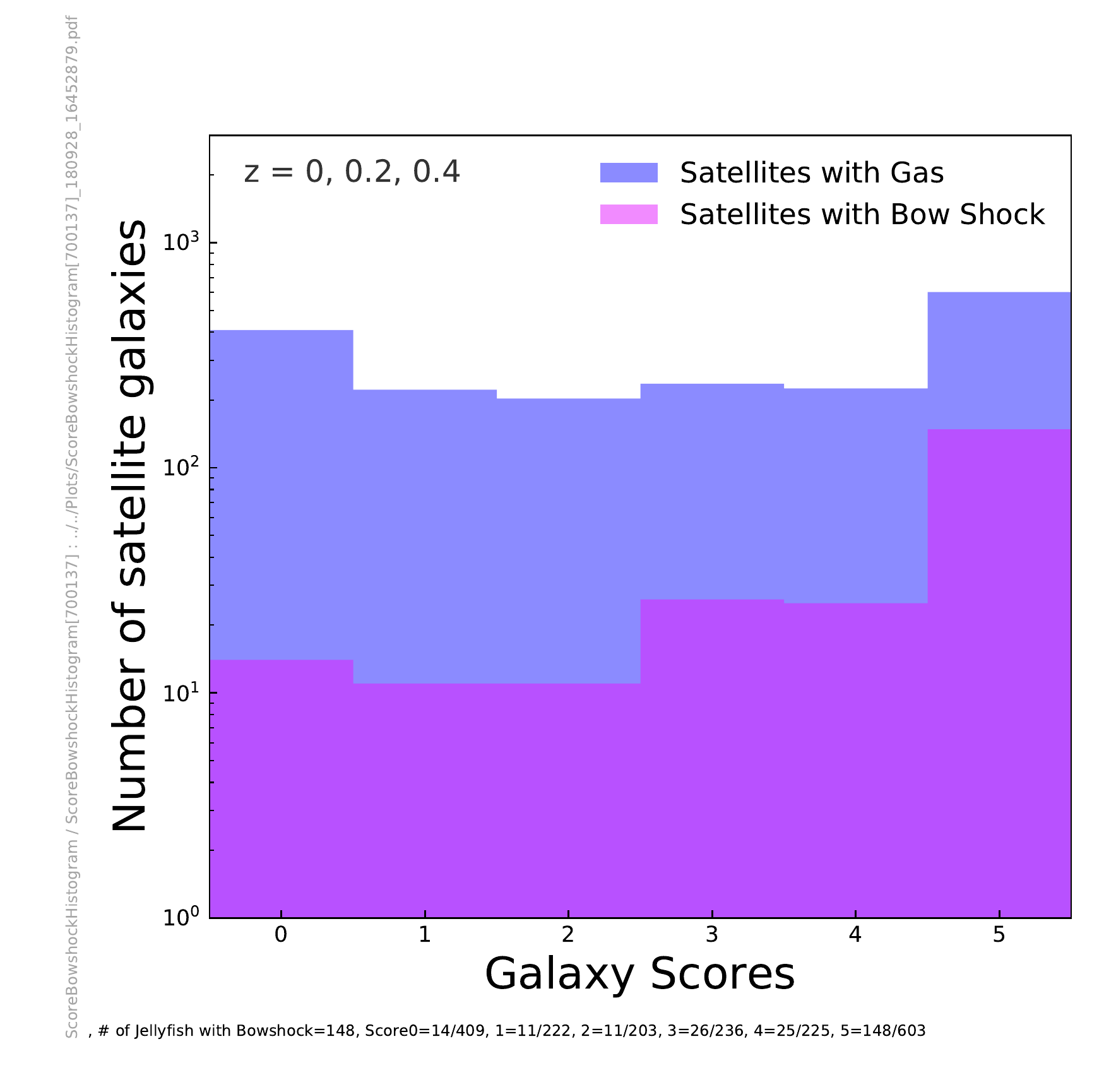}}
    \caption{Numbers of all satellites with gas (in blue) vs.\@ numbers of satellites with gas and with manifest bow shocks in their surroundings (in magenta), as a function of jellyfish score, from the visual classification based on five Inspectors. Jellyfish galaxies are those with score equal to 5. Here we only include three of the four snapshots studied so far, because at the highest redshift snapshot data has not been saved at the particle level to identify bow shocks. The majority of galaxies that exhibit a bow shock are score = 5 galaxies, i.e.\@ jellyfishes.}
    \label{fig:BowShocksStats}
\end{figure}

To quantify the incidence of bow shocks, images like the ones in \cref{fig:BowShocks} have been created for all satellites with gas and hence dubbed, by one Inspector, as exhibiting clear bow shocks or not, without prior knowledge about the presence of a tail from the gas maps -- here we use random projections. The results of this investigation are summarized in \cref{fig:BowShocksStats}. We find that, among the snapshots studied at $z\leq 0.4$, in 63 per cent of the cases, satellites with bow shocks are also jellyfish galaxies, i.e.\@ have extended gaseous tails. Conversely, jellyfish galaxies (score = 5 galaxies in \cref{fig:BowShocksStats}) exhibit bow shocks only in about 25 per cent of the cases. This is a smaller fraction, indeed, than the fraction of jellyfish galaxies that move through the ICM with super-sonic speeds (as from \cref{fig:Mach number vs Ram pressure}). In fact, the longevity of the bow shocks is probably shorter than the survival of the gaseous tails behind the direction of motion of a satellite, the former being more directly dependent on the instantaneous properties of the gas as it is compressed in the front by the satellite motion. Hence, it could be that galaxies dubbed as jellyfish galaxies but without evidence for a bow shock have, nevertheless, gone through a bow-shock phase in the recent past. At the same time, a few satellites that exhibit bow shocks are score = 0 ($1\leq$score$\leq$4) galaxies in 6 (31) per cent of the cases. These rare cases may indicate misclassification or simply a wrong timing in the physical process. 

In fact, if we repeat the same study not with randomly-projected galaxy images but in the optimal configurations (see Discussion in \cref{sec:optimal}), we find that both the total number of identified bow-shocks as well as the number of jellyfish galaxies that also exhibit a bow shock increase (by 60 and 40 per cent respectively). The optimal projection favors the identification of bow-shock features, especially in satellites that in a random configuration would have not been labeled as jellyfish otherwise.

In any case, in practice, if an observer was capable of identifying bow shocks in observed gas distributions, such bow shocks would, with very high probability, be accompanied by gaseous ram-pressure-stripped tails. While the latter could be harder to identify, bow shocks could be more easily identified via specific signatures, for example in spectroscopic gas-line ratios in the region occupied by the bow shock in front of the orbiting satellite \ky{or more directly in X-rays: this has been the case for bow shocks in the the hot intra-cluster medium of merging clusters \citep{Markevitch2002}, albeit admittedly at possibly different densities and thermodynamical properties of the affected intra-cluster gas.}

\section{Summary and Conclusions}
\label{sec:summary}

A growing number of intriguing galaxies that clearly undergo the effects of environmental gas stripping and hence are dubbed `jellyfish' due to their extended trailing gas structures, have been reported in recent years. In this work we have performed, for the first time, a quantitative study of such galaxies, their demographics in the general population of satellites, and their physical properties in the framework of a large-scale, uniform-volume, cosmological, gravity+MHD simulation. 

We have used the highest-resolution $111 \units{Mpc}$ box from the IllustrisTNG project to study the satellite galaxies in massive groups and clusters of $10^{13} \leq \MTC / \MSUN \leq 10^{14.6}\MSUN$ (about 180 hosts in total e.g.\@ at $z=0$). We have examined four snapshots at redshift \zeq{0,0.2,0.4}, and $0.6$ to accumulate a sample of 6\,066 satellites of stellar mass $\ge 10^{9.5}\MSUN$, 2\,610 of which contain some gas (see sample selection in \cref{sec:sample}). 

To identify jellyfish galaxies, a visual inspection was carried out by 5 independent Inspectors who classified images in {\it random projections} of the gaseous and stellar densities of the galaxies based on a set of pre-determined criteria. The main goal was to identify satellite galaxies with clear signatures of ram-pressure stripped gas, i.e.\@ with markedly asymmetric distributions of the projected gas in and around a galaxy, excluding cases of galaxy-galaxy interactions. No distinction has been made as to the gas phase for the inspection: all gas (hot, warm, ionized and star-forming) has been considered. 
The determinations of the Inspectors were tallied and those galaxies, 800 in number, which were unanimously agreed upon were classified as `jellyfish'. \\

Our quantitative findings from TNG100 can be summarized as follows:

\begin{itemize}
\item Within our sample, for satellites with galaxy stellar mass larger than $10^{9.5} \MSUN$ in hosts more massive than $10^{13}\MSUN$ at $z\leq 0.6$, we predict a non-negligible fraction of satellites to exhibit strong evidence for ram-pressure-stripped gas, namely about 13 (31) per cent of all luminous satellites (of satellites still containing gas; see \cref{tab:numbers}). Such evidence can be found in general in both disk-like as well as spheroidal-like galaxies, The stripped gas morphology can be either in the form of thin gas tails that extend to very large distances in comparison to the extension of the galaxy body or of multiple wakes that depart perpendicularly to a stellar and gaseous disk.\\

\item For our stellar mass-limited sample of satellite galaxies, the fraction of jellyfish galaxies increases with host mass, due to stronger environmental forces, and increases with decreasing satellite mass, as a result of the weaker gravitational binding within the satellites. This naturally leads to a decrease in jellyfish fraction with increasing satellite-to-host mass fraction. No strong trends for the jellyfish fraction with redshift have been found, except for a slightly larger fraction of ram-pressure stripped galaxies at the current epoch in the most massive hosts (\cref{fig:Jellyfish fraction}).\\

\item Jellyfish galaxies are found at nearly all radii within the host, but with a notable relative deficiency in the central regions $<0.2\RTC$. Jellyfish galaxies exhibit higher velocities than the general population of satellites. There are roughly equal numbers of jellyfish galaxies on infalling and outgoing trajectories, but of the outgoing satellites with gas, a larger fraction are found to be jellyfish galaxies (\cref{fig:phase-space diagram}).\\

\item TNG galaxies identified as `jellyfish' exhibit higher Mach numbers and experience stronger ram pressures than the general population of satellites with gas. The bulk values of these properties do not depend on whether the jellyfish galaxy is on an infalling or outgoing trajectory, while satellites in more massive hosts experience larger ram pressures than those in smaller hosts (\cref{fig:Mach number vs Ram pressure}). The majority of jellyfish galaxies move at supersonic speeds through the intra-cluster medium.\\	

\item The `tails' of jellyfish galaxies tend to be oriented opposite to the direction of motion (i.e.\@ extend at about 180$^\circ$ with respect to the satellite's velocity), and do \emph{not} favor the direction towards the center of the host (\cref{fig:Tail}).\\

\item Upon inspection of the time-line of each individual galaxy, we find that jellyfish galaxies are very recent infallers, with infall times that essentially never occur more than $2.5-3$ Gyrs in the past. In this respect, jellyfish galaxies are a very special subset of the general satellite populations: satellites that fell into their last host more than 8 Gyrs ago never retain their gas for such a long time (\cref{fig:infallTimes}).\\

\item About 30 per cent of actual jellyfish galaxies are missed or mis-classified because of the limitations of the random orientation, as we determine by inspecting the simulated galaxies in an optimal orientation, namely with the satellite bulk velocity vector lying on the plane of the image. Therefore, all numbers quoted in the paper via the random-projection analysis and any observationally-based estimates are lower limits of the population of galaxies undergoing ram-pressure stripping.\\

\item Finally, we find that bow shocks in front of satellite galaxies are good predictors for the presence of gaseous tails and wakes: among the studied galaxies with gas, 63 per cent of those that exhibit a bow shock in their surrounding gas in random projections are also jellyfish galaxies.\\

\end{itemize}

In conclusion, we have demonstrated that galaxies that orbit in massive galaxy groups and clusters and exhibit markedly asymmetric distributions of gas, or tails, develop naturally in cosmological hydrodynamical simulations. Following the tradition in the observational and theoretical literature, we have dubbed them ``jellyfish'' and we have associated their emergence with ram-pressure stripping. Such a working hypothesis has been confirmed by a series of findings, chiefly: 1. jellyfish galaxies are more frequent at intermediate and large cluster-centric distances; 2. on average they move at larger speeds and with higher Mach numbers in comparison to the general satellite population with the same minimum stellar mass: they are mostly supersonic; 3. they experience on average larger ram pressures; and 4. their tails are generally oriented opposite to the direction of motion, with no connection between the orientation of the tails and the position of the host centers. 
We have also found that the presence of gaseous tails correlates well with the presence of a bow shock in front of the observed galaxy: even if observations could not spatially resolve the ram-pressure stripped structures, the potential for identification based on emission line ratios expected for shock excitation could be used to find candidate jellyfish galaxies. Based on our simulation and analysis, the expected number of jellyfish galaxies within the virial radius of Virgo-like and Fornax-like clusters is on average 3.6 and 2 jellyfish galaxies per cluster, respectively\footnote{The imposed constraints for these predictions read: $1.4 \times 10^{14} \le \MTC / \MSUN \le 4 \times 10^{14}$ and $ 3 \times 10^{13} \le \MTC / \MSUN \le 7 \times 10^{13}$ for the total halo mass of Virgo and Fornax-like hosts; satellites stellar mass exceeding  $10^{9.5}\MSUN$ in stars.}. Such estimate can vary between 1 and 6 or 7 jellyfish examples, respectively, due to host-to-host variations.

We anticipate that this is the first paper in a series of jellyfish investigations with TNG. Future directions will include automatic classifications of jellyfish galaxies and characterization of the gas asymmetries; extension to the TNG300 and TNG50 simulations, with more massive hosts, higher redshifts, larger distances, and better resolved galaxies. Particularly, enhanced spatial and mass resolutions (e.g.\@ with TNG50) will enable us to study the properties of the stripped gas, e.g.\@ whether it is still gravitationally bound to the satellite and its star-forming, density, and temperature states; the morphologies of the stripped satellites; the balance between internal pressure and gravitational binding forces against the ram pressure that leads to gas removal; the quantification of observational biases and comparisons and predictions for observations via mock observations in specific gas phases (e.g.\@ HI, molecular or ionized gas), and ultimately the general implications for the overall picture of satellite quenching in high-density environments.

\section*{Acknowledgements}
The authors would like to thank Benedetta Vulcani, Bianca Poggianti, and Michele Fumagalli for helpful comments on a draft version of this paper and Yen-Ting Lin for useful conversations. It is also a pleasure to thank all the participants of the Ringberg workshop on ``Galaxy Evolution in Groups and Clusters at `low' Redshift, 2017'' for inspiring exchanges. 
SG, through the Flatiron Institute is supported by the Simons Foundation.
M.V. acknowledges support through an MIT RSC award, the support of the Alfred P. Sloan Foundation, and support by NASA ATP grant NNX17AG29G.
The flagship simulations of the IllustrisTNG project used in this work have been run on the HazelHen Cray XC40-system at the High Performance Computing Center Stuttgart as part of project GCS-ILLU of the Gauss centres for Supercomputing(GCS). Ancillary and test runs of the project were also run on the Stampede supercomputer at TACC/XSEDE (allocation AST140063), at the Hydra and Draco supercomputers at the Max Planck Computing and Data Facility, and on the MIT/Harvard computing facilities supported by FAS and MIT MKI.




\bibliographystyle{mnras} 
\bibliography{LEGACY/jellyfishesBib}

\begin{thebibliography}{}
\makeatletter
\relax
\def\mn@urlcharsother{\let\do\@makeother \do\$\do\&\do\#\do\^\do\_\do\%\do\~}
\def\mn@doi{\begingroup\mn@urlcharsother \@ifnextchar [ {\mn@doi@}
  {\mn@doi@[]}}
\def\mn@doi@[#1]#2{\def\@tempa{#1}\ifx\@tempa\@empty \href
  {http://dx.doi.org/#2} {doi:#2}\else \href {http://dx.doi.org/#2} {#1}\fi
  \endgroup}
\def\mn@eprint#1#2{\mn@eprint@#1:#2::\@nil}
\def\mn@eprint@arXiv#1{\href {http://arxiv.org/abs/#1} {{\tt arXiv:#1}}}
\def\mn@eprint@dblp#1{\href {http://dblp.uni-trier.de/rec/bibtex/#1.xml}
  {dblp:#1}}
\def\mn@eprint@#1:#2:#3:#4\@nil{\def\@tempa {#1}\def\@tempb {#2}\def\@tempc
  {#3}\ifx \@tempc \@empty \let \@tempc \@tempb \let \@tempb \@tempa \fi \ifx
  \@tempb \@empty \def\@tempb {arXiv}\fi \@ifundefined
  {mn@eprint@\@tempb}{\@tempb:\@tempc}{\expandafter \expandafter \csname
  mn@eprint@\@tempb\endcsname \expandafter{\@tempc}}}

\bibitem[\protect\citeauthoryear{{Alpaslan} et~al.,}{{Alpaslan}
  et~al.}{2015}]{Gama12015}
{Alpaslan} M.,  et~al., 2015, \mn@doi [\mnras] {10.1093/mnras/stv1176}, \href
  {http://adsabs.harvard.edu/abs/2015MNRAS.451.3249A} {451, 3249}

\bibitem[\protect\citeauthoryear{{Bah{\'e}} \& {McCarthy}}{{Bah{\'e}} \&
  {McCarthy}}{2015}]{Bahe2015}
{Bah{\'e}} Y.~M.,  {McCarthy} I.~G.,  2015, \mn@doi [\mnras]
  {10.1093/mnras/stu2293}, \href
  {http://adsabs.harvard.edu/abs/2015MNRAS.447..969B} {447, 969}

\bibitem[\protect\citeauthoryear{{Bah{\'e}}, {McCarthy}, {Crain}  \&
  {Theuns}}{{Bah{\'e}} et~al.}{2012}]{Bahe2012}
{Bah{\'e}} Y.~M.,  {McCarthy} I.~G.,  {Crain} R.~A.,   {Theuns} T.,  2012,
  \mn@doi [\mnras] {10.1111/j.1365-2966.2012.21292.x}, \href
  {http://adsabs.harvard.edu/abs/2012MNRAS.424.1179B} {424, 1179}

\bibitem[\protect\citeauthoryear{{Bah{\'e}}, {McCarthy}, {Balogh}  \&
  {Font}}{{Bah{\'e}} et~al.}{2013}]{Bahe2013}
{Bah{\'e}} Y.~M.,  {McCarthy} I.~G.,  {Balogh} M.~L.,   {Font} A.~S.,  2013,
  \mn@doi [\mnras] {10.1093/mnras/stt109}, \href
  {http://adsabs.harvard.edu/abs/2013MNRAS.430.3017B} {430, 3017}

\bibitem[\protect\citeauthoryear{Bah{\'{e}} et~al.,}{Bah{\'{e}}
  et~al.}{2016}]{Bahe2016TheHoles}
Bah{\'{e}} Y.~M.,  et~al., 2016, \mn@doi [MNRAS] {10.1093/mnras/stv2674}, 456,
  1115

\bibitem[\protect\citeauthoryear{Barnes et~al.,}{Barnes
  et~al.}{2018}]{Barnes2018EnhancingConduction}
Barnes D.~J.,  et~al., 2018, eprint arXiv:1805.04109

\bibitem[\protect\citeauthoryear{Bekki}{Bekki}{2009}]{Bekki2009Ram-pressureEnvironments}
Bekki K.,  2009, \mn@doi [MNRAS] {10.1111/j.1365-2966.2009.15431.x}, 399, 2221

\bibitem[\protect\citeauthoryear{Bekki}{Bekki}{2014}]{Bekki2014GalacticClusters}
Bekki K.,  2014, \mn@doi [MNRAS] {10.1093/mnras/stt2216}, 438, 444

\bibitem[\protect\citeauthoryear{Boselli \& Gavazzi}{Boselli \&
  Gavazzi}{2006}]{Boselli2006EnvironmentalClusters}
Boselli A.,  Gavazzi G.,  2006, \mn@doi [The Publications of the Astronomical
  Society of the Pacific, Volume 118, Issue 842, pp. 517-559.]
  {10.1086/500691}, 118, 517

\bibitem[\protect\citeauthoryear{{Boselli} et~al.,}{{Boselli}
  et~al.}{2018}]{Boselli2018}
{Boselli} A.,  et~al., 2018, \mn@doi [\aap] {10.1051/0004-6361/201732407},
  \href {http://adsabs.harvard.edu/abs/2018A%26A...614A..56B} {614, A56}

\bibitem[\protect\citeauthoryear{{Buck}, {Macci{\`o}}, {Dutton}, {Obreja}  \&
  {Frings}}{{Buck} et~al.}{2018}]{Buck2018}
{Buck} T.,  {Macci{\`o}} A.~V.,  {Dutton} A.~A.,  {Obreja} A.,   {Frings} J.,
  2018, preprint, \href {http://adsabs.harvard.edu/abs/2018arXiv180404667B} {}
  (\mn@eprint {arXiv} {1804.04667})

\bibitem[\protect\citeauthoryear{Chua, Pillepich, Rodriguez-Gomez,
  Vogelsberger, Bird  \& Hernquist}{Chua
  et~al.}{2017}]{Chua2017SubhaloVariation}
Chua K. T.~E.,  Pillepich A.,  Rodriguez-Gomez V.,  Vogelsberger M.,  Bird S.,
   Hernquist L.,  2017, \mn@doi [Monthly Notices of the Royal Astronomical
  Society] {10.1093/mnras/stx2238}, 472, 4343

\bibitem[\protect\citeauthoryear{Chung, van Gorkom, Kenney  \& Vollmer}{Chung
  et~al.}{2007}]{Chung2007VirgoTails}
Chung A.,  van Gorkom J.~H.,  Kenney J. D.~P.,   Vollmer B.,  2007, \mn@doi
  [ApJ] {10.1086/518034}, 659, L115

\bibitem[\protect\citeauthoryear{{Cortese} et~al.,}{{Cortese}
  et~al.}{2007}]{Cortese2007}
{Cortese} L.,  et~al., 2007, \mn@doi [\mnras]
  {10.1111/j.1365-2966.2006.11369.x}, \href
  {http://adsabs.harvard.edu/abs/2007MNRAS.376..157C} {376, 157}

\bibitem[\protect\citeauthoryear{Couch \& Sharples}{Couch \&
  Sharples}{1987}]{Couch1987A0.31}
Couch W.~J.,  Sharples R.~M.,  1987, \mn@doi [MNRAS] {10.1093/mnras/229.3.423},
  229, 423

\bibitem[\protect\citeauthoryear{{Crain} et~al.,}{{Crain}
  et~al.}{2009}]{Crain2009}
{Crain} R.~A.,  et~al., 2009, \mn@doi [\mnras]
  {10.1111/j.1365-2966.2009.15402.x}, \href
  {http://adsabs.harvard.edu/abs/2009MNRAS.399.1773C} {399, 1773}

\bibitem[\protect\citeauthoryear{Crain et~al.,}{Crain
  et~al.}{2015}]{Crain2015TheVariations}
Crain R.~A.,  et~al., 2015, \mn@doi [MNRAS] {10.1093/mnras/stv725}, 450, 1937

\bibitem[\protect\citeauthoryear{{Davis}, {Efstathiou}, {Frenk}  \&
  {White}}{{Davis} et~al.}{1985}]{Davis1985}
{Davis} M.,  {Efstathiou} G.,  {Frenk} C.~S.,   {White} S.~D.~M.,  1985,
  \mn@doi [\apj] {10.1086/163168}, \href
  {http://adsabs.harvard.edu/abs/1985ApJ...292..371D} {292, 371}

\bibitem[\protect\citeauthoryear{Dressler}{Dressler}{1980}]{Dressler1980AGalaxies}
Dressler A.,  1980, \mn@doi [ApJ Supplement Series] {DOI: 10.1086/190663}, 42,
  565

\bibitem[\protect\citeauthoryear{Dressler \& Gunn}{Dressler \&
  Gunn}{1983}]{Dressler1983SpectroscopyCluster}
Dressler A.,  Gunn J.~E.,  1983, \mn@doi [ApJ] {10.1086/161093}, 270, 7

\bibitem[\protect\citeauthoryear{Ebeling, Stephenson  \& Edge}{Ebeling
  et~al.}{2014}]{Ebeling2014Jellyfish:Clusters}
Ebeling H.,  Stephenson L.~N.,   Edge A.~C.,  2014, \mn@doi [ApJ]
  {10.1088/2041-8205/781/2/L40}, 781, L40

\bibitem[\protect\citeauthoryear{Ellingson, Lin, Yee  \& Carlberg}{Ellingson
  et~al.}{2001}]{Ellingson2001TheInfall}
Ellingson E.,  Lin H.,  Yee H. K.~C.,   Carlberg R.~G.,  2001, \mn@doi [ApJ]
  {10.1086/318423}, 547, 609

\bibitem[\protect\citeauthoryear{{Fattahi}, {Navarro}, {Frenk}, {Oman},
  {Sawala}  \& {Schaller}}{{Fattahi} et~al.}{2018}]{Fattahi2018}
{Fattahi} A.,  {Navarro} J.~F.,  {Frenk} C.~S.,  {Oman} K.~A.,  {Sawala} T.,
  {Schaller} M.,  2018, \mn@doi [\mnras] {10.1093/mnras/sty408}, \href
  {http://adsabs.harvard.edu/abs/2018MNRAS.476.3816F} {476, 3816}

\bibitem[\protect\citeauthoryear{{Forman}, {Schwarz}, {Jones}, {Liller}  \&
  {Fabian}}{{Forman} et~al.}{1979}]{Forman1979}
{Forman} W.,  {Schwarz} J.,  {Jones} C.,  {Liller} W.,   {Fabian} A.~C.,  1979,
  \mn@doi [\apjl] {10.1086/183103}, \href
  {http://adsabs.harvard.edu/abs/1979ApJ...234L..27F} {234, L27}

\bibitem[\protect\citeauthoryear{{Fossati}, {Fumagalli}, {Boselli}, {Gavazzi},
  {Sun}  \& {Wilman}}{{Fossati} et~al.}{2016}]{Fossati2016}
{Fossati} M.,  {Fumagalli} M.,  {Boselli} A.,  {Gavazzi} G.,  {Sun} M.,
  {Wilman} D.~J.,  2016, \mn@doi [\mnras] {10.1093/mnras/stv2400}, \href
  {http://adsabs.harvard.edu/abs/2016MNRAS.455.2028F} {455, 2028}

\bibitem[\protect\citeauthoryear{{Fumagalli}, {Fossati}, {Hau}, {Gavazzi},
  {Bower}, {Sun}  \& {Boselli}}{{Fumagalli} et~al.}{2014}]{Fumagalli2014}
{Fumagalli} M.,  {Fossati} M.,  {Hau} G.~K.~T.,  {Gavazzi} G.,  {Bower} R.,
  {Sun} M.,   {Boselli} A.,  2014, \mn@doi [\mnras] {10.1093/mnras/stu2092},
  \href {http://adsabs.harvard.edu/abs/2014MNRAS.445.4335F} {445, 4335}

\bibitem[\protect\citeauthoryear{{Genel}}{{Genel}}{2016}]{Genel2016}
{Genel} S.,  2016, \mn@doi [\apj] {10.3847/0004-637X/822/2/107}, \href
  {http://adsabs.harvard.edu/abs/2016ApJ...822..107G} {822, 107}

\bibitem[\protect\citeauthoryear{Genel et~al.,}{Genel
  et~al.}{2014}]{Genel2014IntroducingTime}
Genel S.,  et~al., 2014, \mn@doi [MNRAS] {10.1093/mnras/stu1654}, 445, 175

\bibitem[\protect\citeauthoryear{Genel et~al.,}{Genel
  et~al.}{2018}]{Genel2018ARelations}
Genel S.,  et~al., 2018, eprint arXiv:1807.07084

\bibitem[\protect\citeauthoryear{Gullieuszik et~al.,}{Gullieuszik
  et~al.}{2017}]{Gullieuszik2017GASP.JO204}
Gullieuszik M.,  et~al., 2017, \mn@doi [The Astrophysical Journal]
  {10.3847/1538-4357/aa8322}, 846, 27

\bibitem[\protect\citeauthoryear{Gunn \& Gott}{Gunn \&
  Gott}{1972}]{Gunn1972OnEvolution}
Gunn J.~E.,  Gott J.~Richard I.,  1972, \mn@doi [ApJ] {10.1086/151605}, 176, 1

\bibitem[\protect\citeauthoryear{Haynes, Giovanelli  \& Chincarini}{Haynes
  et~al.}{1984}]{Haynes1984TheGalaxies}
Haynes M.~P.,  Giovanelli R.,   Chincarini G.~L.,  1984, \mn@doi [ARAA]
  {10.1146/annurev.aa.22.090184.002305}, 22, 445

\bibitem[\protect\citeauthoryear{He{\ss} \& Springel}{He{\ss} \&
  Springel}{2012}]{He2012GasStudy}
He{\ss} S.,  Springel V.,  2012, \mn@doi [Monthly Notices of the Royal
  Astronomical Society] {10.1111/j.1365-2966.2012.21819.x}, 426, 3112

\bibitem[\protect\citeauthoryear{{J{\'a}chym}, {Combes}, {Cortese}, {Sun}  \&
  {Kenney}}{{J{\'a}chym} et~al.}{2014}]{Jachym2014}
{J{\'a}chym} P.,  {Combes} F.,  {Cortese} L.,  {Sun} M.,   {Kenney} J.~D.~P.,
  2014, \mn@doi [\apj] {10.1088/0004-637X/792/1/11}, \href
  {http://adsabs.harvard.edu/abs/2014ApJ...792...11J} {792, 11}

\bibitem[\protect\citeauthoryear{{J{\'a}chym} et~al.,}{{J{\'a}chym}
  et~al.}{2017}]{Jachym2017}
{J{\'a}chym} P.,  et~al., 2017, \mn@doi [\apj] {10.3847/1538-4357/aa6af5},
  \href {http://adsabs.harvard.edu/abs/2017ApJ...839..114J} {839, 114}

\bibitem[\protect\citeauthoryear{Jaff{\'{e}}, Smith, Candlish, Poggianti, Sheen
   \& Verheijen}{Jaff{\'{e}} et~al.}{2015}]{Jaffe2015BUDHIESGalaxies}
Jaff{\'{e}} Y.~L.,  Smith R.,  Candlish G.~N.,  Poggianti B.~M.,  Sheen Y.-K.,
   Verheijen M. A.~W.,  2015, \mn@doi [MNRAS] {10.1093/mnras/stv100}, 448, 1715

\bibitem[\protect\citeauthoryear{Jaff{\'{e}} et~al.,}{Jaff{\'{e}}
  et~al.}{2018}]{Jaffe2018GASP.Clusters}
Jaff{\'{e}} Y.~L.,  et~al., 2018, \mn@doi [MNRAS] {10.1093/mnras/sty500}, 476,
  4753

\bibitem[\protect\citeauthoryear{Jung, Choi, Wong, Kimm, Chung  \& Yi}{Jung
  et~al.}{2018}]{Jung2018}
Jung S.~L.,  Choi H.,  Wong O.~I.,  Kimm T.,  Chung A.,   Yi S.~K.,  2018, The
  Astrophysical Journal, 865, 156

\bibitem[\protect\citeauthoryear{{Kauffmann}, {White}, {Heckman}, {M{\'e}nard},
  {Brinchmann}, {Charlot}, {Tremonti}  \& {Brinkmann}}{{Kauffmann}
  et~al.}{2004}]{Kauffmann2004}
{Kauffmann} G.,  {White} S.~D.~M.,  {Heckman} T.~M.,  {M{\'e}nard} B.,
  {Brinchmann} J.,  {Charlot} S.,  {Tremonti} C.,   {Brinkmann} J.,  2004,
  \mn@doi [\mnras] {10.1111/j.1365-2966.2004.08117.x}, \href
  {http://adsabs.harvard.edu/abs/2004MNRAS.353..713K} {353, 713}

\bibitem[\protect\citeauthoryear{{Kenney}, {van Gorkom}  \& {Vollmer}}{{Kenney}
  et~al.}{2004}]{Kenney2004}
{Kenney} J.~D.~P.,  {van Gorkom} J.~H.,   {Vollmer} B.,  2004, \mn@doi [\aj]
  {10.1086/420805}, \href {http://adsabs.harvard.edu/abs/2004AJ....127.3361K}
  {127, 3361}

\bibitem[\protect\citeauthoryear{{Larson}, {Tinsley}  \& {Caldwell}}{{Larson}
  et~al.}{1980}]{Larson1980}
{Larson} R.~B.,  {Tinsley} B.~M.,   {Caldwell} C.~N.,  1980, \mn@doi [\apj]
  {10.1086/157917}, \href {http://adsabs.harvard.edu/abs/1980ApJ...237..692L}
  {237, 692}

\bibitem[\protect\citeauthoryear{{Marinacci} et~al.,}{{Marinacci}
  et~al.}{2018}]{Marinacci2018}
{Marinacci} F.,  et~al., 2018, \mn@doi [\mnras] {10.1093/mnras/sty2206}, \href
  {http://adsabs.harvard.edu/abs/2018MNRAS.480.5113M} {480, 5113}

\bibitem[\protect\citeauthoryear{{Markevitch}, {Gonzalez}, {David},
  {Vikhlinin}, {Murray}, {Forman}, {Jones}  \& {Tucker}}{{Markevitch}
  et~al.}{2002}]{Markevitch2002}
{Markevitch} M.,  {Gonzalez} A.~H.,  {David} L.,  {Vikhlinin} A.,  {Murray} S.,
   {Forman} W.,  {Jones} C.,   {Tucker} W.,  2002, \mn@doi [\apj]
  {10.1086/339619}, \href
  {https://ui.adsabs.harvard.edu/#abs/2002ApJ...567L..27M} {567, L27}

\bibitem[\protect\citeauthoryear{McPartland, Ebeling, Roediger  \&
  Blumenthal}{McPartland et~al.}{2016}]{McPartland2016Jellyfish:Clusters}
McPartland C.,  Ebeling H.,  Roediger E.,   Blumenthal K.,  2016, \mn@doi
  [MNRAS] {10.1093/mnras/stv2508}, 455, 2994

\bibitem[\protect\citeauthoryear{{Mistani} et~al.,}{{Mistani}
  et~al.}{2016}]{Mistani2016}
{Mistani} P.~A.,  et~al., 2016, \mn@doi [\mnras] {10.1093/mnras/stv2435}, \href
  {http://adsabs.harvard.edu/abs/2016MNRAS.455.2323M} {455, 2323}

\bibitem[\protect\citeauthoryear{Moore, Katz, Lake, Dressler  \& Oemler}{Moore
  et~al.}{1995}]{Moore1995GalaxyGalaxies}
Moore B.,  Katz N.,  Lake G.,  Dressler A.,   Oemler A.,  1995, \mn@doi
  [Nature] {10.1038/379613a0}, 379, 613

\bibitem[\protect\citeauthoryear{Naiman et~al.,}{Naiman
  et~al.}{2018}]{Naiman2018FirstEuropium}
Naiman J.~P.,  et~al., 2018, \mn@doi [MNRAS] {10.1093/mnras/sty618}, 477, 1206

\bibitem[\protect\citeauthoryear{Nelson et~al.,}{Nelson
  et~al.}{2018a}]{Nelson2018FirstBimodality}
Nelson D.,  et~al., 2018a, \mn@doi [MNRAS] {10.1093/mnras/stx3040}, 475, 624

\bibitem[\protect\citeauthoryear{Nelson et~al.,}{Nelson
  et~al.}{2018b}]{Nelson2018TheIllustrisTNG}
Nelson D.,  et~al., 2018b, \mn@doi [MNRAS] {10.1093/mnras/sty656}, 477, 450

\bibitem[\protect\citeauthoryear{{Owen}, {Keel}, {Wang}, {Ledlow}  \&
  {Morrison}}{{Owen} et~al.}{2006}]{Owen2006}
{Owen} F.~N.,  {Keel} W.~C.,  {Wang} Q.~D.,  {Ledlow} M.~J.,   {Morrison}
  G.~E.,  2006, \mn@doi [\aj] {10.1086/500573}, \href
  {http://adsabs.harvard.edu/abs/2006AJ....131.1974O} {131, 1974}

\bibitem[\protect\citeauthoryear{{Peng} et~al.,}{{Peng}
  et~al.}{2010}]{Peng2010}
{Peng} Y.-j.,  et~al., 2010, \mn@doi [\apj] {10.1088/0004-637X/721/1/193},
  \href {http://adsabs.harvard.edu/abs/2010ApJ...721..193P} {721, 193}

\bibitem[\protect\citeauthoryear{Pillepich et~al.,}{Pillepich
  et~al.}{2018a}]{Pillepich2018SimulatingModel}
Pillepich A.,  et~al., 2018a, \mn@doi [MNRAS] {10.1093/mnras/stx2656}, 473,
  4077

\bibitem[\protect\citeauthoryear{Pillepich et~al.,}{Pillepich
  et~al.}{2018b}]{Pillepich2018FirstGalaxies}
Pillepich A.,  et~al., 2018b, \mn@doi [MNRAS] {10.1093/mnras/stx3112}, 475, 648

\bibitem[\protect\citeauthoryear{Poggianti et~al.,}{Poggianti
  et~al.}{2016}]{Poggianti2016JELLYFISHREDSHIFT}
Poggianti B.~M.,  et~al., 2016, \mn@doi [The Astronomical Journal]
  {10.3847/0004-6256/151/3/78}, 151, 78

\bibitem[\protect\citeauthoryear{{Poggianti} et~al.,}{{Poggianti}
  et~al.}{2017}]{Poggianti2017}
{Poggianti} B.~M.,  et~al., 2017, \mn@doi [\apj] {10.3847/1538-4357/aa78ed},
  \href {http://adsabs.harvard.edu/abs/2017ApJ...844...48P} {844, 48}

\bibitem[\protect\citeauthoryear{{Randall}, {Nulsen}, {Forman}, {Jones},
  {Machacek}, {Murray}  \& {Maughan}}{{Randall} et~al.}{2008}]{Randall2008}
{Randall} S.,  {Nulsen} P.,  {Forman} W.~R.,  {Jones} C.,  {Machacek} M.,
  {Murray} S.~S.,   {Maughan} B.,  2008, \mn@doi [\apj] {10.1086/592324}, \href
  {http://adsabs.harvard.edu/abs/2008ApJ...688..208R} {688, 208}

\bibitem[\protect\citeauthoryear{Rhee, Smith, Choi, Yi, Jaff{\'{e}}, Candlish
  \& S{\'{a}}nchez-J{\'{a}}nssen}{Rhee et~al.}{2017}]{Rhee2017Phase-spaceLoss}
Rhee J.,  Smith R.,  Choi H.,  Yi S.~K.,  Jaff{\'{e}} Y.,  Candlish G.,
  S{\'{a}}nchez-J{\'{a}}nssen R.,  2017, \mn@doi [ApJ]
  {10.3847/1538-4357/aa6d6c}, 843, 128

\bibitem[\protect\citeauthoryear{Rodriguez-Gomez et~al.,}{Rodriguez-Gomez
  et~al.}{2015}]{Rodriguez-Gomez2015TheModels}
Rodriguez-Gomez V.,  et~al., 2015, \mn@doi [Monthly Notices of the Royal
  Astronomical Society] {10.1093/mnras/stv264}, 449, 49

\bibitem[\protect\citeauthoryear{Roediger \& Br{\"{u}}ggen}{Roediger \&
  Br{\"{u}}ggen}{2007}]{Roediger2007RamDisc}
Roediger E.,  Br{\"{u}}ggen M.,  2007, \mn@doi [MNRAS]
  {10.1111/j.1365-2966.2007.12241.x}, 380, 1399

\bibitem[\protect\citeauthoryear{Roediger et~al.,}{Roediger
  et~al.}{2015a}]{Roediger2015STRIPPEDELLIPTICALS}
Roediger E.,  et~al., 2015a, \mn@doi [ApJ] {10.1088/0004-637X/806/1/103}, 806,
  103

\bibitem[\protect\citeauthoryear{Roediger et~al.,}{Roediger
  et~al.}{2015b}]{Roediger2015STRIPPEDM89}
Roediger E.,  et~al., 2015b, \mn@doi [ApJ] {10.1088/0004-637X/806/1/104}, 806,
  104

\bibitem[\protect\citeauthoryear{{Sales} et~al.,}{{Sales}
  et~al.}{2015}]{Sales2015}
{Sales} L.~V.,  et~al., 2015, \mn@doi [\mnras] {10.1093/mnrasl/slu173}, \href
  {http://adsabs.harvard.edu/abs/2015MNRAS.447L...6S} {447, L6}

\bibitem[\protect\citeauthoryear{{Schaal} \& {Springel}}{{Schaal} \&
  {Springel}}{2015}]{Schaal2015}
{Schaal} K.,  {Springel} V.,  2015, \mn@doi [\mnras] {10.1093/mnras/stu2386},
  \href {http://adsabs.harvard.edu/abs/2015MNRAS.446.3992S} {446, 3992}

\bibitem[\protect\citeauthoryear{Schaal et~al.,}{Schaal
  et~al.}{2016}]{Schaal2016ShockUniverse}
Schaal K.,  et~al., 2016, \mn@doi [MNRAS] {10.1093/mnras/stw1587}, 461, 4441

\bibitem[\protect\citeauthoryear{Schaye et~al.,}{Schaye
  et~al.}{2015}]{Schaye2015TheEnvironments}
Schaye J.,  et~al., 2015, \mn@doi [MNRAS] {10.1093/mnras/stu2058}, 446, 521

\bibitem[\protect\citeauthoryear{Sheen et~al.,}{Sheen et~al.}{2017}]{Sheen2017}
Sheen Y.-K.,  et~al., 2017, \mn@doi [ApJ] {10.3847/2041-8213/aa6d79}, 840, L7

\bibitem[\protect\citeauthoryear{{Sijacki}, {Vogelsberger}, {Kere{\v s}},
  {Springel}  \& {Hernquist}}{{Sijacki} et~al.}{2012}]{Sijacki2012}
{Sijacki} D.,  {Vogelsberger} M.,  {Kere{\v s}} D.,  {Springel} V.,
  {Hernquist} L.,  2012, \mn@doi [\mnras] {10.1111/j.1365-2966.2012.21466.x},
  \href {http://adsabs.harvard.edu/abs/2012MNRAS.424.2999S} {424, 2999}

\bibitem[\protect\citeauthoryear{{Simpson}, {Grand}, {G{\'o}mez}, {Marinacci},
  {Pakmor}, {Springel}, {Campbell}  \& {Frenk}}{{Simpson}
  et~al.}{2018}]{Simpson2018}
{Simpson} C.~M.,  {Grand} R.~J.~J.,  {G{\'o}mez} F.~A.,  {Marinacci} F.,
  {Pakmor} R.,  {Springel} V.,  {Campbell} D.~J.~R.,   {Frenk} C.~S.,  2018,
  \mn@doi [\mnras] {10.1093/mnras/sty774}, \href
  {http://adsabs.harvard.edu/abs/2018MNRAS.478..548S} {478, 548}

\bibitem[\protect\citeauthoryear{Smith, Duc, Candlish, Fellhauer, Sheen  \&
  Gibson}{Smith et~al.}{2013}]{Smith2013TheGalaxies}
Smith R.,  Duc P.~A.,  Candlish G.~N.,  Fellhauer M.,  Sheen Y.-K.,   Gibson
  B.~K.,  2013, \mn@doi [MNRAS] {10.1093/mnras/stt1619}, 436, 839

\bibitem[\protect\citeauthoryear{Springel}{Springel}{2010}]{Springel2010}
Springel V.,  2010, \mn@doi [MNRAS] {10.1111/j.1365-2966.2009.15715.x}, 401,
  791

\bibitem[\protect\citeauthoryear{{Springel}, {White}, {Tormen}  \&
  {Kauffmann}}{{Springel} et~al.}{2001}]{SpringelPopulating0}
{Springel} V.,  {White} S.~D.~M.,  {Tormen} G.,   {Kauffmann} G.,  2001,
  \mn@doi [\mnras] {10.1046/j.1365-8711.2001.04912.x}, \href
  {http://adsabs.harvard.edu/abs/2001MNRAS.328..726S} {328, 726}

\bibitem[\protect\citeauthoryear{Springel et~al.,}{Springel
  et~al.}{2018}]{Springel2018FirstClustering}
Springel V.,  et~al., 2018, \mn@doi [MNRAS] {10.1093/mnras/stx3304}, 475, 676

\bibitem[\protect\citeauthoryear{Steinhauser, Schindler  \&
  Springel}{Steinhauser et~al.}{2016}]{Steinhauser2016SimulationsInteractions}
Steinhauser D.,  Schindler S.,   Springel V.,  2016, \mn@doi [Astronomy {\&}
  Astrophysics] {10.1051/0004-6361/201527705}, 591, A51

\bibitem[\protect\citeauthoryear{{Su} et~al.,}{{Su} et~al.}{2017}]{Su2017}
{Su} Y.,  et~al., 2017, \mn@doi [\apj] {10.3847/1538-4357/834/1/74}, \href
  {http://adsabs.harvard.edu/abs/2017ApJ...834...74S} {834, 74}

\bibitem[\protect\citeauthoryear{{Tonnesen} \& {Bryan}}{{Tonnesen} \&
  {Bryan}}{2010}]{Tonnesen2010}
{Tonnesen} S.,  {Bryan} G.~L.,  2010, \mn@doi [\apj]
  {10.1088/0004-637X/709/2/1203}, \href
  {http://adsabs.harvard.edu/abs/2010ApJ...709.1203T} {709, 1203}

\bibitem[\protect\citeauthoryear{{Tonnesen} \& {Stone}}{{Tonnesen} \&
  {Stone}}{2014}]{Tonnesen2014}
{Tonnesen} S.,  {Stone} J.,  2014, \mn@doi [\apj]
  {10.1088/0004-637X/795/2/148}, \href
  {http://adsabs.harvard.edu/abs/2014ApJ...795..148T} {795, 148}

\bibitem[\protect\citeauthoryear{Toomre \& Toomre}{Toomre \&
  Toomre}{1972}]{Toomre1972GalacticTails}
Toomre A.,  Toomre J.,  1972, \mn@doi [The Astrophysical Journal]
  {10.1086/151823}, 178, 623

\bibitem[\protect\citeauthoryear{Torrey, Vogelsberger, Genel, Sijacki, Springel
   \& Hernquist}{Torrey et~al.}{2014}]{Torrey2014AValidation}
Torrey P.,  Vogelsberger M.,  Genel S.,  Sijacki D.,  Springel V.,   Hernquist
  L.,  2014, \mn@doi [Monthly Notices of the Royal Astronomical Society]
  {10.1093/mnras/stt2295}, 438, 1985

\bibitem[\protect\citeauthoryear{Torrey et~al.,}{Torrey
  et~al.}{2018}]{Torrey2018SimilarRelation}
Torrey P.,  et~al., 2018, \mn@doi [MNRAS: Letters] {10.1093/mnrasl/sly031},
  477, L16

\bibitem[\protect\citeauthoryear{Vogelsberger, Genel, Sijacki, Torrey, Springel
   \& Hernquist}{Vogelsberger et~al.}{2013}]{Vogelsberger2013APhysics}
Vogelsberger M.,  Genel S.,  Sijacki D.,  Torrey P.,  Springel V.,   Hernquist
  L.,  2013, \mn@doi [Monthly Notices of the Royal Astronomical Society]
  {10.1093/mnras/stt1789}, 436, 3031

\bibitem[\protect\citeauthoryear{Vogelsberger et~al.,}{Vogelsberger
  et~al.}{2014a}]{Vogelsberger2014IntroducingUniverse}
Vogelsberger M.,  et~al., 2014a, \mn@doi [MNRAS] {10.1093/mnras/stu1536}, 444,
  1518

\bibitem[\protect\citeauthoryear{Vogelsberger et~al.,}{Vogelsberger
  et~al.}{2014b}]{Vogelsberger2014PropertiesSimulation}
Vogelsberger M.,  et~al., 2014b, \mn@doi [Nature] {10.1038/nature13316}, 509,
  177

\bibitem[\protect\citeauthoryear{Vogelsberger et~al.,}{Vogelsberger
  et~al.}{2018}]{Vogelsberger2018TheSimulations}
Vogelsberger M.,  et~al., 2018, \mn@doi [MNRAS] {10.1093/mnras/stx2955}, 474,
  2073

\bibitem[\protect\citeauthoryear{{Vollmer}, {Cayatte}, {Boselli}, {Balkowski}
  \& {Duschl}}{{Vollmer} et~al.}{1999}]{Vollmer1999}
{Vollmer} B.,  {Cayatte} V.,  {Boselli} A.,  {Balkowski} C.,   {Duschl} W.~J.,
  1999, \aap, \href {http://adsabs.harvard.edu/abs/1999A%26A...349..411V} {349,
  411}

\bibitem[\protect\citeauthoryear{{Vulcani} et~al.,}{{Vulcani}
  et~al.}{2015}]{Vulcani2015}
{Vulcani} B.,  et~al., 2015, \mn@doi [\apj] {10.1088/0004-637X/814/2/161},
  \href {http://adsabs.harvard.edu/abs/2015ApJ...814..161V} {814, 161}

\bibitem[\protect\citeauthoryear{Vulcani et~al.,}{Vulcani
  et~al.}{2018}]{Vulcani2018GASPFormation}
Vulcani B.,  et~al., 2018, \mn@doi [Monthly Notices of the Royal Astronomical
  Society] {10.1093/mnras/sty2095}, 480, 3152

\bibitem[\protect\citeauthoryear{{Weinberger} et~al.,}{{Weinberger}
  et~al.}{2017}]{Weinberger2017}
{Weinberger} R.,  et~al., 2017, \mn@doi [\mnras] {10.1093/mnras/stw2944}, \href
  {http://adsabs.harvard.edu/abs/2017MNRAS.465.3291W} {465, 3291}

\bibitem[\protect\citeauthoryear{Weinberger et~al.,}{Weinberger
  et~al.}{2018}]{Weinberger2018SupermassiveSimulation}
Weinberger R.,  et~al., 2018, \mn@doi [MNRAS] {10.1093/mnras/sty1733}, 479,
  4056

\bibitem[\protect\citeauthoryear{Yagi et~al.,}{Yagi
  et~al.}{2010}]{Yagi2010ACluster}
Yagi M.,  et~al., 2010, \mn@doi [Astronomical Journal]
  {10.1088/0004-6256/140/6/1814}, 140, 1814

\bibitem[\protect\citeauthoryear{{Yoshida} et~al.,}{{Yoshida}
  et~al.}{2002}]{Yoshida2002}
{Yoshida} M.,  et~al., 2002, \mn@doi [\apj] {10.1086/338353}, \href
  {http://adsabs.harvard.edu/abs/2002ApJ...567..118Y} {567, 118}

\bibitem[\protect\citeauthoryear{Zinger, Dekel, Kravtsov  \& Nagai}{Zinger
  et~al.}{2018}]{Zinger2018QuenchingClusters}
Zinger E.,  Dekel A.,  Kravtsov A.~V.,   Nagai D.,  2018, \mn@doi [MNRAS]
  {10.1093/mnras/stx3329}, 475, 3654

\makeatother
\end{thebibliography}




\appendix

\section{Demographics of our simulated samples}
\begin{figure*}
	\centering
	\adjustbox{trim={0.043\width} {0.024\height} {0.003\width} {0.4\height},clip}%
	{\includegraphics[width=1.05\textwidth,height=1.05\textheight, keepaspectratio=true]{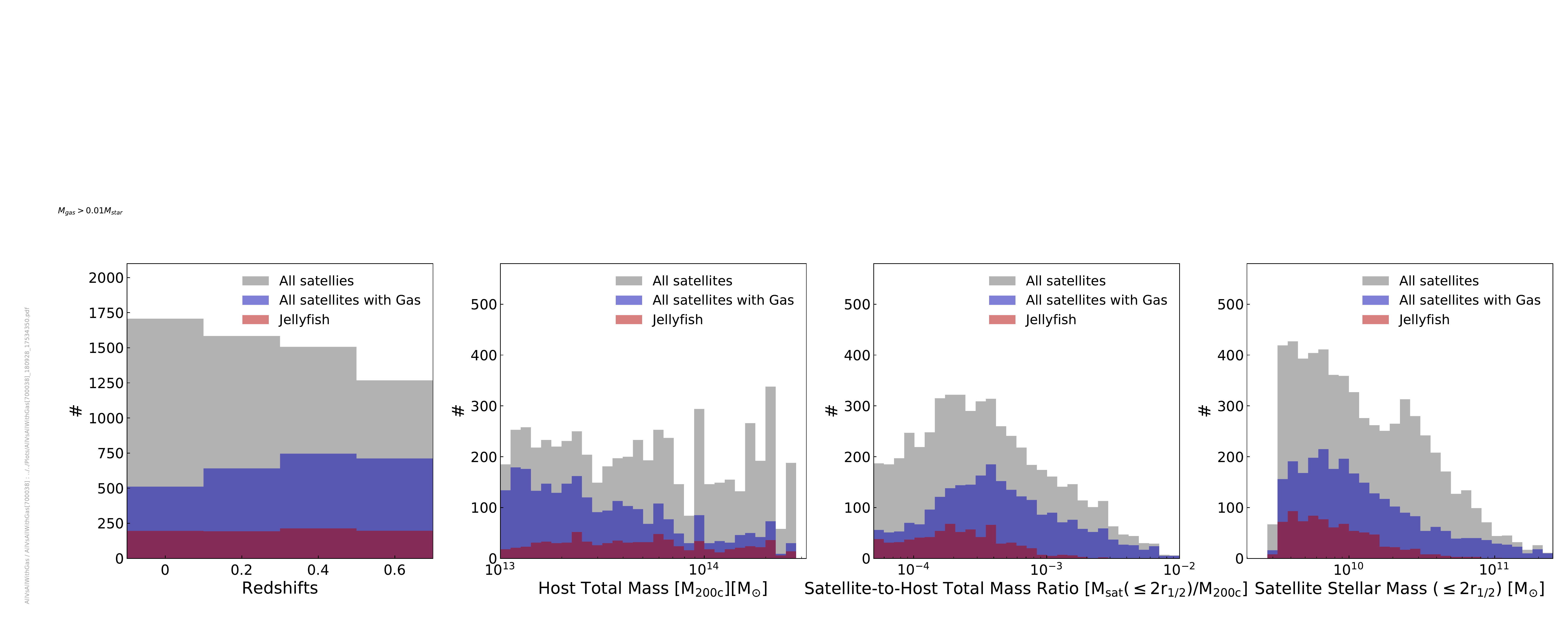}}
   \caption{Distributions of all satellites (gray), all satellites with gas (blue), and jellyfish galaxies (magenta) in our TNG100 sample. Jellyfish galaxies are those with score = 5 according to our visual classification. In all panels, unless otherwise stated we include all galaxies in our sample, with stellar mass larger than $10^{9.5}\MSUN$, in hosts with total mass larger than $10^{13}\MSUN$ and smaller than $\sim 10^{14.6}\MSUN$, and at all the four snapshots considered, i.e.\@ $z \simeq 0.0, 0.2, 0.4$, and 0.6. Trends are shown as function of redshift, host mass, host-to-satellite total mass ratio, and satellite stellar mass, from left to right.}
    \label{fig:Comparison_All_Gas_Jellyfish}
\end{figure*}

\begin{figure*}
   	\centering
    \adjustbox{trim={0.0\width} {0.01\height} {0.01\width} {0.0\height},clip}%
	{\includegraphics[width=0.51\columnwidth]{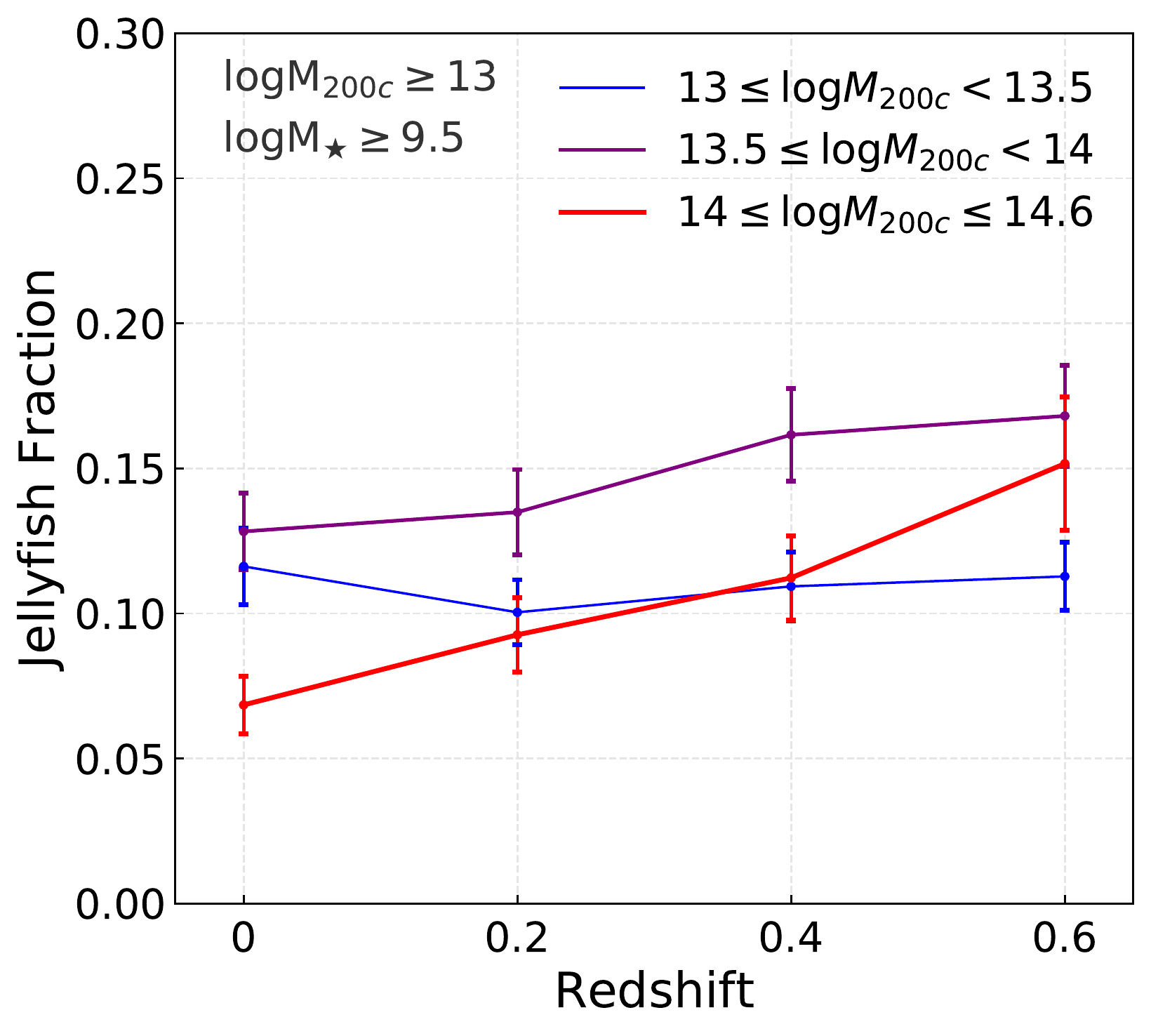}}
    \adjustbox{trim={0.0\width} {0.01\height} {0.01\width} {0.0\height},clip}%
    {\includegraphics[width=0.51\columnwidth]{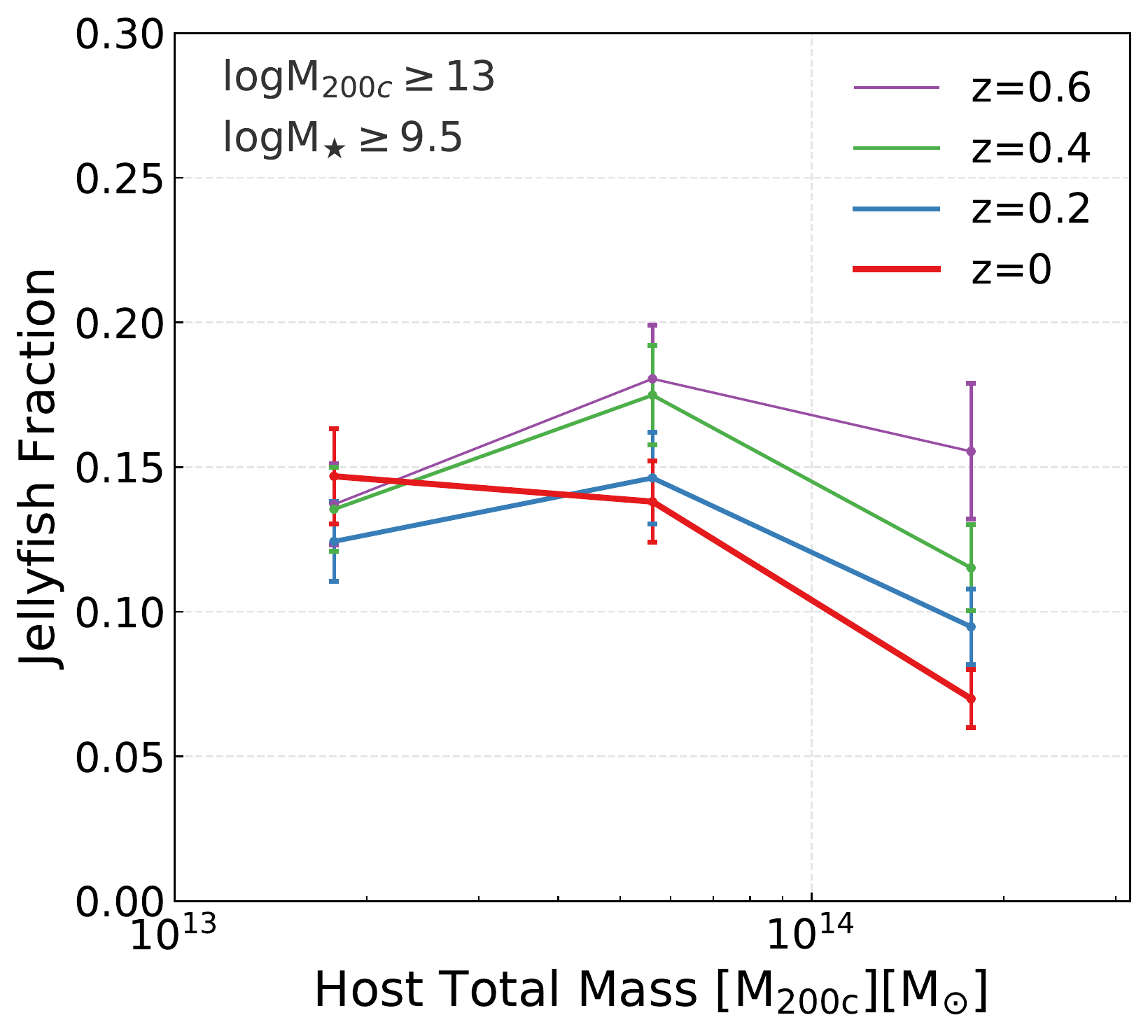}}
    \adjustbox{trim={0.0\width} {0.01\height} {0.01\width} {0.0\height},clip}%
    {\includegraphics[width=0.53\columnwidth]{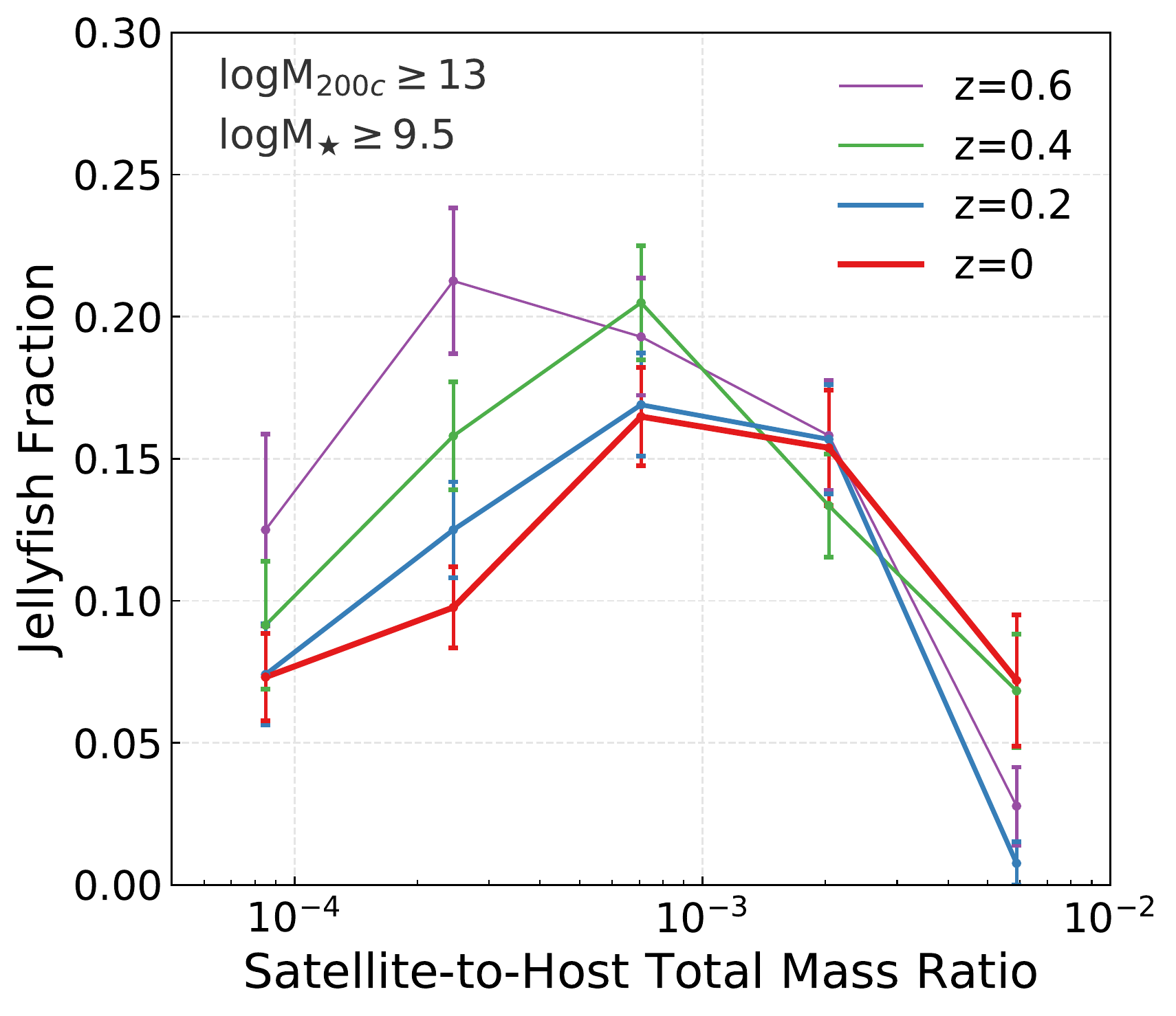}}
    \adjustbox{trim={0.0\width} {0.01\height} {0.03\width} {0.0\height},clip}%
    {\includegraphics[width=0.53\columnwidth]{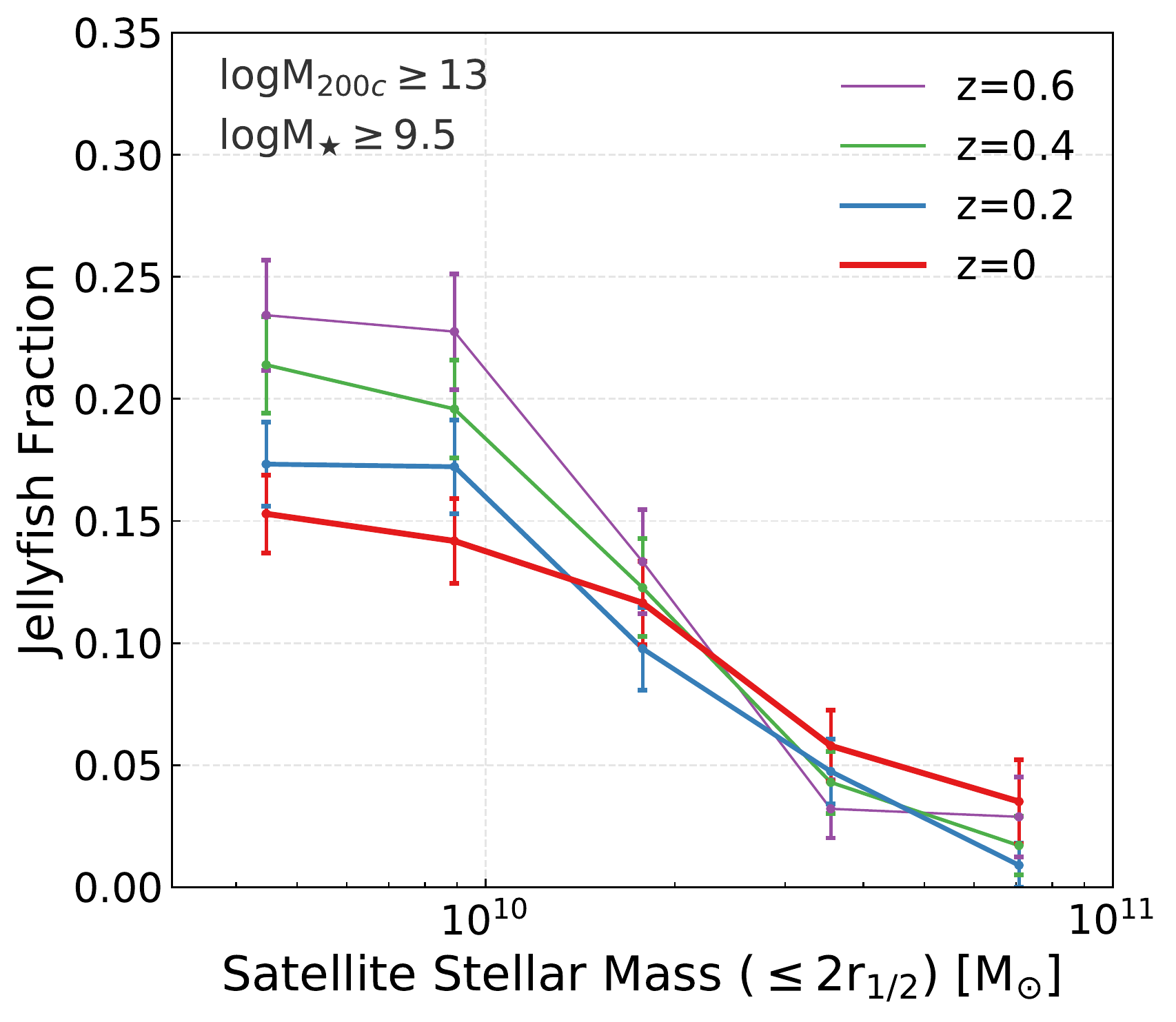}}
    \caption{The fractions of jellyfish with respect to the whole parent sample, i.e.\@ all satellites with at least $10^{9.5}\MSUN$ in stars, irrespective of their gas content. This is analogous to \cref{fig:Jellyfish fraction}, but there the fractions are given within the sample of satellites with at least 1 per cent of gas fraction. Trends are shown as a function of redshift, host mass, host-to-satellite total mass ($\leq 2 \, \rm r_{\rm 1/2}$) ratio, and satellite stellar mass, from left to right.
    No monotonic trends in the jellyfish fractions can be seen as function of host mass and host-to-satellite mass ratio, in contrast to the case where the fraction is taken by considering only galaxies with at least some gas as in \cref{fig:Jellyfish fraction}.}
    \label{fig:Jellyfish fraction All}
\end{figure*}

\begin{figure}
    \adjustbox{trim={0.08\width} {0.06\height} {0.01\width} {0.06\height},clip}%
	{\includegraphics[width=0.4\textwidth,height=0.4\textheight, keepaspectratio=true]{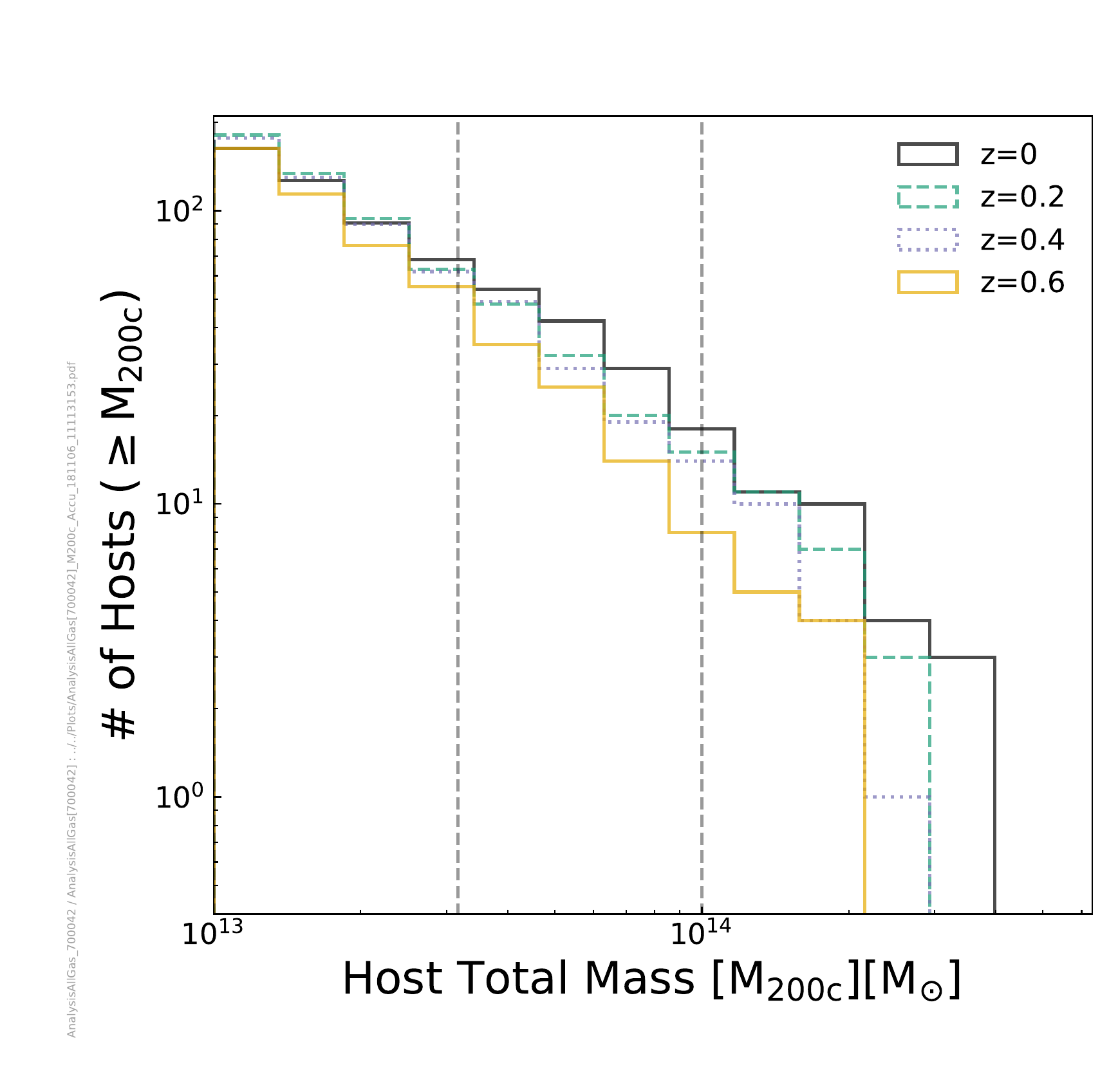}}
   \caption{\ky{Cumulative distribution of hosts as function of their total mass. Vertical dashed lines identify the bins of host mass adopted e.g. in the top-left panel of \cref{fig:Jellyfish fraction} and in the left-most panel of \cref{fig:Jellyfish fraction All}. As we are working with a volume-limited simulation, the number of massive hosts increases with decreasing redshift, as it is apparent when comparing the \zeq{0.6} and \zeq{0} host mass functions, especially at the highest-mass end ($\geq 10^{14}\MSUN$). }
   }
    \label{fig:Host mass distribution}
\end{figure}

In \cref{fig:Comparison_All_Gas_Jellyfish}, we spell out the demographics of all the samples of simulated galaxies extracted from the TNG100 simulation and discussed in the paper. Histograms of all satellites, all satellites with gas, and jellyfish galaxies are provided, as a function of redshift, host mass, satellite stellar mass, and satellite-to-host mass ratio. 
Of particular interest are the deviations between the distributions of all satellites vs.\@ all satellites with gas, of which we comment in the main text.
\cref{fig:Jellyfish fraction All} provides the frequency of jellyfish galaxies from the parent sample of all satellite galaxies (i.e.\@ objects with non-vanishing stellar mass) irrespective of their gas content. These fractions are obtained from the ratios between the histograms of jellyfish and of all satellites of \cref{fig:Comparison_All_Gas_Jellyfish}. The jellyfish fractions with respect to all satellites are different from those with respect to all satellites with gas that are shown in \cref{fig:Jellyfish fraction}.
\ky{\cref{fig:Host mass distribution} shows the cumulative number distribution of hosts in bins of redshift. The distribution of host total mass overall shifts to larger numbers of more massive hosts at lower redshifts: namely, the number of hosts with mass larger than $10^{14} \MSUN$ is generally higher at lower redshifts and it is highest at $z=0$.}

\section{Jellyfish galaxy}

\begin{figure*}
	\includegraphics[width=\textwidth,height=0.95\textheight, keepaspectratio=true]{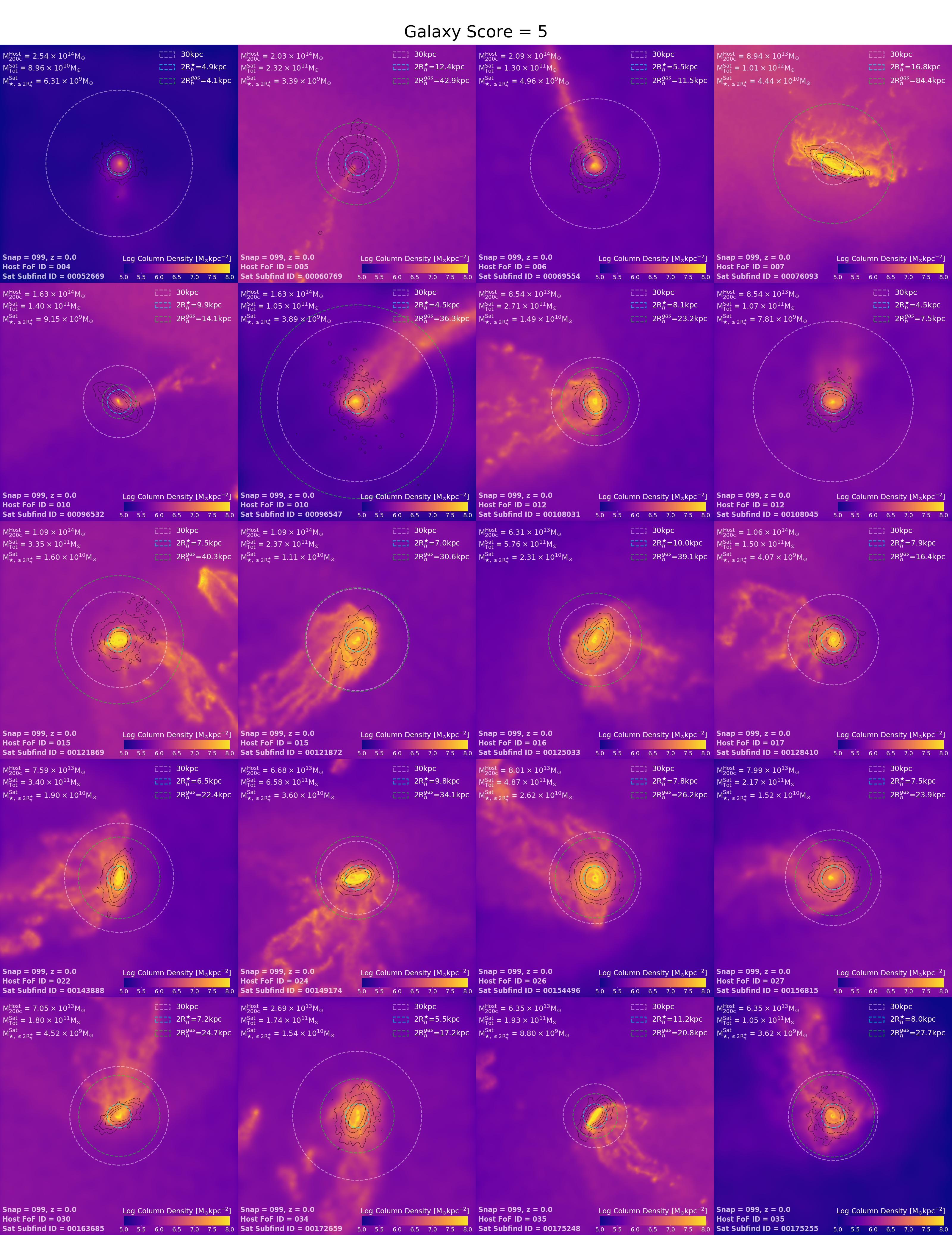}
   \caption{Gallery of a selection of our jellyfish galaxies, i.e.\@ satellite galaxies as determined in this work, in random projections.}
    \label{fig:Jellyfish Galley 1}
\end{figure*}

\begin{figure*}
	\includegraphics[width=\textwidth,height=0.95\textheight, keepaspectratio=true]{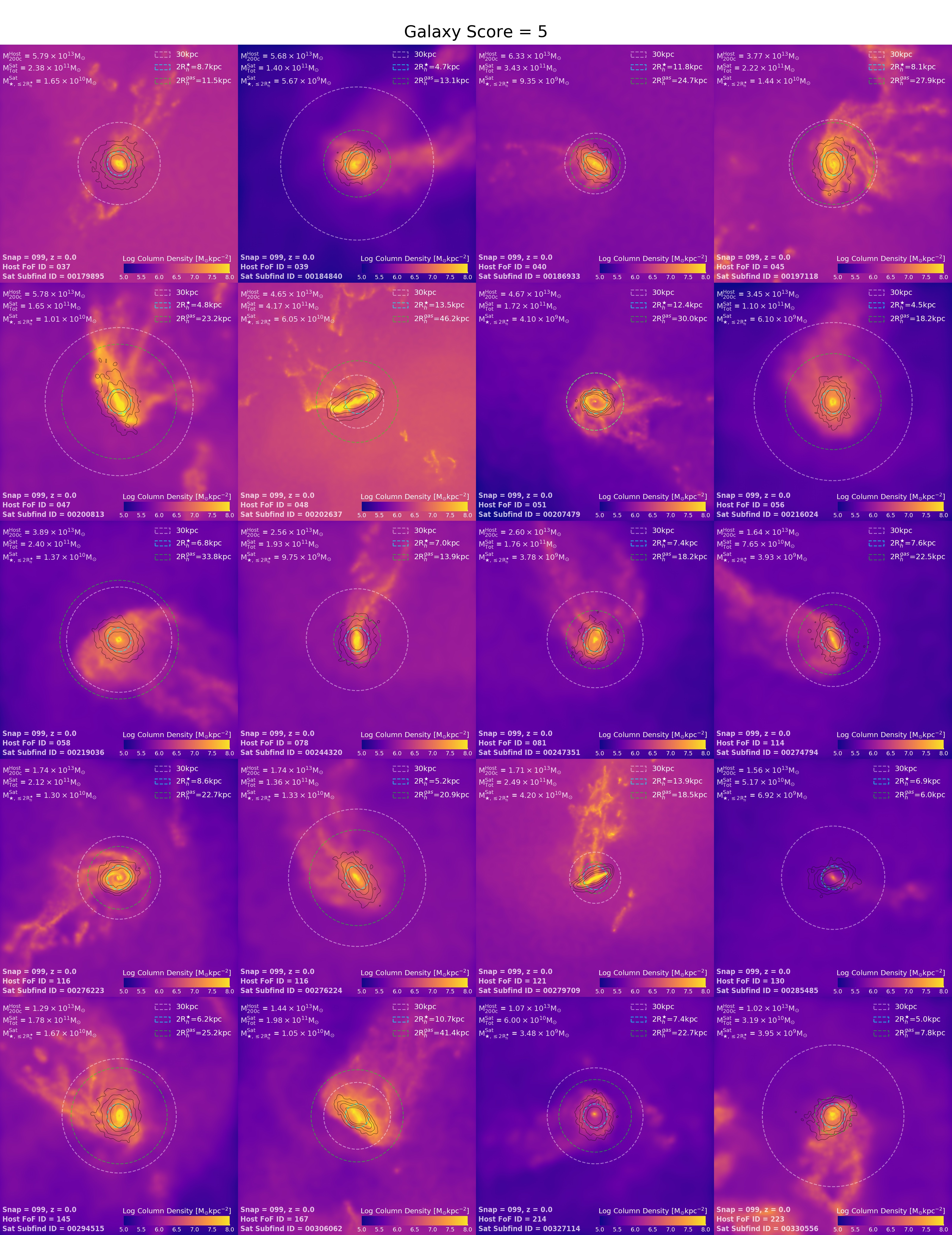}
   \caption{Cont: Gallery of a selection of our jellyfish galaxies, i.e.\@ satellite galaxies as determined in this work.}
    \label{fig:Jellyfish Galley 2}
\end{figure*}

A random set of TNG galaxies dubbed as `jellyfish' (score = 5 galaxies) is showcased in \cref{fig:Jellyfish Galley 1} and \cref{fig:Jellyfish Galley 2}, in random projections. The color maps denote gas column densities with the configurations adopted and described in \cref{sec:3visual}.

\bsp	
\label{lastpage}
\end{document}